 \documentclass{article}

\usepackage{epsfig} 
\usepackage{amsmath}
\usepackage{amsfonts}
\usepackage{graphicx}

 \usepackage{amssymb}

\date{\today}

\newcommand{\insertplot}[5]{\begin{figure}
 \hfill\hbox to 0.05in{\vbox to #5in{\vfill
 \inputplot{#1}{#4}{#5}}\hfill}
 \hfill\vspace{-.1in}
 \caption{#2}\label{#3}
 \end{figure}}
 \newcommand{\inputplot}[3]{
 \special{ps: plotfile #1}
}
\newcounter{fig}

\renewcommand{\c}{\gamma}
\renewcommand{\d}{\delta}  
\newcommand{\g}{\nu}

\renewcommand{\t}{\theta}

\newcommand{\ee}{\end{equation}}
\newcommand{\eea}{\end{eqnarray}}
\newcommand{\be}{\begin{equation}}
\newcommand{\bea}{\begin{eqnarray}}

\newcommand{\e}{\mu}

\begin{document}

\title{ 
Gravitating vortons
as ring solitons in general relativity
} 
  
\author{{\large Jutta Kunz,} 
{\large Eugen Radu} 
and {\large Bintoro Subagyo}  \\ 
 {\small  Institut f\"ur Physik, Universit\"at Oldenburg, Postfach 2503
D-26111 Oldenburg, Germany} }
\maketitle

\begin{abstract}  
Vortons can be viewed as (flat space-) field theory analogs of  black rings in general
relativity.
They are made from loops of  
vortices,
being sustained against collapse by the centrifugal force.
In this work we discuss such configurations
in the global version of Witten's 
$U(1)\times U(1)$ theory.
We first consider solutions in a flat spacetime background
and show their nonuniqueness.
The inclusion of gravity leads to new features.
In particular, an ergoregion can occur. 
Also, similar to boson stars, we show that the vortons exist 
only in a limited frequency range.
The coupling to gravity gives rise to a spiral-like frequency 
dependence of the mass and charge.
New solutions of the model describing  'semitopological vortons' and  'di-vortons'
are also discussed. 
\end{abstract}

\section{Introduction}

Vortons are certainly among the most interesting particle-like solutions in 
field theory.
Within a rather informal definition,
they are 
 ring-like solitons
in some nonlinear field theory which is supposed to possess also string-like 
vortex solutions.
Then the heuristic construction of a vorton involves
taking a piece of finite length of such a vortex,
bending it and gluing its ends
to form a loop.
The resulting configuration --the vorton, necessarily caries a nonzero angular momentum,
such that the tension of the loop is compensated by the  centrifugal force.
This stationary ring does not radiate classically, 
and at large distances looks like a point particle with 
quantized charge and angular momentum. 

Vortons were initially proposed
25 years ago  
\cite{Davis:1988jq,Davis:1988ip,Davis:1988ij}
in the context of the local $U(1)\times U(1)$ theory of superconducting cosmic strings
of Witten \cite{Witten:1984eb}. 
This model possesses vortex solutions 
with a non-dissipative current flowing along the string due 
 to a complex scalar field which vanishes asymptotically.
Vortons are closed loops of such superconducting strings,
carrying both current and charge.
However, vortons may exist in a variety of other theories as well,
ranging from cosmology
\cite{Brandenberger:1996zp}
to QCD \cite{Buckley:2002mx} and condensed matter systems \cite{Battye:2001ec}.
One can  speculate that vortons may perhaps even exist 
in the Standard Model of Particle Physics.

The study of vortons has been a fruitful avenue of research since its inception. 
Some of the ideas which appeared in this context
have found interesting applications in other domains of physics.
For example, vortons share the same heuristic construction
with black rings in $d\geq 5$ general relativity 
\cite{Emparan:2001wn,Emparan:2007wm},
with vortices playing the role of black strings there.
Moreover, a general formalism proposed by Carter in \cite{Carter:1989xk}
to describe the properties of large radius vortons
has found recently beautiful applications and  generalizations in the physics
of blackfolds in higher dimensional Einstein gravity  \cite{Emparan:2009at}.

At the same time, different from black rings in general relativity\footnote{Note,
however, that the
black rings are known in closed form for $d=5$ spacetime dimensions only \cite{Emparan:2001wn}.  
The  $d>5$ black rings were studied based on 
analytical solutions valid in the large radius limit \cite{Emparan:2007wm} 
or on numerical nonperturbative solutions \cite{Kleihaus:2012xh}. 
},
there are no closed form vorton solutions.
Moreover, even the numerical study of such configurations is  difficult, due to the existence 
of multiple scales in the problem and the complicated 
behaviour of the matter functions.

In fact, to our knowledge there is no explicit construction
in the literature of vorton solutions
in a model containing gauge fields.
However, by now, there are a number of independent studies of 
vortons in the global version of Witten's $U(1)\times U(1)$ model, 
which contains only two complex scalars.
The numerical study of such solutions has started 
with the work of Lemperier and Shellard in Ref. \cite{Lemperiere:2003yt}.
There, the full hyperbolic evolution problem for the
fields, with the initial data corresponding to a vortex loop has been considered.
The results in \cite{Lemperiere:2003yt}
 suggested the existence of
vortons which are stable against perturbations. 
However, not so many
details  on the properties of such configurations 
were given there.
 In addition,  Lemperier and Shellard did not consider precisely the rigid version of
Witten's model, but, in order to improve the numerics, added 
an extra Q-ball type interaction term.

The existence of stationary, axially symmetric vortons in Witten''s model
has been reported in  Ref. \cite{Radu:2008pp}.
There, the solutions have been constructed 
by solving a set of elliptic partial differential equations with 
suitable boundary conditions\footnote{An independent construction of these solutions has
been reported in \cite{Grandclement:2009ju},
based on a very different treatment of the numerical problem.}.
The configurations in \cite{Radu:2008pp}
possess a nonzero value of the angular momentum and have ring profiles,
describing thick vortons.
Also, the results there point towards the existence of
some analogies between vortons
and spinning Q-balls
in a simpler model with a single scalar field with a non-renormalizable potential 
\cite{Kleihaus:2005me,volkov}.

Large radius vortons have been reported in \cite{Battye:2008mm}
for a different choice of the parameters of Witten's model
and a different numerical approach than the one employed in \cite{Radu:2008pp}.
Ref. \cite{Battye:2008mm} has investigated
also the numerical stability of such configurations with respect to
non-axially symmetric perturbations.
The results there show that (at least) thin vortons
possess a pinching instability.

The main purpose of this work is to enlarge our understanding of 
vortons by including the effects of gravity, at a nonperturbative level.
This is a legitimate task since 
the geometry in the vorton's core 
may exhibit large deviations from the flat spacetime limit.
Also, as shown by a number of other models  
\cite{Kaup:1968zz,Ruffini:1969qy,Volkov:1998cc},
the inclusion of gravity usually leads to a variety of new effects.
Indeed, as we shall see, this is also the case for vortons.
For example, they may
possess an ergoregion, with the associated instability.
Also, gravitating vortons exhibit 
a rather different type of frequency dependence. 
 
However, as a first step towards this task, we shall start with a
study of vortons in a flat spacetime background.
The solutions are easier to study
in this limit, while
some of their basic properties remain unaffected by the inclusion of gravity.
 Based on the heuristic construction of vortons as pieces of vortices,
 we propose a classification
 of the resulting objects in terms of an integer $k$
 related to the winding along a torus generator.
 Then it follows that vortons can be viewed as
 the second member ($k=1$) of a general family of  
 solutions.
 An unexpected result reported here is 
 the nonuniqueness of vortons,
 that is the existence of 
different solutions
for the same input parameters.
Apart from solutions already known in the literature,
 new types of configurations,
in particular those describing di-vortons,
are also shown here.

This paper is organized as follows. 
In Section 2 we recall Witten's gravitating global model, 
and give a precise formulation of the problem
we propose to solve.
We continue in Section 3 with a discussion of the properties of vortons
in a flat spacetime background. 
The effects of gravity on the properties of vortons are discussed in Section 4.
Two kinds of solutions are described there, corresponding to  
`semitopological' vortons and `true' vortons.
Section 5 gives our conclusions together with a discussion of possible generalizations
and open problems.
In the Appendices we present the system of
differential equations solved in this work (A),  the components of the
energy-momentum tensor (B), 
and the spherically symmetric limit of our solutions (C).

\section{The model}

\subsection{The action and field equations}

Witten's ungauged model of superconducting cosmic strings \cite{Witten:1984eb} 
 contains two complex scalar fields  $\phi$ and $\sigma$,
with the following Lagrangian density\footnote{Note that, as opposed to 
previous studies 
\cite{Lemperiere:2003yt,Radu:2008pp,Battye:2008mm}, 
we are using in this work a mostly plus signature for the spacetime metric.}
\begin{eqnarray} 
\label{Ls}
 L_s=\partial_\mu \phi^\ast \partial^\mu \phi + 
\partial_\mu \sigma^\ast \partial^\mu \sigma +U(|\phi|,|\sigma|),
\end{eqnarray}

 \setlength{\unitlength}{1cm}
\begin{picture}(8,6)
\label{fig-potential}
\put(-0.5,0.0){\epsfig{file=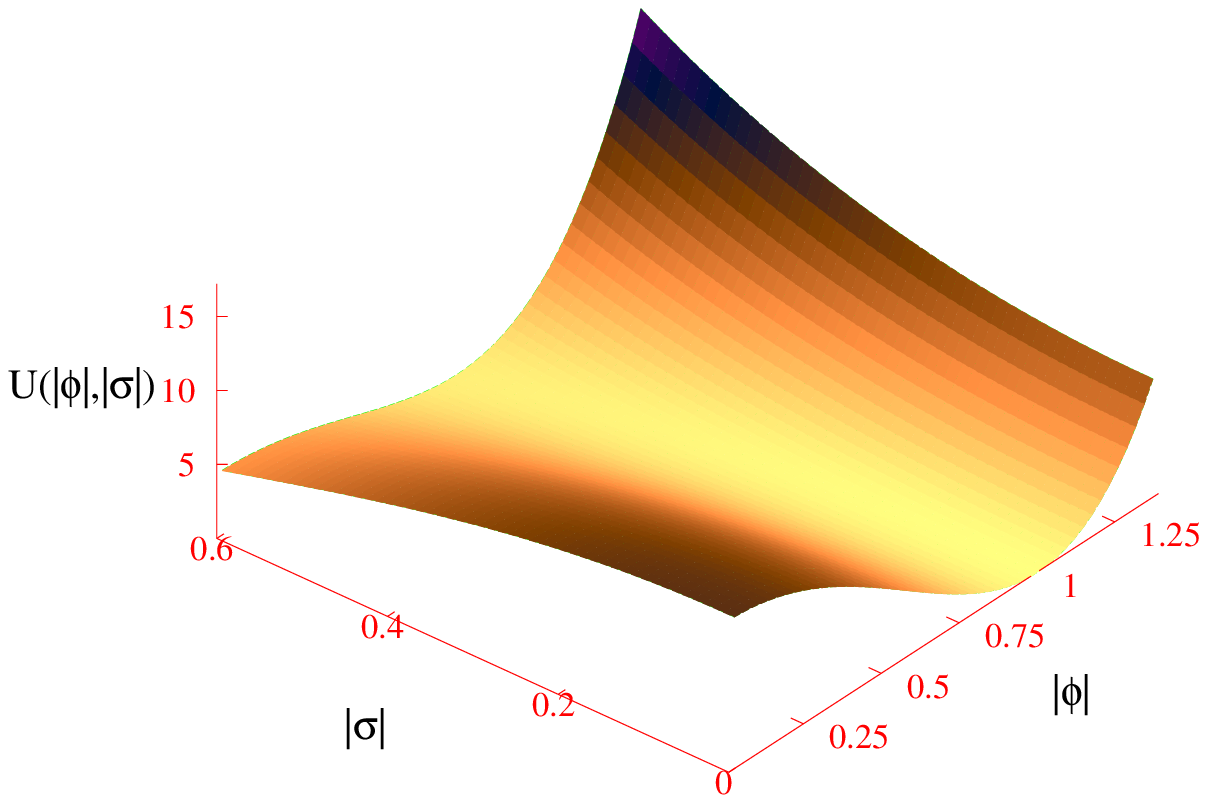,width=8cm}}
\put(8,0.0){\epsfig{file=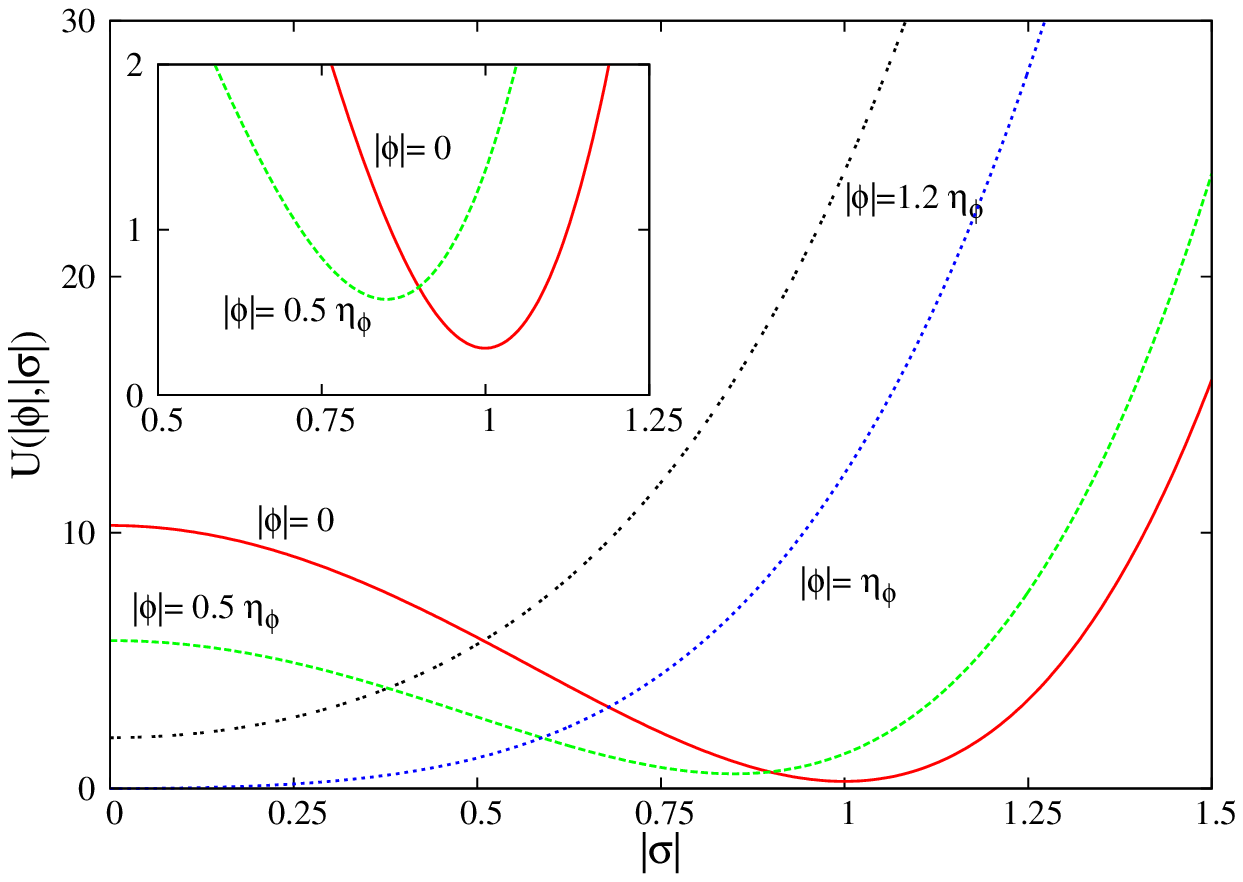,width=8cm}}
\end{picture}
\\
\\
{\small {\bf Figure 1.} The potential $U(|\phi|,|\sigma|)$ of the model is shown for a set of parameters
with $\eta_\phi=\eta_\sigma=1$, $\lambda_\phi=41$, $\lambda_\sigma=40$, $\gamma=22$. 
One can notice the existence of a single global minimum at 
$|\phi|=\eta_\phi$, $|\sigma|=0$.} 
\vspace{0.5cm}
\\
where $U(|\phi|,|\sigma|)$ is the scalar fields potential 
\begin{eqnarray}                                                          
\label{potential}
U(|\phi|,|\sigma|)=
\frac{1}{4}\lambda_\phi (|\phi|^2-\eta_\phi^2)^2
+\frac{1}{4}\lambda_\sigma |\sigma|^2(|\sigma|^2-2\eta_\sigma^2) 
+\gamma |\phi|^2 |\sigma|^2.
\end{eqnarray}

In the above potential, $\lambda_\phi,\lambda_\sigma,\eta_\phi,\eta_\sigma$ and $\gamma$ 
are positive constants which are not specified apriori.
A minimal value of the potential is achieved for  
$|\phi|=\eta_\phi$, $|\sigma|=0$, in which case
$U(|\phi|,|\sigma|)=0$ (see Figure 1).
 This minimum is globally unique and stable if 
\begin{eqnarray}  
\label{conds}
4\gamma^2>\lambda_\sigma\lambda_\phi~~{\rm and}~~\gamma \eta_\phi^2>\frac12\,\lambda_\sigma\eta_\sigma^2.
\end{eqnarray}
 
Then the perturbative spectrum 
of the field excitations around the vacuum state $|\phi|=\eta_\phi$, $|\sigma|=0$ 
consists of two massless 
Goldstone particles, corresponding to excitations of the phases of the fields,
and of two Higgs bosons with the masses
\begin{eqnarray} 
\label{MASSES}
M_\phi=\sqrt{\lambda_\phi}\eta_\phi,~~~~~
M_\sigma=\sqrt{\gamma \eta_\phi^2-\frac12\,\lambda_\sigma\eta_\sigma^2}.
\end{eqnarray}

When coupling to gravity, the 
  dynamics of the solutions  
is governed by the following action
\begin{eqnarray} 
\label{action}
S=\int d^4 x \sqrt{-g}\left(\frac{R}{16\pi G}-L_s \right)~,
\end{eqnarray}
where $g$ is determinant of the spacetime metric $g_{\mu\nu}$, 
$G$ is Newton's constant and $R$
is the curvature scalar.
%
As usual in general relativity, variation of the action with respect to the metric
leads to the Einstein equations
\begin{equation}
E_{\mu\nu}= R_{\mu\nu}-\frac{1}{2}g_{\mu\nu}R - 8 \pi G~ T_{\mu\nu}=0
\ , 
\label{Einstein}
\end{equation}
with 
the stress-energy tensor $T_{\mu\nu}$
\begin{eqnarray}
\label{Tik}
T_{\mu \nu} &=& \phantom{-} g_{\mu \nu} L_s 
-2 \frac{ \partial L_s}{\partial g^{\mu\nu}}
 \\
&=&
\left( \phi_{, \, \mu}^*\phi_{, \, \nu}
+\phi_{, \, \nu}^*\phi_{, \, \mu}
+\sigma_{, \, \mu}^*\sigma_{, \, \nu}
+\sigma_{, \, \nu}^*\sigma_{, \, \mu}
 \right )
-g_{\mu\nu} \big[\partial_\alpha \phi^\ast \partial^\alpha \phi + 
\partial_\alpha \sigma^\ast \partial^\alpha \sigma +U(|\phi|,|\sigma|) \big] .
\label{tmunu} \end{eqnarray}
Variation of (\ref{action}) with respect to the scalar fields
leads to the matter equations,
\begin{equation}
\left(\Box+\frac{\partial U }{\partial\left|\phi\right|^2}\right)\phi=0 \ ,
~~
\left(\Box+\frac{\partial U }{\partial\left|\sigma\right|^2}\right)\sigma=0 \ ,
\label{scalar-eqs}
\end{equation}
where $\Box$ represents the covariant d'Alembert operator. 

We close this part by noting the existence of several interesting limits 
of the considered model.
First, for a real field $\phi$ and  a choice 
$\lambda_\sigma=0$ in the potential (\ref{potential}),
one recovers the model proposed by Friedberg, Lee and Sirlin (FLS hereafter)  in \cite{Friedberg:1976me}, with
\begin{eqnarray} 
\label{Ls-FLS}
 L_s=\partial_\mu \phi  \partial^\mu \phi + 
\partial_\mu \sigma^\ast \partial^\mu \sigma +\frac{1}{4}\lambda_\phi (|\phi|^2-\eta_\phi^2)^2
+\gamma |\phi|^2 |\sigma|^2.
\end{eqnarray}
As shown in \cite{Friedberg:1976me}, this model possesses spherically symmetric, nontopological
soliton solutions. In Section 3 we shall give numerical evidence for the existence of
spinning, axially symmetric generalizations of those configurations. 

Another important limit 
is found in the special case $\eta_\sigma=\eta_\phi$ by redefining the parameters in the potential as
$\lambda_\phi=\lambda_\sigma=\beta$,
 and $\eta=\frac{1}{2}\beta+\gamma_0$.
Then, after  taking the limit $\beta \to \infty$,
the theory (\ref{Ls}) becomes a sigma model
for two complex fields $\phi,\sigma$ with  
\begin{eqnarray} 
\label{Ls-sigma}
 L_s=\partial_\mu \phi^\ast \partial^\mu \phi  + 
\partial_\mu \sigma^\ast \partial^\mu \sigma +\gamma_0 |\phi|^2|\sigma|^2,
\end{eqnarray}
subject to the constraint
\begin{eqnarray} 
\label{constr}
\nonumber
|\phi|^2+|\sigma|^2=\eta^2_\phi.
\end{eqnarray} 
Note that this model has less degrees of freedom and no free
parameters, since the value of $\gamma_0$ in (\ref{Ls-sigma}) can be changed by rescaling the spacetime coordinates.
Vorton solutions of this model (although for a different formulation)
have been discussed in \cite{Battye:2001ec}. 
 
Finally, one should remark that
the system described by (\ref{action})
possesses an interesting  limit with $\phi\equiv \eta_\phi$, $i.e.$ one scalar field only:
\begin{eqnarray} 
\label{Ls-BS}
 L_s=  
\partial_\mu \sigma^\ast \partial^\mu \sigma + M_{\sigma}^2 |\sigma|^2
+\frac{1}{4}\lambda_\sigma |\sigma|^4.
\end{eqnarray}
Different from the above cases, no flat spacetime solitons
exist in this case\footnote{Here we consider strictly positive values for the parameter $\lambda_\sigma$.
However, flat spacetime solitons are found for a potential
unbounded from below, $\lambda_\sigma<0$, in which case they are likely to be unstable.}.
However, physically interesting solutions are found when gravity is included.
These are the gravitating boson stars with a quartic self-interaction potential 
originally discussed in \cite{Colpi:1986ye}. The field $\sigma$ then
necessarily
possesses a harmonic time dependence
(note that the limit $\lambda_\sigma=0$ is allowed for such solutions).
Their existence can be attributed  to the balance between the dispersive effect
due to the wave character of the complex scalar field and 
 the self-gravitating effect.

\subsection{Global charges}

Our solutions are globally regular
($i.e.$ without an event horizon or conical singularities),
and stationary. Also, they approach asymptotically the Minkowski spacetime background. 
Then their mass $M$ and  angular momentum $J$
can be obtained from the respective Komar expressions \cite{wald},
\begin{equation}
\label{komar}
{M} = \frac{1}{{4\pi G}} \int_{\Sigma}
 R_{\mu\nu}n^\mu\xi^\nu dV,~~
 {\cal J} =  -\frac{1}{{8\pi G}} \int_{\Sigma}
 R_{\mu\nu}n^\mu\eta^\nu dV.
\end{equation}
Here $\Sigma$ denotes an asymptotically flat spacelike hypersurface,
$n^\mu$ is normal to $\Sigma$ with $n_\mu n^\mu = -1$,
$dV$ is the natural volume element on $\Sigma$,
$\xi$ denotes an asymptotically timelike Killing vector field
and $\eta$ an asymptotically spacelike Killing vector field (corresponding to azimuthal symmetry)
\cite{wald}.
After replacing in (\ref{komar}) the Ricci tensor by the
stress-energy tensor (via the Einstein equations (\ref{Einstein})), one finds  
\begin{eqnarray}
 \label{komarM2}
&&
M
= \, 2 \int_{\Sigma} \left(  T_{\mu \nu} 
-\frac{1}{2} \, g_{\mu\nu} \, T_{\gamma}^{\ \gamma}
 \right) n^{\mu }\xi^{\nu} dV \ ,
 \\
&&
 \label{komarJ2}
{\cal J} = -
 \int_{\Sigma} \left(  T_{\mu \nu}
-\frac{1}{2} \, g_{\mu\nu} \, T_{\gamma}^{\ \gamma}
 \right) n^{\mu }\eta^{\nu} dV \ .
\end{eqnarray}

 The Lagrangian (\ref{Ls}) of the scalar fields  has a global $U(1)\times U(1)$ symmetry, 
$\phi \to e^{i\alpha_1} \phi$, 
$\sigma \to e^{i\alpha_2} \sigma$,
which leads to two conserved conserved currents
\begin{eqnarray}
j^{\mu}_{(\phi)} =  - i \left( \phi^* \partial^{\mu} \phi 
 - \phi \partial^{\mu}\phi ^* \right) \ , 
~~
 j^{\mu}_{(\sigma)} =  - i \left( \sigma^* \partial^{\mu} \sigma 
 - \sigma \partial^{\mu}\sigma ^* \right) \ , 
 ~~~{\rm with}~~~~
j^{\mu} _{(\phi);\, \mu}  =  0,~~j^{\mu} _{(\sigma) ; \, \mu}  =  0.
\end{eqnarray}
This implies  two conserved Noether charges,
\begin{eqnarray}
Q_{(\phi)} =  \int_{\Sigma} j^{\mu}_{(\phi)}n_\mu dV, 
~~
Q_{(\sigma)} =  \int_{\Sigma} j^{\mu}_{(\sigma)}n_\mu dV.
\end{eqnarray}

\subsection{The axially symmetric ansatz}
 
The line-element for the solutions in this work  possesses  
two commuting Killing vector fields,
$\xi$ and $\eta$, with
\begin{equation}
\xi=\partial_t \ , \ \ \ \eta=\partial_{\varphi},
\end{equation}
in a system of adapted coordinates.
General relativity solutions with these symmetries are usually studied 
within a metric ansatz
\begin{eqnarray} 
\label{metric-i}
ds^2=e^{2U(\rho,z)}\left(e^{2k(\rho,z)}(d\rho^2+dz^2)+P^2(\rho,z)d\varphi^2 \right)
-e^{-2U(\rho,z)} (dt+\Omega(\rho,z)d \varphi)^2,
\end{eqnarray}
where $(\rho,z)$ correspond asymptotically to the usual cylindrical coordinates.
 In the  (electro-)vacuum case, it is always possible 
to set $P\equiv \rho$,
such that only three independent metric functions appear in the equations,
and $(\rho,z)$ become  the canonical Weyl coordinates \cite{book}.
For more general matter sources, however,
the generic metric ansatz (\ref{metric-i}) with four independent functions is needed.
A variety of interesting general relativity solutions have been constructed within this framework \cite{book}, 
some of them describing toroidal configurations\footnote{Note, however, 
that the known toroidal solutions in four dimensional general relativity do not include black holes.
Indeed, for $d=4$
there are a number of {\it no go}
results
forbidding, under very general
assumptions,
a nonspherical topology of the horizon
for an asymptotically flat black hole  \cite{d4}.} (see $e.g.$
\cite{BW,NE}).

One might wonder whether in the numerical study of 
gravitating ring-shaped solitons 
other coordinate systems would be better suited.
For example, the use of ring coordinates seems to be an obvious choice, 
since the required symmetry of the configurations is intrinsic in this case.
However,  even in the absence of
back reaction,
we have met severe numerical
difficulties
in the implementation of a numerical scheme for these coordinates\footnote{For example, 
 the whole of spatial infinity maps in this case to
a single point, which leads to bad numerical convergence.
Also, the ring radius is an input parameter which should be specified {\it apriori};
however, this did not appear to be possible within our treatment of the problem.}.

The solutions in this work are constructed 
by using `quasi-isotropic' spherical  coordinates $(r,\theta)$,
defined by the coordinate transformation in (\ref{metric-i})
\begin{eqnarray} 
\label{ct}
 \rho=r\sin \theta,~~z=r\cos \theta~,
\end{eqnarray}
and possessing the usual range $0\leq r<\infty$, $0\leq \theta \leq \pi$.
The corresponding line-element{\footnote{A similar line element 
(in particular the same metric gauge choice for the $(r,\theta)$-part
of the metric) has also been used in
previous numerical work on spinning solutions in general relativity, 
starting with the early work \cite{Schunck:1996wa,Kleihaus:2001kg}.
} has an ADM-like 
form and reads
\begin{eqnarray} 
\label{metric}
ds^2=\frac{m}{f}(dr^2+r^2d\theta^2)+\frac{l}{f}r^2\sin^2\theta (d\varphi-\frac{W}{r} dt)^2-f dt^2~.
\end{eqnarray}
The four metric functions $f$, $l$, $m$ and $W$
are functions of the variables $r$ and $\theta$ only, chosen such that
the trivial angular and radial dependence of the line element is already factorized.
The relation between the metric functions in the above line-element  and those
 which enter (\ref{metric-i})
is
$f=e^{2U}P^2/(e^{4U}P^2-\Omega^2)$,
$l \sin^2 \theta=P^2$,
$m=P^2/(e^{4U}P^2-\Omega^2)$
and 
$W/r=\Omega/(e^{4U}P^2-\Omega^2)$.
The symmetry axis of the spacetime is given by $\eta=0$.
It corresponds to the $z$-axis with $\theta=0,\pi$.
The Minkowski spacetime background is approached for $r\to \infty$, with $f=l=m=1$, $W=0$.


The ansatz for the scalar fields which we shall use is  similar
to that proposed in the previous work \cite{Radu:2008pp}, with\footnote{Note
that (\ref{scalar-ansatz}) is not the general 
ansatz compatible with the symmetries of the line element (\ref{metric}).
In principle, one can add an extra $(r,\theta)$-dependent phase to the scalar field 
$\sigma$ by taking
\begin{eqnarray} 
\nonumber
\sigma=|\sigma(r,\theta)|e^{i(\Psi(r,\theta)+ n \varphi+w t)}=(Z_1(r,\theta)+i Z_2(r,\theta))e^{i(n \varphi+w t)}
\end{eqnarray}
with $Z_1,Z_2$ real functions.
(Solutions with such a complex scalar field were constructed in \cite{Radu:2008pp,Shnir:2011gr}
for a Q-ball model in a fixed Minkowski spacetime background.)
Also, one can easily verify that $Z_2=0$
is a consistent truncation of the full model.
}
\begin{eqnarray} 
\label{scalar-ansatz}
\phi=X(r,\theta)+i Y(r,\theta), ~~\sigma=Z(r,\theta)e^{i(n \varphi+w t)}
\end{eqnarray}
where $X,Y,Z$ are real functions,
and $w$ and $n$ are real constants.
Single-valuedness of the scalar field $\sigma$ requires
\begin{equation}
\sigma(\varphi)=\sigma(2\pi + \varphi) \ , 
\end{equation}
thus the constant $n$ must be an integer,
$i.e.$, $n=0, \pm 1, \pm 2, \dots$~.
As usual in the literature, we shall refer to $n$ as rotational quantum number.
 Without any loss of generality, we shall take in this work $w>0,n\geq 0$.
 
We close this part by noting that $\xi=\partial_t$ and $\eta=\partial_{\varphi}$
are $not$ symmetries of the solutions
in this work, 
the system being  only {\it effectively} stationary\footnote{Thus, one can say that
the configurations in this work do not possess Killing vectors.
The scalar field $\sigma$ depends in a nontrivial way on the Killing coordinates $\varphi$
and $t$, though the corresponding energy-momentum tensor
is still compatible with the spacetime symmetries.
This is, however, a generic feature of $d=4$ gravitating spinning configurations 
possessing Q-ball features
(see also Refs. \cite{Hartmann:2010pm,Dias:2011at} for examples of $d=5$
boson stars and black holes with only one Killing vector).
}.

\subsection{The equations and the issue of the constraints}
After inserting the ansatz (\ref{metric}), (\ref{scalar-ansatz})
into the field equations (\ref{Einstein}), (\ref{scalar-eqs})
we find a  system of seven coupled partial differential equations
that needs to be solved.
These are four equations for the metric functions $f,l,m,W$ (which are found by 
taking a suitable combination of the Einstein equations 
$E_r^r+E_\theta^\theta=0$,
$E_\varphi^\varphi=0$,
$E_t^t=0$ and $E_\varphi^t=0$, see Appendix A)
and three equations for the functions $X,Y$ and $Z$ describing the scalar fields.
  The equations
are of the form
\begin{eqnarray} 
\nabla^2 {\cal F}_i={\cal J}_i
\end{eqnarray} 
where ${\cal F}=(f,l,m,W;X,Y,Z)$, 
$\nabla^2=\frac{1}{r^2}\partial_r(r^2 \partial_r)+\frac{1}{r^2 \sin\theta }\partial_\theta(\sin\theta  \partial_\theta)$,
and ${\cal J}_i$ are `source' terms depending on the functions ${\cal F}$ and their first derivatives
(the explicit form of the equations is presented in Appendix A). 

Apart from these, there are two more Einstein equations
  $E_\theta^r =0,~E_r^r-E_\theta^\theta  =0$,
  which are not solved in practice. 
Following an argument originally proposed in \cite{Wiseman:2002zc}, 
one can, however, show that
the identities $\nabla_\mu E^{\mu r} =0$ and $\nabla_\mu E^{\mu \theta}=0$, 
imply the Cauchy-Riemann relations
\begin{eqnarray}
\partial_{\bar r} {\cal P}_2  +
\partial_\theta {\cal P}_1  
= 0 ,~~
 \partial_{\bar r} {\cal P}_1  
-\partial_{\theta} {\cal P}_2
~= 0 ,
\end{eqnarray}
with ${\cal P}_1=\sqrt{-g} E^r_\theta$, ${\cal P}_2=\sqrt{-g}r(E^r_r-E^\theta_\theta)/2$
and $d\bar r=\frac{dr}{r }$.
Therefore the weighted constraints  $E_\theta^r$ and $E_r^r-E_\theta^\theta$ 
still satisfy Laplace equations in $(\bar r,\theta)$ variables.
Then they 
are fulfilled, when one of them is satisfied on the boundary and the other 
at a single point
\cite{Wiseman:2002zc}. 
From the boundary  conditions 
(\ref{bc-r0})-(\ref{bc-Pi2})
 we are imposing,
it turns out that this is the case for our solutions,
 $i.e.$ the numerical scheme is consistent.

\subsection{Mass, angular momentum and charge} 

The mass $M$ and the angular momentum $J$ can be evaluated based on the general
Komar expressions (\ref{komar}).
There we insert the metric ansatz (\ref{metric}),
with unit vector $n = -\sqrt{f}dt$,
and volume element 
$dV =1/ \sqrt{f} \, \sqrt{-g} \, dr \, d\t \, d\varphi$.
Also,  
we make use of the fact that
both $R_t^t$ and $R_\varphi^t$
can be expressed as total derivatives:
\begin{eqnarray}
\nonumber
\sqrt{-g}R_t^t
&=&
\frac{\partial}{\partial r} 
\left (
r^2 \sin \theta \frac{\sqrt{l}}{2f^2}
\left(
\sin^2 \theta  lW (\partial_r W-\frac{W}{r})
-f \partial_r f
\right)
\right )
+
\frac{\partial}{\partial \theta}
\left (
 \sin \theta \frac{\sqrt{l}}{2f^2}
\left(
\sin^2 \theta lW  \partial_\theta W 
-f \partial_\theta f
\right)
\right ),
\\
\label{totder}
-\sqrt{-g}R_\varphi^t
&=& 
\frac{\partial}{\partial r} 
\left (
\frac{r^2 l^{3/2}\sin^3\theta}{2f^2}(r\partial_r W -W)
\right )
+
\frac{\partial}{\partial \theta}
\left (
\frac{r l^{3/2}\sin^3\theta}{2f^2} \partial_\theta W
\right ),
\end{eqnarray}
(where $\sqrt{-g}=r^2 \sin \theta \frac{\sqrt{l}m}{f}$).
Then it follows that  $M$ and $J$ 
can be read off the asymptotic expansion of the metric functions $f$ 
and $W$, respectively
\begin{eqnarray}
f = 1- \frac{2MG}{r} + O\left( \frac{1}{r^2} \right) \ , \ \ \
W= \frac{2JG}{r^2} + O\left( \frac{1}{r^3} \right)\ ,
\label{MQasym1} 
\end{eqnarray}
$i.e.$ after putting (\ref{totder}) into  (\ref{komar}) one finds
\begin{eqnarray}
M=\frac{1}{2G} \lim_{r \rightarrow \infty} r^2\partial_r \, f 
\ , \ \ \  J=\frac{1}{2G} \lim _{r \rightarrow \infty} r^2W \ .
\label{MJ2}
\end{eqnarray} 
Alternatively, the mass $M$ and the angular momentum $J$
can be obtained by direct integration of the expressions
(\ref{komarM2}) and (\ref{komarJ2}), 
\begin{eqnarray}              
 \label{tolman}
M
=\phantom{2}\int_{\Sigma} \left( 2 T_{\e}^{\g} 
 - \d _{\e}^{\g} \, T_{\c}^{\c} \right) \, n_{\g} \, \xi^{\e} dV 
=\int \left(2 \, T_t^t -T_{\e}^{\e} \right) \, \sqrt{-g} 
\, dr \, d\t \, d \varphi \ ,
\end{eqnarray}
corresponding to the Tolman mass, and
\begin{eqnarray}
J= -
\int T_{\, \varphi}^{\, t} \, \sqrt{-g} \, dr \, d\t \, d \varphi \ . 
\label{ang2}
\end{eqnarray}
The explicit form of the nonvanishing components of $T_\mu^\nu$
for the  ansatz (\ref{metric}), (\ref{scalar-ansatz}) is given in 
Appendix B.

As discussed above, the global $U(1)\times U(1)$ symmetry 
of the scalar fields may lead to further nonvanishing
 global charges.
For the ansatz of the scalar fields  (\ref{scalar-ansatz}) 
with the scalar field $\phi$ possessing no time dependence,
there is only one nonvanishing conserved scalar charge 
\begin{eqnarray}
 \label{Qc}
 Q_{(\sigma)}&=&- \int j^t_{(\sigma)} \sqrt{-g} dr d\t d\varphi 
\nonumber 
\\
&=& 4 \pi    \int_0^{\infty} dr \int _0^{\pi} d\theta~
r^2 \sin \theta \frac{\sqrt{l}m}{f^2} 
\left( w+
  n\frac{W}{r} \right)Z^2=Q~.
\end{eqnarray}
However, one can easily verify that the angular momentum density 
$T_\varphi^t$
and the temporal component 
of the current 
$j^t_{(\sigma)}$
are proportional, $T_\varphi^t=n j^t_{(\sigma)}$.
Then, 
from Eq.~(\ref{ang2}) for the angular momentum $J$
and Eq.~(\ref{Qc}) for the scalar charge $Q$,
one obtains the important quantization relation for the angular momentum,
\begin{equation}
\label{JnQ}
J=n \, Q ~.
\end{equation}
Thus $Q$ is not an independent global charge, and  the vortons' angular momentum
is carried by the scalar field $\sigma$ only.
Also, it follows that all solutions with $J\neq 0$, thus in particular the vorton solutions, 
cannot be viewed as perturbations
around some static configurations\footnote{A similar result has been 
found for Q-balls and boson star solutions  
\cite{Schunck:1996wa,volkov,Kleihaus:2001kg}.}.
(Note, that (\ref{JnQ}) holds for solutions in a flat spacetime background as well.)
 
We close this part by noting that gravitating vortons have no horizon. Therefore they are
zero entropy objects, without an intrinsic temperature.
 The first
law of thermodynamics \cite{Lee:1991ax} reads in this case 
\begin{equation}
\label{first-law}
dM=w dQ=\frac{w}{n}dJ~.
\end{equation}
\subsection{Boundary conditions} 

In our approach, the solutions are
constructed by solving the set (\ref{eqf})-(\ref{eq2}) 
of seven elliptic partial differential equations
supplemented with suitable boundary conditions
on the boundaries of the domain of integration (see the discussion in Section 2.8).
The choice of appropriate boundary conditions must guarantee 
that the solutions
are globally regular and asymptotically flat,
and that they possesses finite global charges with finite densities\footnote{These boundary
conditions are also implied by a power series expansion of the solutions
on the boundaries of the domain of integration.}.

Thus, at the origin $r=0$ we require 
\begin{eqnarray}
\label{bc-r0} 
\partial_r f|_{r=0}= 
\partial_r l|_{r=0}= 
\partial_r m|_{r=0}= 
W|_{r=0}=0 \ , 
~~
\partial_r X|_{r=0}=  Y|_{r=0}= Z|_{r=0}= 0,
\end{eqnarray}
The boundary conditions at infinity  
are such that the ground state of the model is approached,
\begin{eqnarray}
\label{bc-inf} 
f|_{r \rightarrow \infty} = 
l|_{r \rightarrow \infty} = 
m|_{r \rightarrow \infty} =1,~~
W|_{r \rightarrow \infty} =0 \ , ~~
X| _{r \rightarrow \infty}=\eta_\phi,~~Y| _{r \rightarrow \infty}=Z| _{r \rightarrow \infty}=0.
\end{eqnarray}
On the symmetry axis ($\t=0$ and $\t=\pi$), 
we require the boundary conditions
\begin{eqnarray}
\label{bc-axis}  
\partial_{\t} f|_{\t=0,\pi}= 
\partial_{\t} l|_{\t=0,\pi}= 
\partial_{\t} m|_{\t=0,\pi}= 
\partial_{\t} W|_{\t=0,\pi}=0,~~
\partial_{\t} X|_{\t=0,\pi}= 
\partial_{\t} Y|_{\t=0,\pi}= 
Z|_{\t=0,\pi}=0.
\end{eqnarray} 
The elementary flatness condition \cite{book} ($i.e.$ the absence of conical singularities)
 imposes an extra condition on the symmetry axis 
\begin{equation}
\label{regularity}
l|_{\theta=0,\pi}=m|_{\theta=0,\pi}.
\end{equation}

The solutions in this work possess an extra symmetry $w.r.t.$ reflection in the equatorial plane $z=0$ ($\theta=\pi/2$).
Namely, all functions except $Y$ are invariant under the transformation $\theta \to \pi-\theta$,
while $Y$ changes the sign.
We use this symmetry to reduce the domain of integration to $ [0,\infty)\times [0,\pi/2]$.
Then the boundary conditions which are imposed in the equatorial plane are
\begin{eqnarray}
\label{bc-Pi2} 
\partial_{\t} f|_{\t=\pi/2}= 
\partial_{\t} l|_{\t=\pi/2}= 
\partial_{\t} m|_{\t=\pi/2}= 
\partial_{\t} W|_{\t=\pi/2}=0,~~
\partial_{\t} X|_{\t=\pi/2}= 
Y|_{\t=\pi/2}= 
\partial_{\t} Z|_{\t=\pi/2}=0. 
\end{eqnarray} 

\subsection{Scaling properties and input parameters}
A simple inspection of the field equations reveals that the model
possesses seven input parameters: $\{\lambda_\phi,\lambda_\sigma,\eta_\phi,\eta_\sigma,\gamma \}$  and  $\{ w,n \}$.
However, as usual, some
of them can be absorbed into a redefinition of variables
together with suitable rescalings.

Let us start with the simplest case of nongravitating solutions.
Then the scaling  of the fields
\begin{eqnarray} 
\label{s1}
 \phi \to \eta_\phi \phi~,~~~\sigma \to \eta_\phi \sigma~
 \end{eqnarray}
 together with the redefinitions
\begin{eqnarray} 
\label{s2}
\lambda_\phi \to {\lambda_\phi}/{\eta_\phi^2},~~
\lambda_\sigma \to {\lambda_\sigma}/{\eta_\phi^2},~~
\gamma \to {\gamma}/{\eta_\phi^2},~~
\end{eqnarray}
show that one can set $\eta_\phi=1$ in   
all field equations
without loss of generality.
Then  $\eta_\phi^2$ appears as an overall factor in front of the (rescaled-) Lagrangian density.

In the next step, we note that a rescaling of the spacetime coordinates
\begin{eqnarray} 
\label{s3}
x^\mu \to x^\mu/L
\end{eqnarray}
 (with $L>0$ an arbitrary constant) together with a redefinition of several parameters in the potential
\begin{eqnarray} 
\label{s31}
\lambda_\phi \to \lambda_\phi L^2,~~
\lambda_\sigma \to \lambda_\sigma L^2,~~
\gamma \to \gamma L^2,
\end{eqnarray}
leaves the field equations invariant.
This symmetry is used in this work
to set\footnote{ Note that other choices are possible. For example, one can also set $L^2=1/\lambda_\sigma \eta_\phi^2$ or
 $L^2=1/\gamma \eta_\phi^2$. 
 However, 
 the choice (\ref{L})
 leads to a length scale with a transparent physical meaning.
 } 
the length scale of the vortons as given by the mass of the scalar field $\phi$,
\begin{eqnarray}
\label{L}
L=\frac{1}{M_\phi}=\frac{1}{\eta_\phi \sqrt{\lambda_{\phi}}}.
\end{eqnarray} 
This directly reveals the existence of three essential, dimensionless parameters of the nongravitating
vorton model
\begin{eqnarray} 
\label{beta}
\beta_1=\frac{\eta_\sigma}{\eta_\phi},~~
\beta_2=\frac{\lambda_\sigma}{\lambda_\phi},~~
\beta_3=\frac{\gamma}{\lambda_\phi}~.
\end{eqnarray} 
From  (\ref{conds}), the parameters 
 $\beta_i$  should satisfy the conditions
\begin{eqnarray} 
\label{cond1}
4 \beta_3^2>\beta_2,~~\beta_3>\frac{1}{2}\beta_2\beta_1^2.
\end{eqnarray} 
Note that the scalings above are independent of any ansatz.
However, when considering a scalar $\sigma$ of the form (\ref{scalar-ansatz}),
one has to take also 
\begin{eqnarray} 
\label{scale-w}
w \to w/M_\phi,
\end{eqnarray} 
 such that the phase of $\sigma$ remains invariant under the scaling (\ref{s3}),
 with $L$ given by (\ref{L}).

The scalings discussed above hold also when gravity is included.
However, in that case, the presence of Newton's constant $G$ leads to the occurrence of 
one more dimensionless
coupling constant, which, following previous work on solitons coupled to gravity \cite{Volkov:1998cc},
we shall call $\alpha$, with
\begin{eqnarray} 
\label{alpha}
\alpha^2=16 \pi G \eta_\phi^2.
\end{eqnarray} 
Thus the constant $\alpha$ corresponds to the ratio between 
the symmetry breaking scale and the Planck mass $M_{Pl}=1/\sqrt{G}$.
On general grounds, we expect this ratio to be very small;
however, to understand the general pattern of the solutions, 
we shall attempt to investigate the full range of $\alpha$.

Then, for the specific ansatz (\ref{metric}), (\ref{scalar-ansatz}) with the above scalings, 
the equations for gravitating vortons (which are effectively solved)
can be derived by extremizing the following reduced action
\begin{eqnarray} 
\label{Leff}
S_{red}=\int dr d \theta \left({\cal L}_g-\alpha^2 {\cal L}_s \right),
\end{eqnarray} 
with (here we define  $(\nabla A)\cdot (\nabla B)=
A_{,r}B_{,r}+\frac{1}{r^2}A_{,\theta}B_{,\theta}$):
\begin{eqnarray} 
\nonumber
{\cal L}_g&=&r^2 \sin \theta \sqrt{l} 
 \bigg [
\frac{1}{2 lm}(\nabla l)\cdot (\nabla m)
+r^2 \sin \theta\frac{l}{2 f^2}(\nabla W)^2
-\frac{1}{2f^2}(\nabla f)^2
\\
\label{Leff-gs}
&&{~~~~~~~~~~~~}
-\frac{1}{r l}(l_{,r}+\frac{\cot \theta}{r}l_{,\theta})
+\frac{1}{r m}(m_{,r}+\frac{\cot \theta}{r}m_{,\theta})
 \bigg ],
\\
\nonumber
{\cal L}_s&=& r^2 \sin \theta\frac{m \sqrt{l}}{f}
\bigg [
\frac{f}{m}( (\nabla X)^2+(\nabla Y)^2+(\nabla Z)^2 )
+\frac{1}{4}(X^2+Y^2-1)^2
+\frac{\beta_2}{4}Z^2(Z^2-2\beta_1^2) 
\\
\nonumber
&&{~~~~~~~~~~~~}
+\beta_3 Z^2(X^2+Y^2)
-\frac{1}{f}(w+\frac{n W}{r})^2 Z^2
+\frac{n^2 }{r^2 \sin^2 \theta }\frac{f}{l}Z^2
\bigg].
\end{eqnarray} 
In the above relations we use a dimensionless radial variable $r$
and a dimensionless frequency $w$ given in units set by $M_\phi$. 
These conventions we shall use for all numerical data reported in this work.
Also, in all plots, the total mass $M$ 
is given in units of $4\pi M_{\phi}/\lambda_\phi$,
while the total angular momentum $J$ and the charge $Q$ are given in units of $4\pi/\lambda_\phi$.
The components $T_t^t$ and $T_\varphi^t$ of the energy momentum tensor
are given in units of 
$M_\phi^4/\lambda_\phi$
and
$M_\phi^3/\lambda_\phi$,
respectively.

\subsection{The numerical method }
 
The set of seven coupled non-linear
elliptic partial differential equations
for the functions ${\cal F}=(X,Y, Z; f,l,m,W)$
has been solved numerically  
subject to the boundary conditions   (\ref{bc-r0})-(\ref{bc-Pi2}).
In a first stage, a new compactified radial variable $x$ is introduced
instead of $r$, such that the 
semi-infinite region $[0,\infty)$ is mapped to the finite interval $[0,1]$.
(This avoids the use of a cutoff radius.)
 There
are various possibilities for such a mapping, 
a flexible enough choice being $ x=r/(1+r)$.
This involves the following substitutions in the differential equations
\begin{eqnarray}
r {\cal F}_{,r}   \longrightarrow    (1-x) {\cal F}_{,x} 
~~~{\rm and }~~~
r^2   {\cal F}_{,rr}   \longrightarrow
(1-x)^2    {\cal F}_{,xx}
  - 2 (1-x){\cal F}_{,x}.
\end{eqnarray} 
Next, the equations for these functions
are discretized on a grid in $x$ and $\theta$. 
Here we have considered various grid choices,  
the number of grid points ranging between $280 \times 30$ and $150 \times 70$. 
The grid covers the integration region
$0\leq x \leq 1$ and $0\leq \theta \leq \pi/2$.   

All numerical calculations have been 
performed by using the professional package FIDISOL/CADSOL \cite{schoen}.
%
This code requests the system of nonlinear partial differential equations 
in the form
$
P(x,\theta;{\cal F};{\cal F}_{x},{\cal F}_{\theta};
{\cal F}_{x \theta},{\cal F}_{xx},{\cal F}_{\theta\theta})=0,
$
subject to a set of boundary conditions on a rectangular domain. 
Besides  that, FIDISOL/CADSOL requires
 the Jacobian matrices for the equations $w.r.t.$ the
 functions ${\cal F}$ and their first and second derivatives,
the boundary conditions, as well as some initial guess for the functions ${\cal F}$. 
The solver uses a Newton-Raphson method,
which requires to have a good first guess in order
to start a successful iteration procedure.
This software package provides also error estimates for each function,
which allows to judge the quality of the computed solution.
The typical  numerical error\footnote{However,  
the errors
increase dramatically when studying solutions close to the limits 
of the domain of existence in the parameter space.} 
for the solutions reported in this work is estimated to be of the order of $10^{-3}$. 
A detailed description of  the numerical method and explicit examples are provided in
\cite{schoen} (see also $e.g.$ the Appendix in \cite{Kleihaus:2011yq} for a discussion within a physical problem).
As a further check of numerics, we have verified that the families of solutions 
with a varying frequency satisfy  with a very good accuracy
the first law of
thermodynamics (\ref{first-law}).

In this scheme, there are six input parameters: the  
parameters $\beta_i$ in the potential of the scalar fields, the (scaled) frequency $w$ and
the winding number $n$ in the ansatz for the scalar field $\sigma$,
and also the constant $\alpha$ as given by (\ref{alpha}).
The number of nodes $k$ of $|\phi|$
in the equatorial plane (which
provides a classification of the solutions, see the discussion below),
as well as the global charges (mass and angular momentum) are computed from the numerical
solution.

In constructing gravitating solutions,
we start with flat spacetime configurations 
as initial guess ($i.e.$ taking $f=l=m=1$, $W=0$).
Then we  slowly increase the value of $\alpha$.
When the iterations converge we obtain,
by repeating this procedure,
solutions for higher values of $\alpha$. 
A similar procedure is used to investigate the dependence on other parameters of the problem,
starting instead with solutions
with a given $\alpha$.
Also, in some of the calculations, we interpolate the resulting
configurations on points between the chosen grid points,
and then use these for a new guess on a finer grid.

As a general remark, we have found the study of vortons 
more difficult than a number of other problems which
we studied with a similar approach.
First, the present system possesses a large parameter space, 
and the solutions exist only in some regions there 
(and thus not for arbitrary values of $\{ \beta_i,w,n,\alpha \}$).
Moreover, 
rather good initial profiles
are usually required for the convergence of the numerical iteration.
Next, the profiles of the scalar functions exhibit large gradients in a small region
around the equatorial plane, which requires a careful choice of the grid.
Finally, the problem is further complicated by the possible existence of
several different solutions for the same input parameters 
(see the discussion below).

\section{Solutions in flat space: nongravitating vortons}

\subsection{The heuristic construction of a vorton 
and general features}

As mentioned in the introduction, vortons are essentially loops made of 
pieces of vortices.
Remarkably,
this simple heuristic construction leads to a number of 
predictions about their qualitative properties.
This is why 
\newpage
 \setlength{\unitlength}{1cm}
\begin{picture}(8,6)
\put(0,0.0){\epsfig{file=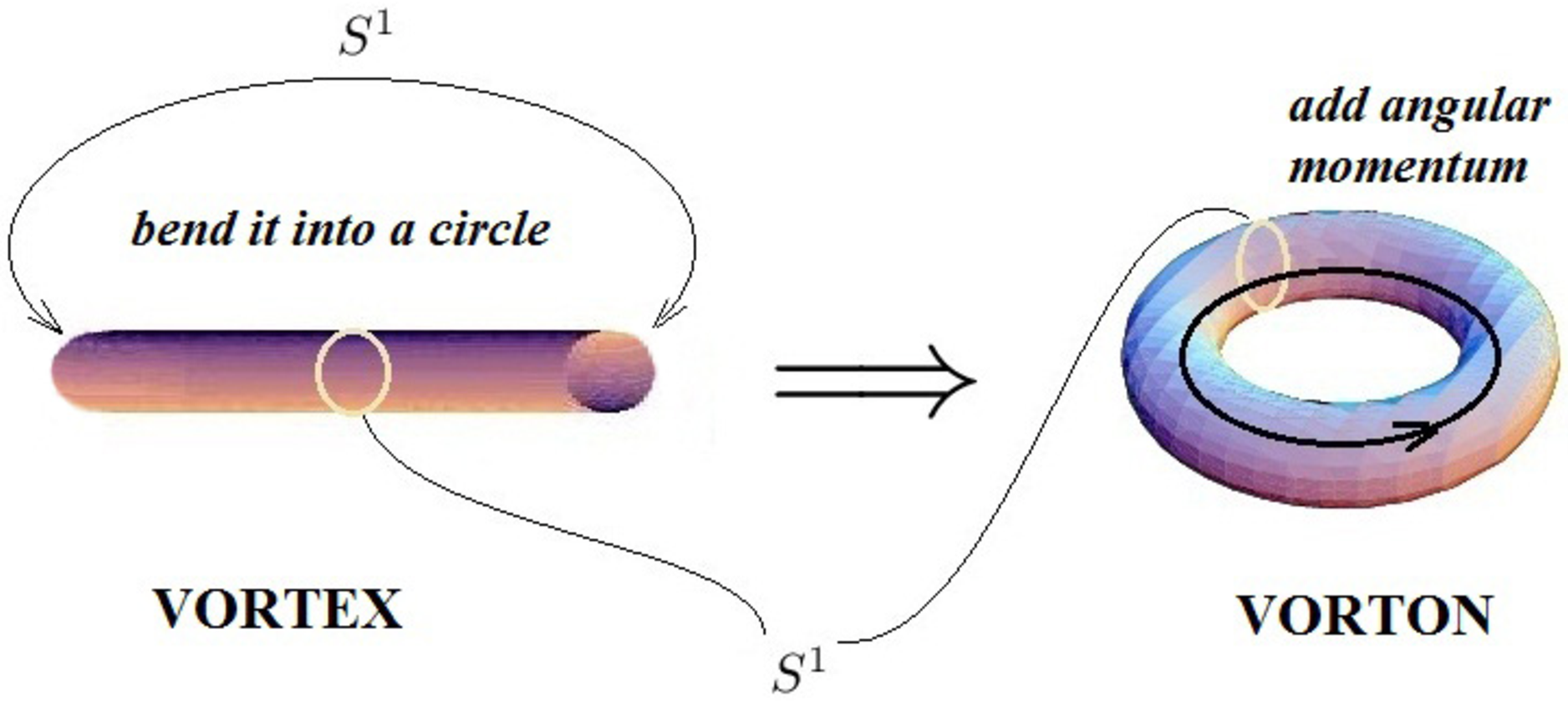,width=14cm}} 
\end{picture}
\\
\\
{\small {\bf Figure 2.} The heuristic construction of a vorton starting 
with a piece of a vortex. } 
\vspace{0.5cm}
\\
 we shall  start with a discussion 
of this point (more details on that, including relevant plots,
 are given in Section 6 of Ref. \cite{Radu:2008pp}).

As discussed by several authors (see $e.g.$
 \cite{Davis:1988ip,Lemperiere:2002en,Hartmann:2008yr}), 
Witten's model of superconducting strings
admits cylindrically symmetric, vortex-type solutions.
 For a cylindrically symmetric line element with $ds^2=d\rho^2+\rho^2 d\psi^2+dz^2-dt^2$,
 the appropriate ansatz for the scalar fields reads
 \be
\label{vortices}  
\phi=|\phi(\rho)|e^{i k \psi},~~~~~~~~
\sigma =|\sigma(\rho)|e^{i(k_z z-w t)}. 
\ee
As usual, the single-valuedness of the scalar fields imposes $k$ to be an integer,
$k=0,\pm 1, \dots$, while there are no apriori restrictions
for $k_z,w$ in (\ref{vortices}).

A study of the asymptotic form of the solutions
 reveals that the scalar field $\phi$
approaches the vacuum expectation value ($v.e.v.$) for $\rho\to \infty$; as $\rho \to 0$,  $|\phi|$ behaves  as $\rho^k$
(thus it vanishes on the $z-$axis, unless $k=0$).
At the same time, $\sigma$ is always nonzero at $\rho=0$ (where $|\sigma|$ takes its maximal value) 
and vanishes as $\rho\to \infty$.
(This is why  $\phi$ and $\sigma$
are usually refered to as vortex and condensate fields, respectively.)
Then the phase of $\phi$ changes
by $2\pi k$ after one revolution around the vortex, while the phase of $\sigma$
increases along the vortex.

Solutions with this behaviour can  be constructed by using a standard ordinary differential equations
solver;
typical profiles for $|\phi|$, $|\sigma|$ and a discussion of their basic properties
can be found in \cite{Lemperiere:2002en}.
These configurations possess a nonzero momentum along the vortex,
with $T_z^t=k_z w |\sigma|^2$.
Also, their energy density is finite everywhere.
However, for $k \neq 0$, the total mass per unit length
exhibits a logarithmic divergence originating in the $\phi$-sector
of the model\footnote{Note that this divergence is
cured when
the $\phi-$field is gauged.}.

The heuristic construction of a vorton involves taking a piece of 
length $L$ of such a vortex and identifying its ends to make a loop (see Figure 2).
Therefore the central axis of the vortex where the field $\phi$
vanishes (for $k \neq 0$), becomes a circle of radius $R=L/2\pi$.
Then the
 coordinate $z$ along the vortex becomes periodic, 
being replaced by the azimuthal angle $\varphi$.
This implies that the constant $k_z$ in (\ref{vortices}) also becomes quantized, $k_z=2\pi n/L$.
In such a heuristic construction, 
the momentum $T_z^t$ along the vortex
will then circulate along the loop, thus implying a nonzero angular momentum
density.
Then the corresponding centrifugal force will
compensate the tension of the loop,
thereby producing an equilibrium configuration.
 
The vorton would inherit some basic properties of the
 initial vortex configuration. 
For example, for $k\neq 0$, 
the magnitude of the field $\phi$ would vanish on a circle of radius
$R$ in the equatorial plane (the initial position of the vortex).
At the same time, the maxima of the  condensate scalar field $\sigma$
would be located on the same circle at $\theta=\pi/2$.
Also, it is clear that a configuration generated in this way would possess
an intrinsic toroidal symmetry,
the angle $\psi$ for the vortex  becoming an $S^1$ generator of the 
resulting torus, see Figure 2.

Based on this heuristic construction, a description of large radius vortons
has been proposed in \cite{Carter:1989xk},
within an approximation of the vortons'
cross section by that of an infinite string carrying a current\footnote{However,
as discussed in \cite{Battye:2008mm},
this approximation fails to provide a good quantitative description,
at least for the  vorton solutions investigated in that work.
This is presumably due to the fact that the initial vortices
have an infinite energy per unit length.
However, a similar description of large radius vortons
provides excellent agreement with the full numerical
simulation for $d=2+1$ dimensions \cite{Battye:2008zh}.}.

From the above analysis, it is clear that
  the resulting toroidal configurations can be described  by the following 
axially symmetric ansatz,
\be                              \label{anz}
\phi=|\phi(r,\theta)|e^{i  \psi_1(r,\theta)}=X(r,\theta)+iY(r,\theta),~~~~~~
\sigma=|\sigma(r,\theta)|e^{i(n\varphi+w t)}=Z(r,\theta)e^{i(n\varphi+w t)}.
\ee
Then any axially symmetric vorton configuration can be characterized by a 
pair of integers.
Apart from the azimuthal winding number $n$
which enters the ansatz (\ref{anz}) explicitly,
they inherit\footnote{Note, that the number $k$
can be defined also for non-axially symmetric configurations.} also the integer $k$ of the initial vortex configuration  (\ref{vortices}).
This can be seen as follows:
the amplitude of the scalar field $\phi$ should vanish for some value of $r$ at $\theta=\pi/2$,
while its phase twists by $2 \pi k$ after one revolution around
  the boundary of the half-plane consisting of the $z-$axis (with $z=r\cos \theta$) and a
 half-circle at infinity.
 
Then, following \cite{Radu:2008pp}, one can formally
 define  a `topological charge',
 which is the product of the two integers characterizing the $S^1\times S^1$
 topology of the torus
\be                                 \label{topol}
N=n k.
\ee
However, one should emphasize that,  for the model 
 considered in this paper, the solutions do not describe topological solitons and thus (\ref{topol}) 
 is not a topologically invariant quantity.
$N=n k$ becomes a topological charge only  in the the sigma-model limit (\ref{Ls-sigma})
(see Ref. \cite{Battye:2001ec}).

Unfortunately, in the absence of exact solutions,
it is extremely difficult to follow
the details of the above heuristic construction\footnote{Remarkably,
 a similar heuristic construction  could be demonstrated in
  $d=5$  Einstein gravity \cite{Emparan:2006mm}. The large radius black rings there are essentially boosted black strings.}.
 Moreover, while the numerical construction of the initial vortices does not pose special problems (since
 one deals with ordinary differential equations),
 the systematic study of vortons represents a numerical challenge\footnote{The correspondence 
between vortices and vortons could be studied in detail for planar {\it kinky} vortons,
$i.e.$ solutions in $2+1$ dimensions \cite{Battye:2008zh}.
The starting configuration represents a kink string,
which is known in closed form.
This leads to a number of exact results and greatly simplifies both the analytical
and numerical treatment of the problem.
}.  
An approximate
form of the solutions can be found on the boundaries of the domain of integration only.
For example, the leading terms in the asymptotic form of the solutions, as $r\to \infty$, read
(restoring the dimensionful variables)
\begin{eqnarray} 
\label{asympt}
\psi_1=\frac{A \sin \theta}{r^2}+\dots,~
|\phi|=1-\frac{1}{\eta_\phi \lambda_\phi}\frac{A^2}{r^6}(1+3\cos^2\theta)+\dots
+\frac{C}{r}e^{-M_\phi r}+\dots,~
|\sigma|=\frac{B}{r}e^{-\sqrt{M_\sigma^2-w^2}r}+\dots,~~{~~}
\end{eqnarray} 
with $X=|\phi|\cos \psi_1$, $Y=|\phi|\sin\psi_1$ 
and $A,B,C$
arbitrary constants.
 Note, that finiteness of the mass imposes an upper bound on the frequency,  
 \begin{eqnarray} 
\label{max-w}
w^2<w_{max}^2=\gamma \eta_\phi^2-\frac12\,\lambda_\sigma\eta_\sigma^2=M_\sigma^2~.
\end{eqnarray} 
When expressed in terms of the dimensionless parameters,
the above condition reads
 \begin{eqnarray} 
\label{max-w1}
\beta_3>w^2+\frac{1}{2}\beta_2 \beta_1^2,
\end{eqnarray}
a form which helps to understand the critical behaviour of the numerical solutions.

Also, as proven in \cite{Radu:2008pp},
 vortons in a flat spacetime background satisfy 
the virial identity
\begin{eqnarray} 
\label{virial}
{\cal T}+{\cal U}=\frac{3 w}{2} Q 
\end{eqnarray} 
with\footnote{Here the operator $\nabla$ is expressed with respect to the three dimensional
euclidean metric.}
\begin{eqnarray}
\label{virial1}
{\cal T}= \int d^3 x 
\left (
|\nabla \phi|^2+|\nabla \sigma|^2
\right ),
~~{\cal U}=\int d^3 x~U(|\phi|,|\sigma|),~~
Q =2 w \int d^3 x~|\sigma|^2~,
\end{eqnarray}

\newpage
\setlength{\unitlength}{1cm}
\begin{picture}(15,20.85)
\put(-1,2){\epsfig{file=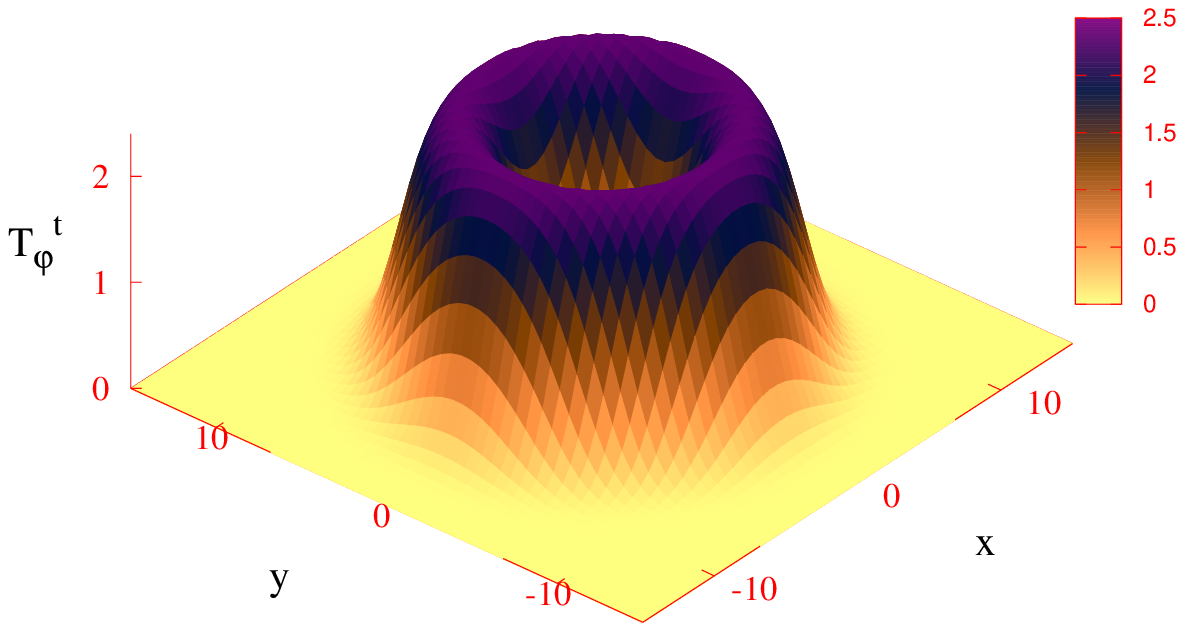,width=7.5cm}}
\put(7,2){\epsfig{file=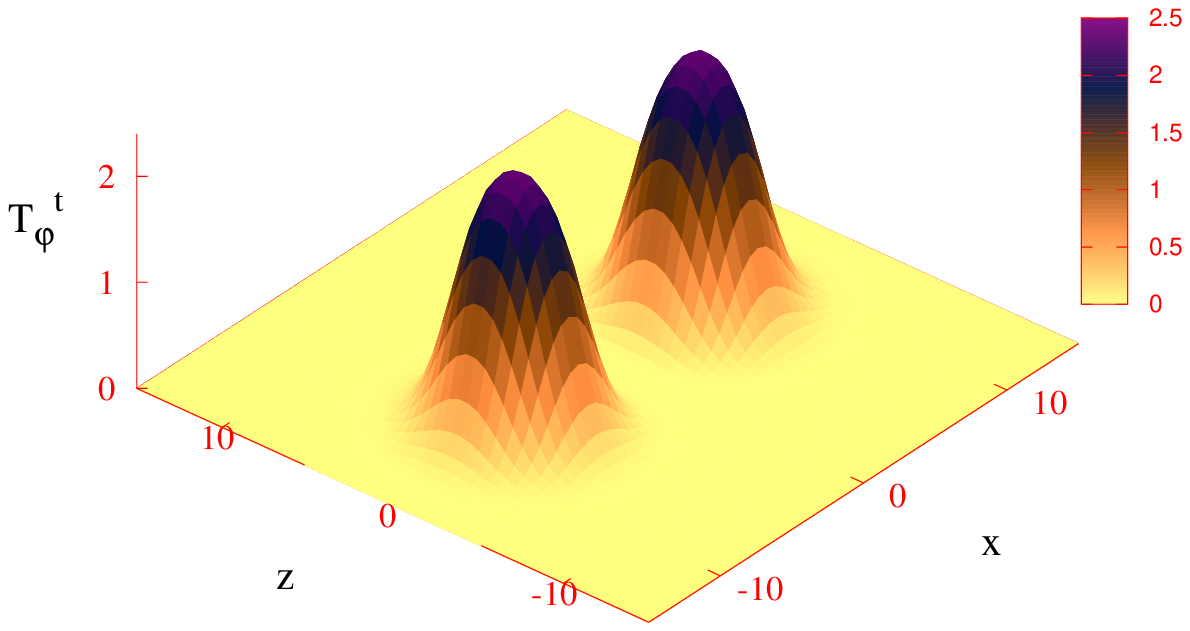,width=7.5cm}}
\put(-1,7){\epsfig{file=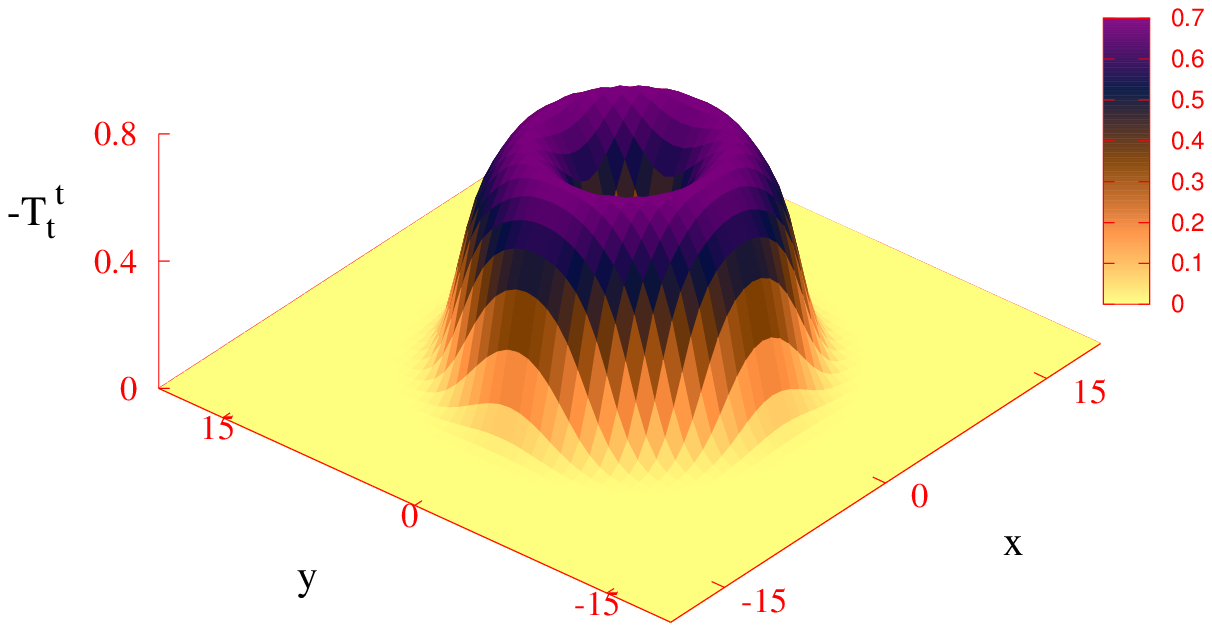,width=7.5cm}}
\put(7,7){\epsfig{file=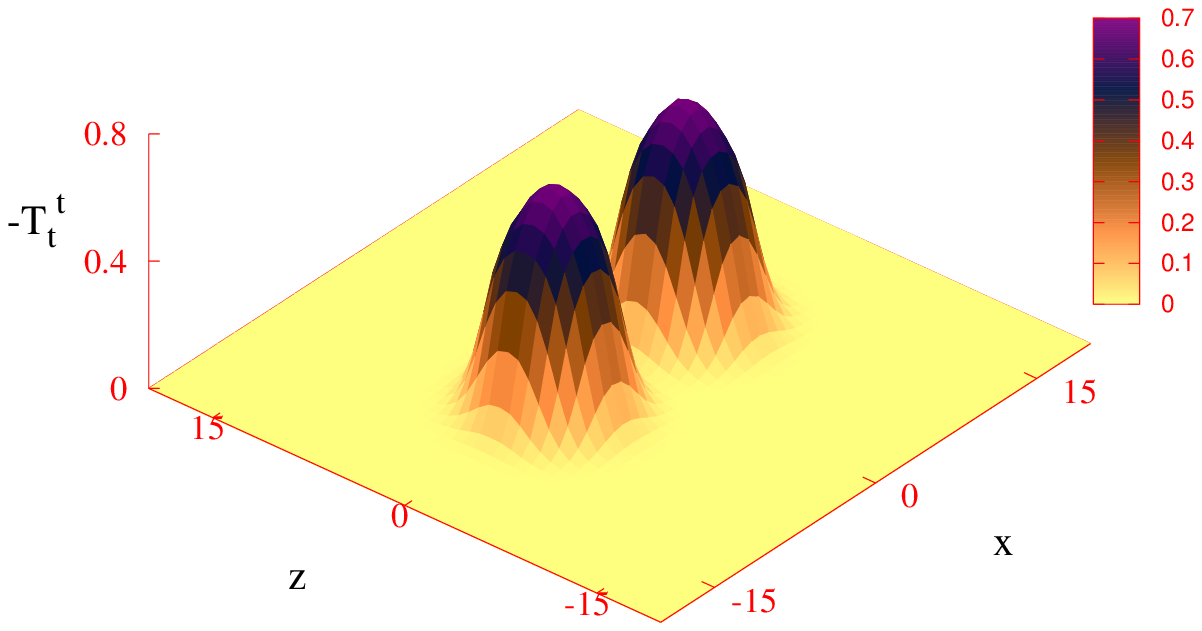,width=7.5cm}}
\put(-1,12){\epsfig{file=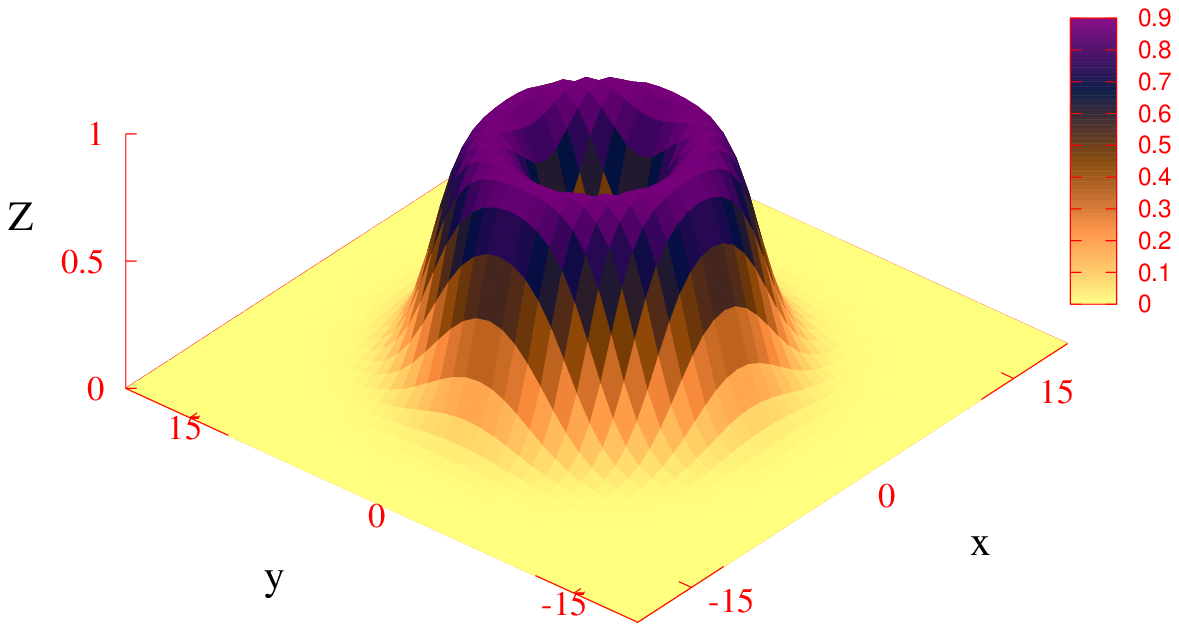,width=7.5cm}}
\put(7,12){\epsfig{file=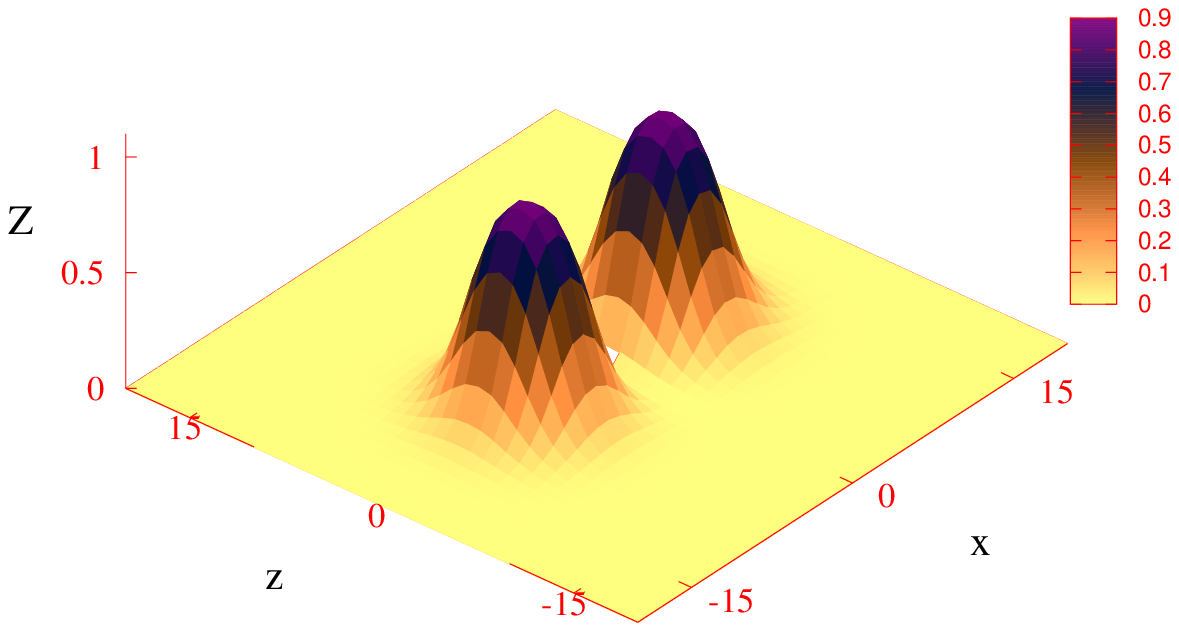,width=7.5cm}}
\put(-1,17){\epsfig{file=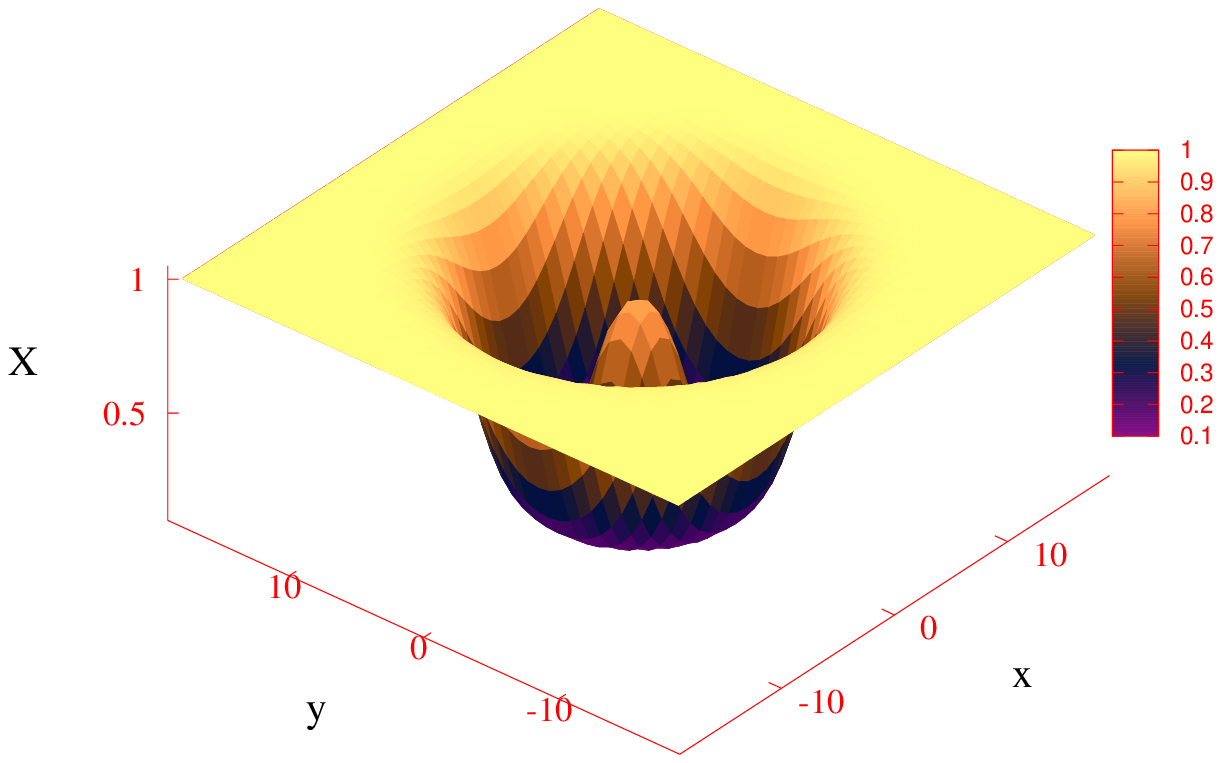,width=7.5cm}}
\put(7,17){\epsfig{file=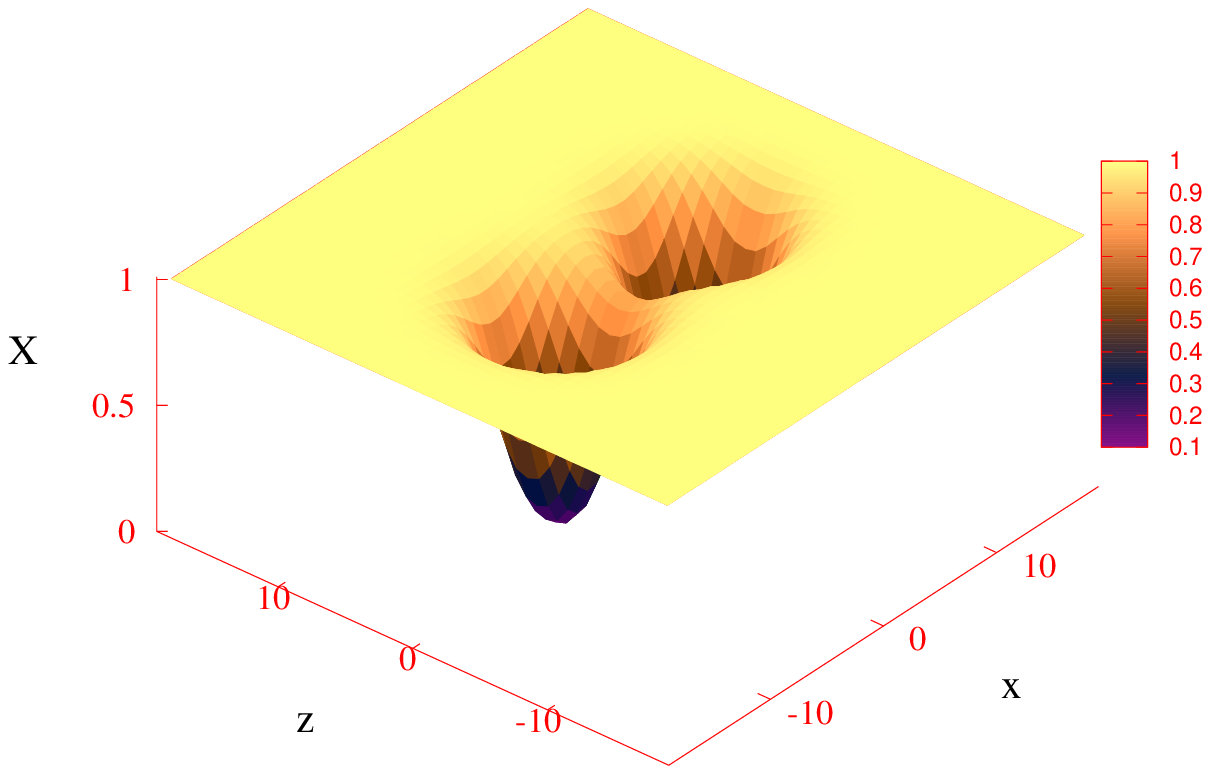,width=7.5cm}}  
\end{picture}
 \vspace*{-1.cm}
\\
{\small {\bf Figure 3.} The scalar functions $X$, $Z$,
the energy density $-T_t^t$ and the angular momentum density $T_\varphi^t$ are shown for 
a typical $k=0$ `semitopological' vorton. 
The left plots are for the $z=0$ plane, while the right plots are for a plane containing the axis of
symmetry.
The parameters of the solution are
$w=0.75$,  $n=2$ and  $\beta_1=0.71,~\beta_2=2.22,~\beta_3=1.7$. } 
 
\newpage
\setlength{\unitlength}{1cm}
\begin{picture}(8,6)
\put(-0.5,0.0){\epsfig{file=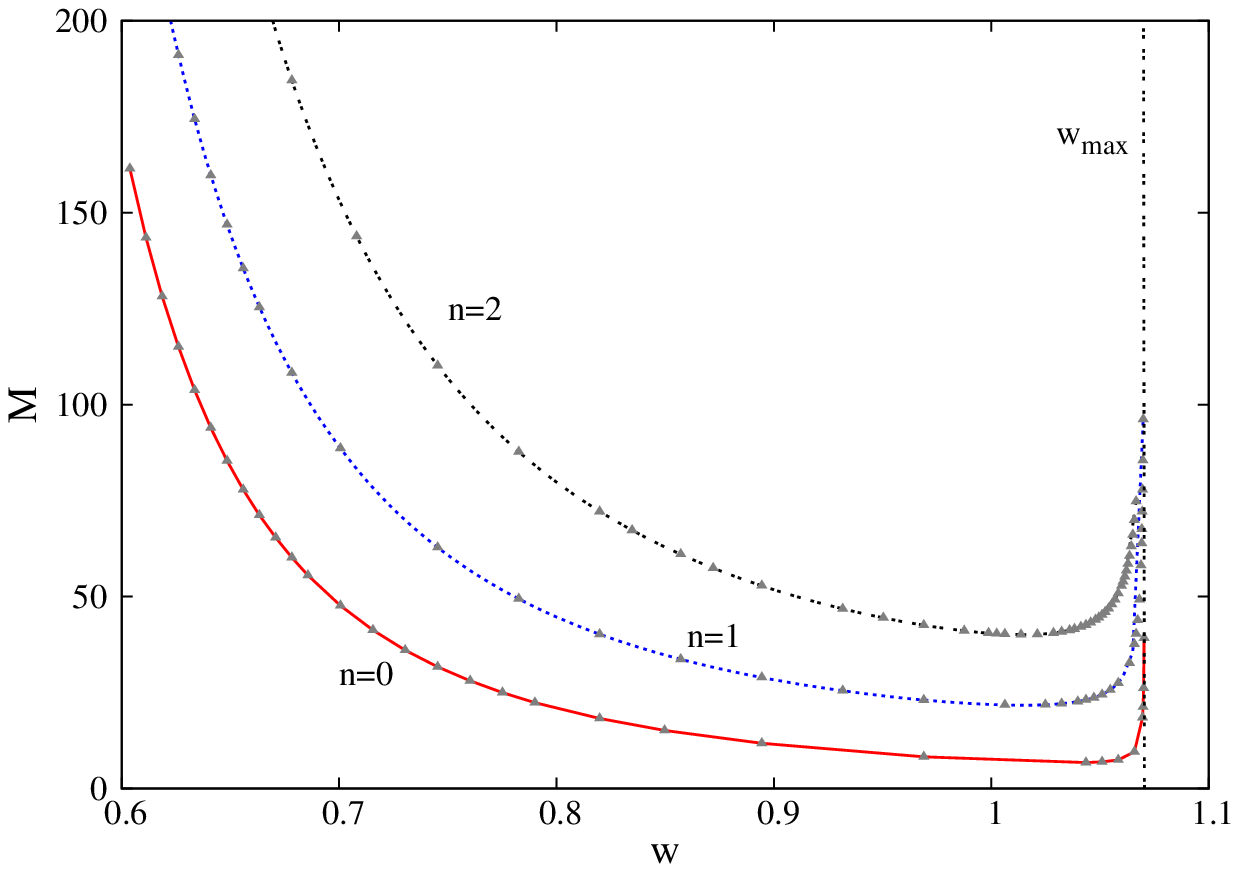,width=8cm}}
\put(8,0.0){\epsfig{file=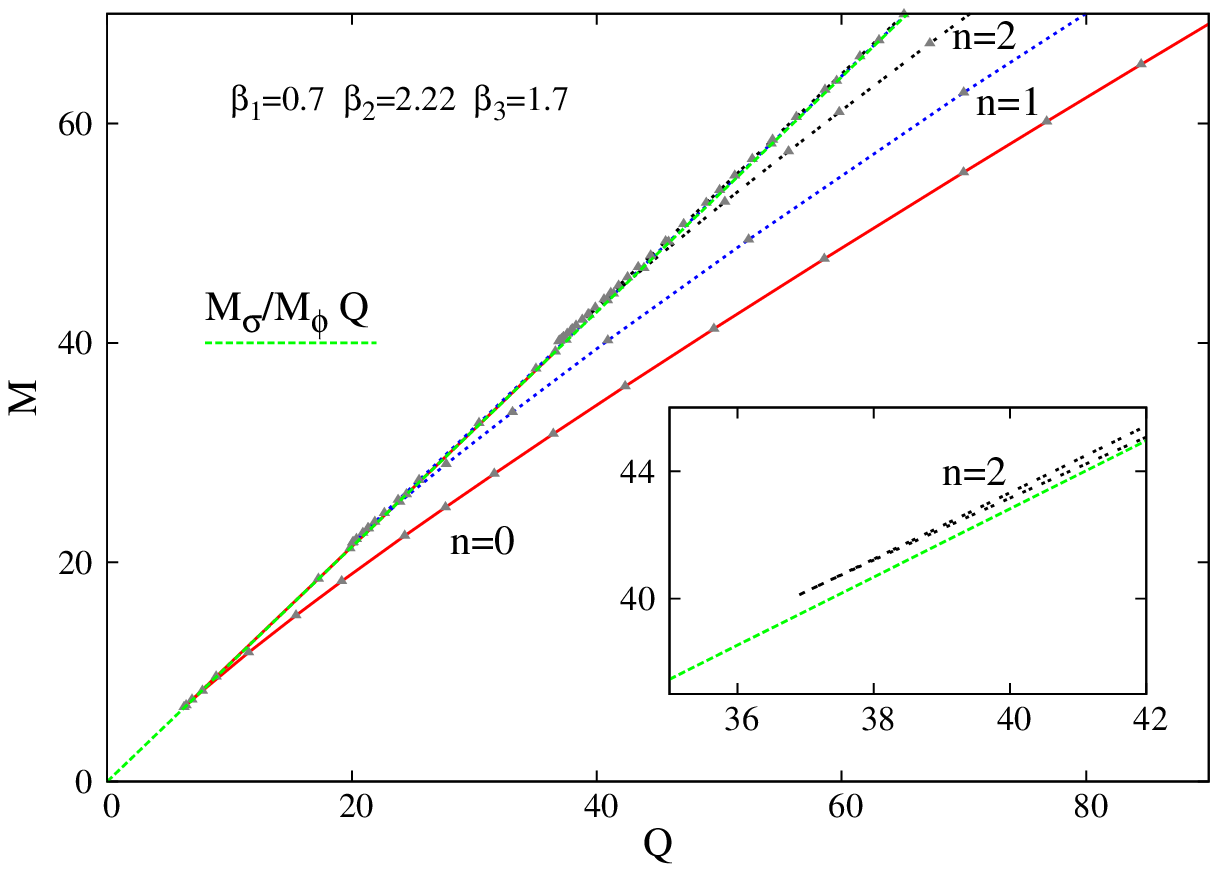,width=8cm}}
\end{picture} 
\\ 
\\
{\small {\bf Figure 4.} {\it Left:} The mass  of $k=0$ `semitopological' vortons in a flat spacetime background
is shown as a function of the frequency for several values of the azimuthal winding number $n$.
The vertical line corresponds to the maximal value of the frequency as given by (\ref{max-w}).
{\it Right:} The $M(Q)$ diagram for the same solutions.
Also shown is the mass for $Q$ free
bosons, $M = M_\sigma/M_\phi Q$
as given in units set by the scalar field $\phi$.
Note, that in all figures in this work exhibiting results for families 
of solutions, the large dots represent the
data points.}  
\vspace*{0.8cm}
\\
which clearly shows that the solutions owe their existence to the
harmonic time dependence of the condensate field $\sigma$.
Note, that (\ref{virial}) holds for generic solutions of the model, without any symmetry assumption.
The mass of the solutions can be written in terms of the above quantities 
as
\begin{eqnarray}
\nonumber
M={\cal T}+\frac{1}{2}w Q+{\cal U}.
\end{eqnarray}

For spinning axially symmetric solutions, a more suggestive form of (\ref{virial})
is found by replacing $Q=J/n$.  
Then one can see that  the angular momentum of the solutions is bounded from below, $i.e.$
the limit $J\to 0$
is not allowed.
The relation (\ref{virial}) is also used to verify the accuracy of the numerical results.

A rather similar virial identity holds for $Q-$ball solutions \cite{Radu:2008pp}.
These are solitonic solutions in a model with a single complex scalar field which   vanishes at infinity (like $\sigma$)
and possesses a non-renormalizable potential.
This  suggests to view 
the toroidal solutions
as $\phi-$dressed
$Q-$balls made of the complex $\sigma$-field,
where the
non-renormalizable $|\sigma|^6$ term in the Q-ball potential is replaced by the interaction with
the scalar field $\phi$.
As we shall see, the
numerical results give some support for this analogy.
However, the vortons possess also some special features
which singularize them.


\subsection{$k=0$ `semitopological' vortons}

These are the simplest solutions of the model studied in this work.
In a heuristic construction, they are found by
starting with loops made of vortices possessing $k=0$ 
in the vortex ansatz (\ref{vortices}).
This implies that the scalar field $\phi$
is purely real, $i.e.$ $Y\equiv 0$.

Moreover, since $\phi$ does not vanish
at the core of the vortex,
then, for the corresponding axially symmetric solution,
$X$ should be a nodeless function\footnote{Indeed, for all solutions
with a real $\phi$ which we could construct,
the function $X(r,\theta)$ was strictly positive.
}.

The azimuthal winding number
$n$ is non-trivial in this case, but the `topological charge' $N = k n$ as defined by (\ref{topol}) is zero, and so strictly
speaking these solutions should not be called vortons.
Then in what follows we shall call these configurations 
`semitopological' vortons.

Also, in some sense they represent axially symmetric generalizations\footnote{The
 FLS model \cite{Friedberg:1976me} is recovered for a vanishing $\sigma$-potential in (\ref{potential}).
 However, we have found that the basic properties of the more general solutions here
 hold also in that limit.} of the
spherically symmetric  

\newpage
\setlength{\unitlength}{1cm}
\begin{picture}(8,6)
\put(-0.5,0.0){\epsfig{file=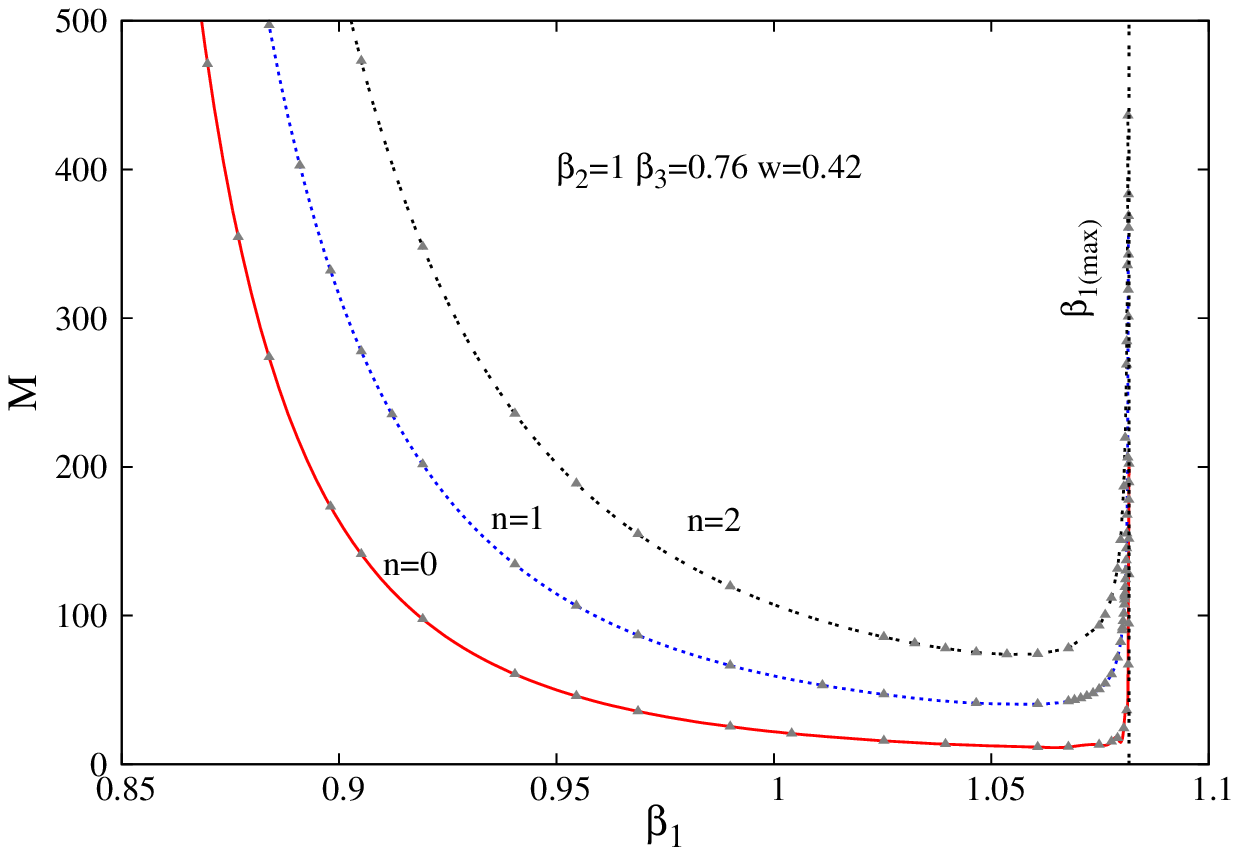,width=8cm}}
\put(8.2,0.0){\epsfig{file=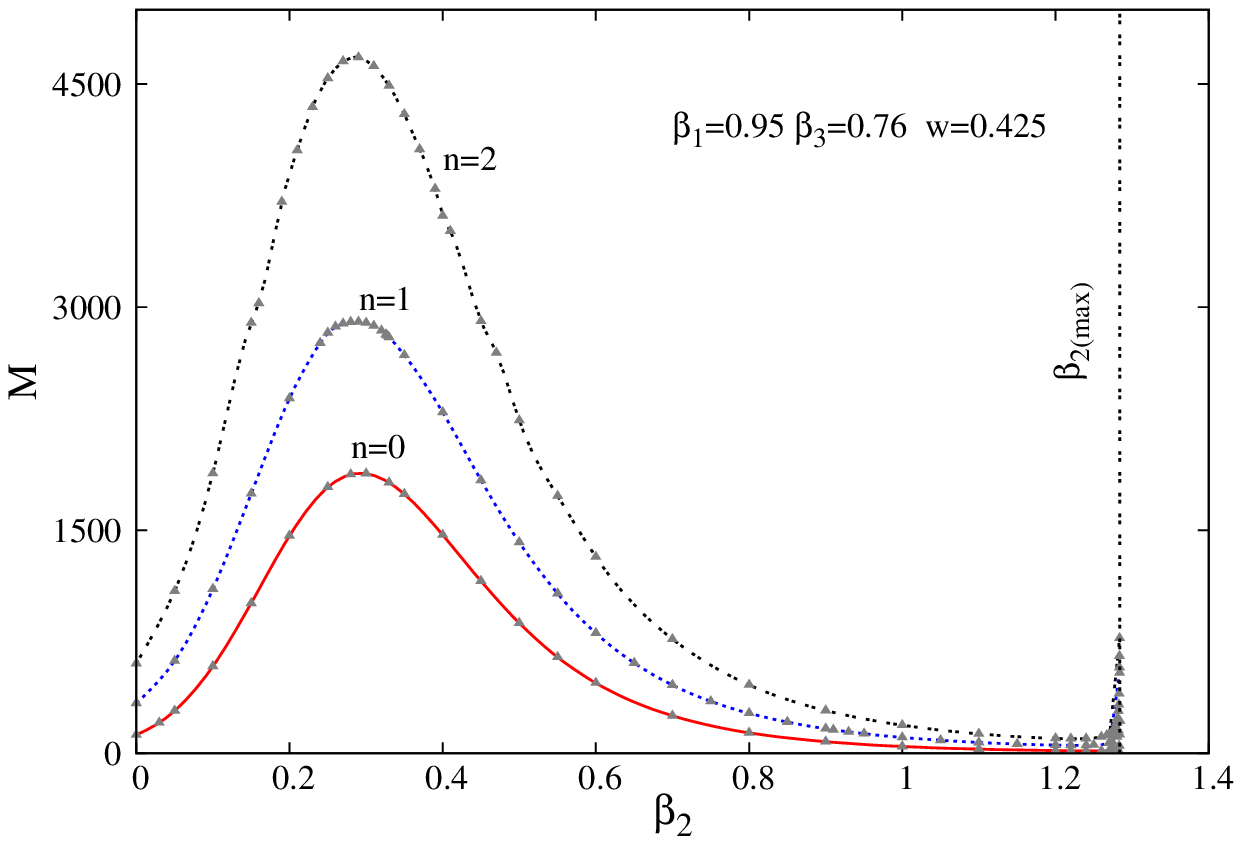,width=8cm}}
\end{picture} 
\\ 
\\
{\small {\bf Figure 5.}  The dependence of the $k=0$ solutions
on the parameters $\beta_1$, $\beta_2$
is shown for several values of the winding number $n$.
}  
\\
\\
\setlength{\unitlength}{1cm}
\begin{picture}(8,6)
\put( 0.,0.0){\epsfig{file=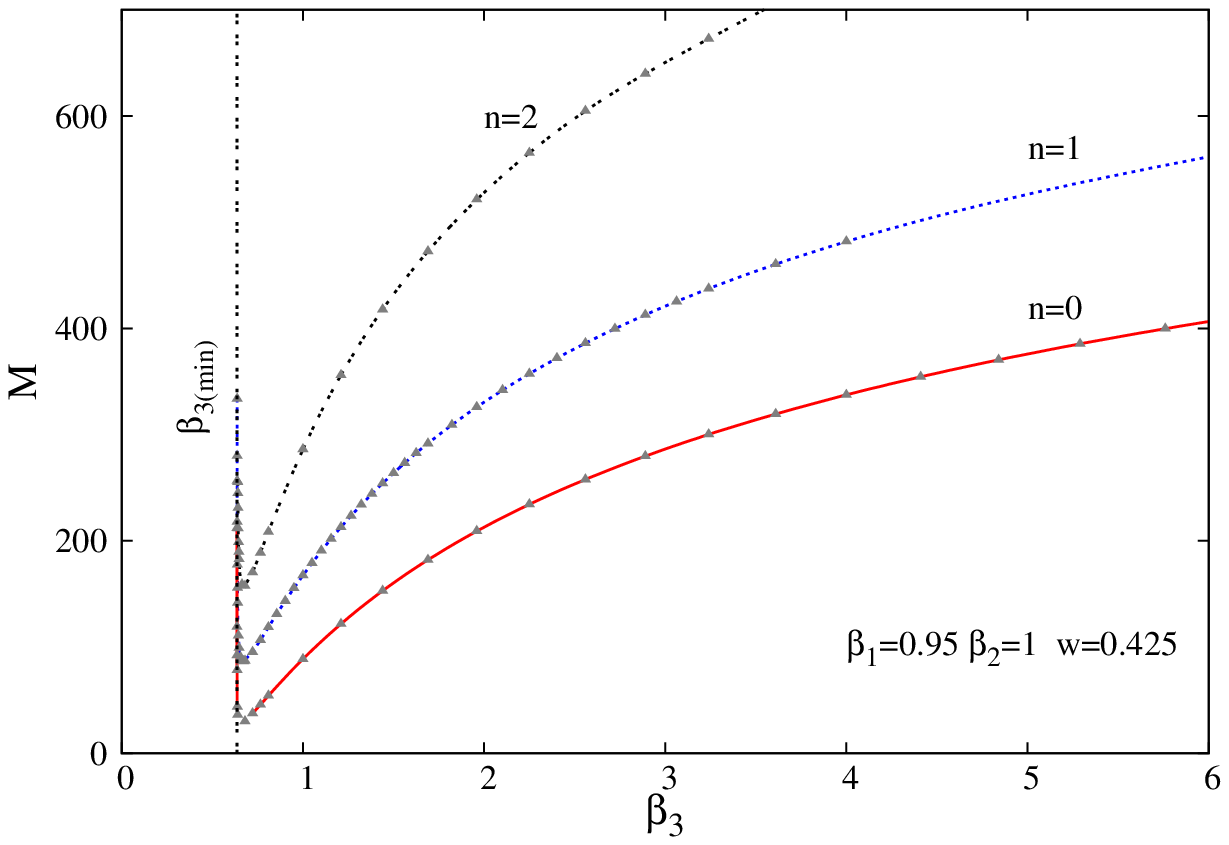,width=8cm}}
\put(8.7,0.0){\epsfig{file=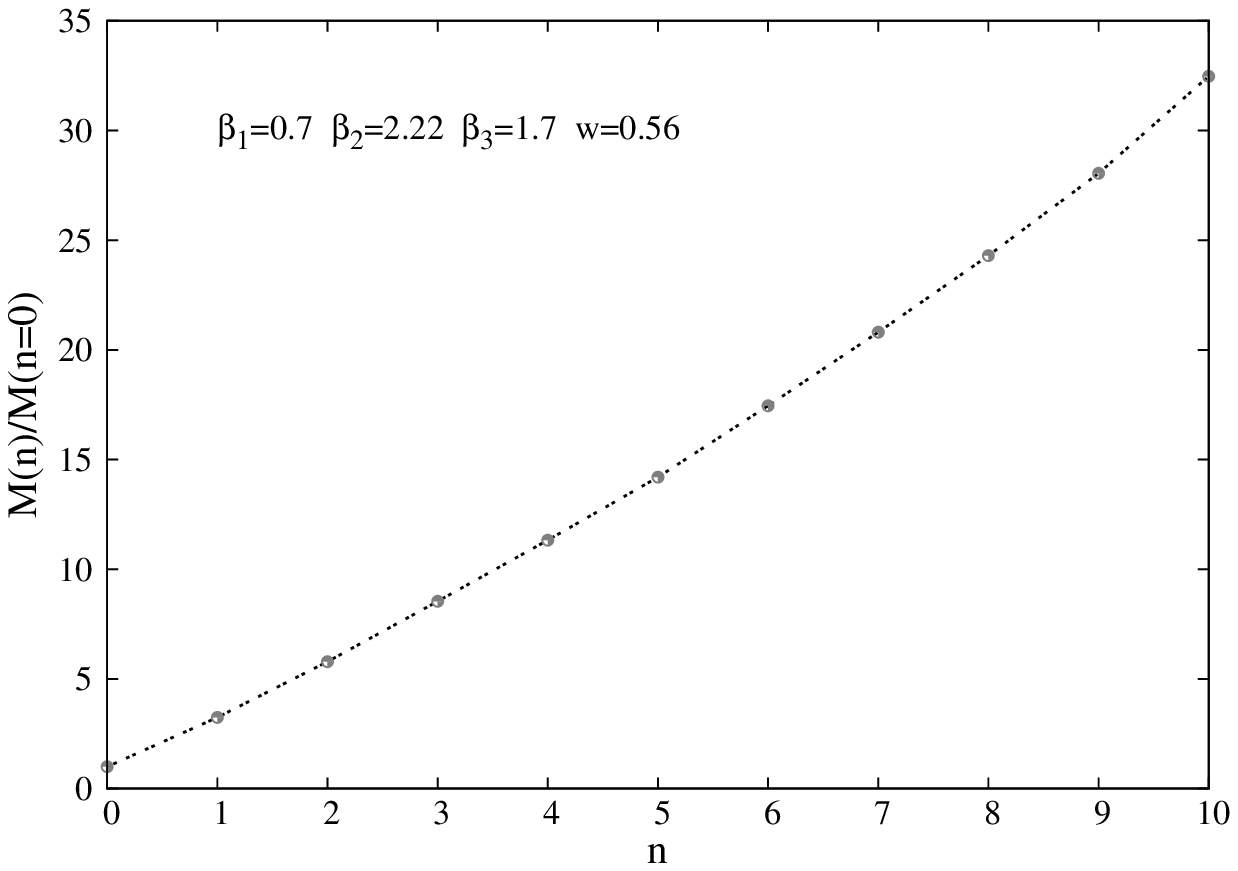,width=8cm}}
\end{picture} 
\\ 
\\
{\small {\bf Figure 6.} {\it Left:} The dependence of the $k=0$  solutions
on the parameter  $\beta_3$,  
is shown for several values of the winding number $n$. 
Here and in Figure 5,
the vertical lines correspond to the limiting values of $\beta_i$ as given by (\ref{max-w1}).
{\it Right:} The mass of typical $k=0$ vorton solutions is
shown as a function of the azimuthal winding number $n$ for fixed parameters of the potential
and a given frequency.
}  
 \vspace*{0.7cm}
 \\
solitons studied  by Friedberg, Lee and Sirlin  in \cite{Friedberg:1976me}.

For $n\neq 0$, the $k=0$ configurations satisfy the same boundary conditions as in 
  (\ref{bc-r0}),
  (\ref{bc-inf}),
  (\ref{bc-axis}), 
  (\ref{bc-Pi2})
  (with $Y\equiv 0$ this time).  
The profiles of a typical solution are shown in Figure 3, together with the corresponding
distributions for 
the energy density $-T_t^t$ and the angular momentum density $T_\varphi^t$.
There we plot the quantities for a slice corresponding to the  $z=0$ plane (left panel) and also for the $y=0$ plane 
(with $x,y,z$ the usual cartesian coordinates).

One can see that the functions $X,Z$
possess  a pronounced angular dependence, with a nonzero minimum
of $X$ located in the $\vartheta=\pi/2$ plane $(z=0)$, at some value $r=R$.
There the function $Z$ approaches the maximal value, 
and likewise the energy and angular momentum densities.
Based on that, we can define formally an effective ring radius $R$ also for the $k=0$
solutions\footnote{Alternatively, following \cite{japan}
one can define a mean radius of the configurations
$R=\frac{\int r j^t_{(\sigma)}\sqrt{-g}d^3 x}{\int j^t_{(\sigma)}\sqrt{-g}d^3 x}$
.}.
The energy density  typically exhibits a torus-like
structure. (This becomes apparent by considering surfaces of constant
energy density of $e.g.$ half the respective maximal value of the energy density.)

Here we note that the $k=0$ configurations are obtained easily, 
and with a better numerical accuracy than any other solutions
in this work.
Therefore it was possible to investigate some regions of the five-dimensional parameter space in a systematic way.
Starting with the dependence on the frequency, we note that the behaviour found in \cite{Kleihaus:2005me}
for flat spacetime spinning Q-ball solutions is recovered, see Figure 4.
(Similar results have been found also for other sets
of input parameters.)
As one can see, the solutions exist for a limited range of frequency, $w_{min}<w<w_{max}$ (with $w_{max}=M_{\sigma}/M_{\phi}$
for the scaling used in this work, see (\ref{max-w1})).
Both mass and charge (or, equivalently $J=nQ$) diverge at the limits of this interval.
They assume minimal values at some intermediate value $w_{cr}$.
Then a diagram $M(Q)$
shows the existence of two branches, see Figure 4 (right).
For the same $Q$, 
the solution with $w<w_{cr}$
has lower mass than the corresponding one with $w>w_{cr}$.
 Moreover, those configurations  which have a smaller mass than   a
collection of non-interacting field quanta with the same charge  are expected to be
stable with respect to decay to the free particles\footnote{This conjecture is based also on 
similarities we have found between the $k=0$ solutions and the spinning
Q-balls in a model with a single complex scalar field.
In that case, as discussed in \cite{stable-Qballs} (though for a different 
formulation of the problem),
the $n= 0,1$ solitons are stable for some range of frequency, both at the linear and non-linear levels.
}, 
at least for small values of $n$.

The dependence of the solutions on the parameter $\beta_1$ is exhibited in Figure 5 (left).
One can see  that
the solutions exist only for a limited interval, $\beta_{1(min)}<\beta_1<\beta_{1(max)}$, 
and the 
mass  diverges as the limits are approached.
Interestingly, the value of $\beta_{1(max)}$ does not depend on the winding number $n$,
with $\beta_{1(max)}=\sqrt{ {2(\beta_3-w^2)}/{\beta_2}}$, see the relation (\ref{max-w1}).

When studying instead the dependence on $\beta_2$, one first notices  the existence of axially symmetric generalizations of the 
FLS spherically symmetric solitons \cite{Friedberg:1976me}, which are found for $\beta_2 \to 0$.
Also, the mass  seems to diverge when a critical, maximal value of $\beta_2$
is approached, see Figure 5 (right).
As
implied by (\ref{max-w1}), 
this maximal value is given by $2(\beta_3-w^2)/\beta_1^2$ 
(and thus again it does not depend on $n$).

We have performed a similar study also for $\beta_3$, see Figure 6 (left).
One can notice the existence of a minimal value of this parameter, $\beta_{3(min)}=w^2+\frac{1}{2}\beta_2\beta_1^2$,
see (\ref{max-w1}).
As  $\beta_{3}\to \beta_{3(min)}$ the mass 
diverges.
At the same time, no upper bound on $\beta_3$ appears to exist, at least for the set of parameters we have considered.
Finally, in Figure 6 (right), we plot the dependence of mass on the winding number $n$ 
(where one can notice an almost linear relation).
Also, as expected, the angular dependence of the scalar functions
increases with the winding number $n$ and the location of the
biggest angular splitting moves further outward, as well as the global maximum of the energy density.

Note also that for all cases in Figures 5, 6, the charge $Q$ shows a similar pattern.
(Here we recall that in all plots in this work the global charges are expressed in natural
units set by $\eta_{\phi}, \lambda_{\phi}$.)

The  $k=0$ solutions possess also an interesting
 spherically symmetric limit, in which case $n=0$ and 
$X(r,\theta)=X(r)$ and $Z(r,\theta)=Z(r)$ (with $Z(0)\neq 0$ this time).
Some properties of these static solutions are discussed in a more general context in Appendix C.
Here we mention that, as seen in Figures 4-6, they share the basic properties
of the spinning solutions.

 We close this part by remarking that
since the study of the  $k=0$ solutions does not pose special numerical difficulties,
and also  the corresponding vortices possess a finite mass per unit length, 
these may
open the possibility to work out the details of 
the heuristic construction described in Section 3.1.
We hope to report elsewhere on that.

\subsection{$k=1$  vortons}

The distinguishing feature of these solutions is that the vortex field
$\phi$ has a zero on a circle in the $\theta=\pi/2$ plane ($i.e.$ at $z=0$).
The profiles of a typical solution are shown in Figure 7.
There the plots are again given for the $(xy)$ and $(xz)$ planes.
(Note, that the $X,Y,Z,|\phi|$ profiles are shown in Figure 15 in terms of cylindrical coordinates $(\rho,z)$.)
We also recall that $x,y,z$ correspond to dimensionless coordinates, the length scale being set by $1/M_{\phi}$). 

A description of the generic properties
of the scalar functions can be made as follows.
At $r=0$, $Z$ vanishes, while $X$ has a local (negative) extremum.
Near the origin, the functions $X, Y, Z$ show a large 
\newpage
\setlength{\unitlength}{1cm}
\begin{picture}(15,20.85) 
\put(-1,1){\epsfig{file=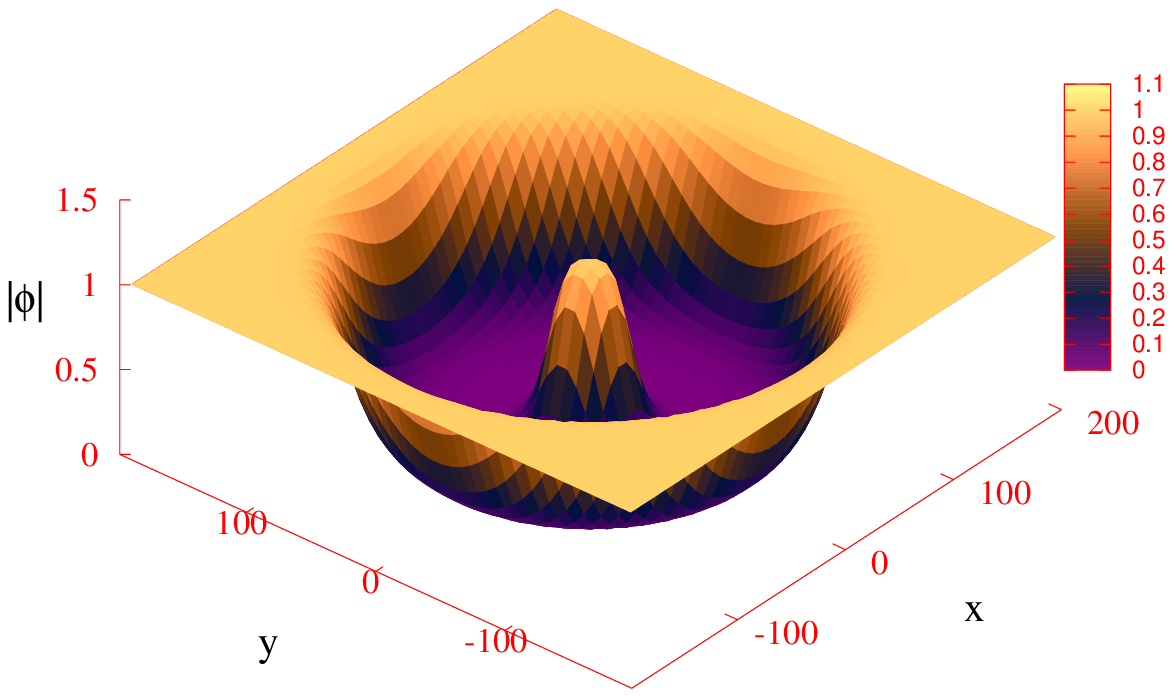,width=7.5cm}}
\put(7,1){\epsfig{file=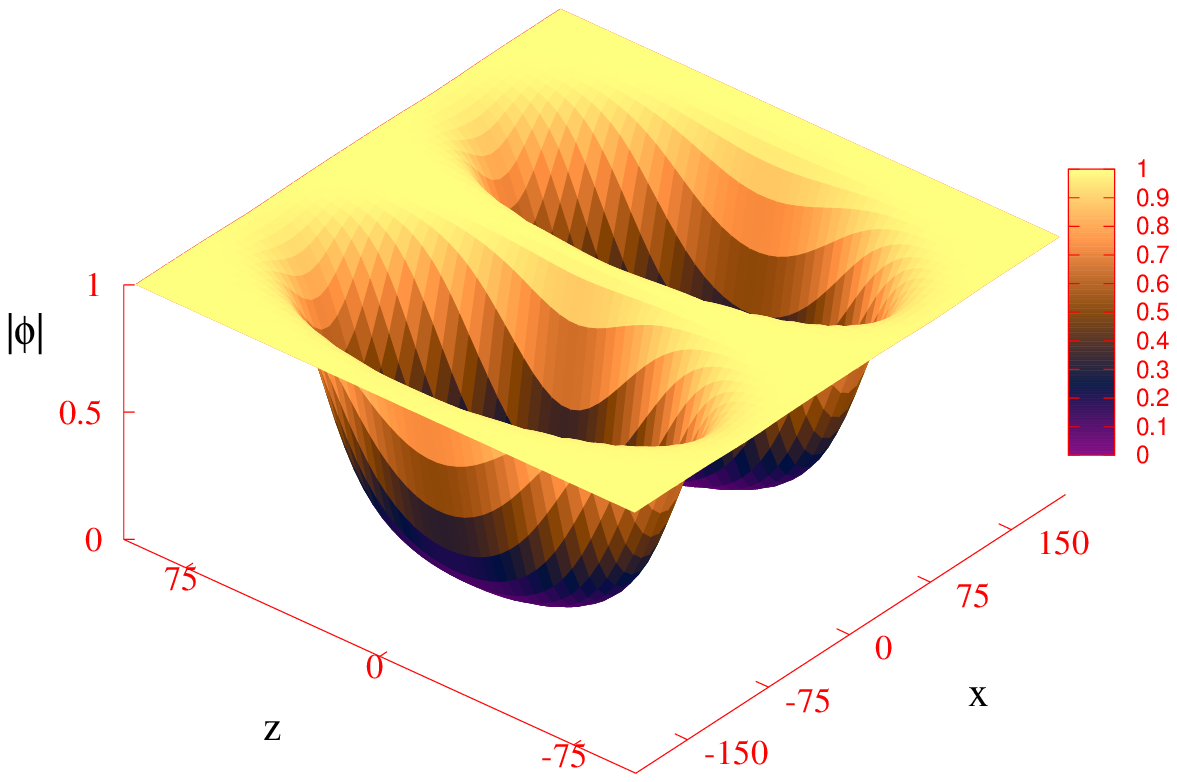,width=7.5cm}}
\put(-1,6.5){\epsfig{file=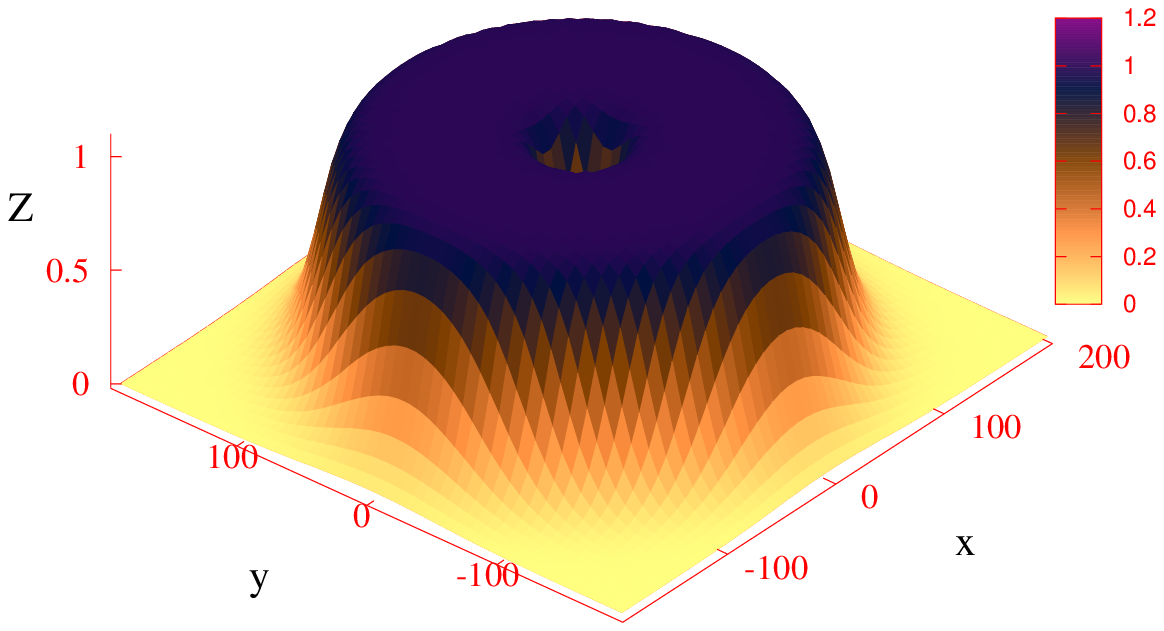,width=7.5cm}}
\put(7,6.5){\epsfig{file=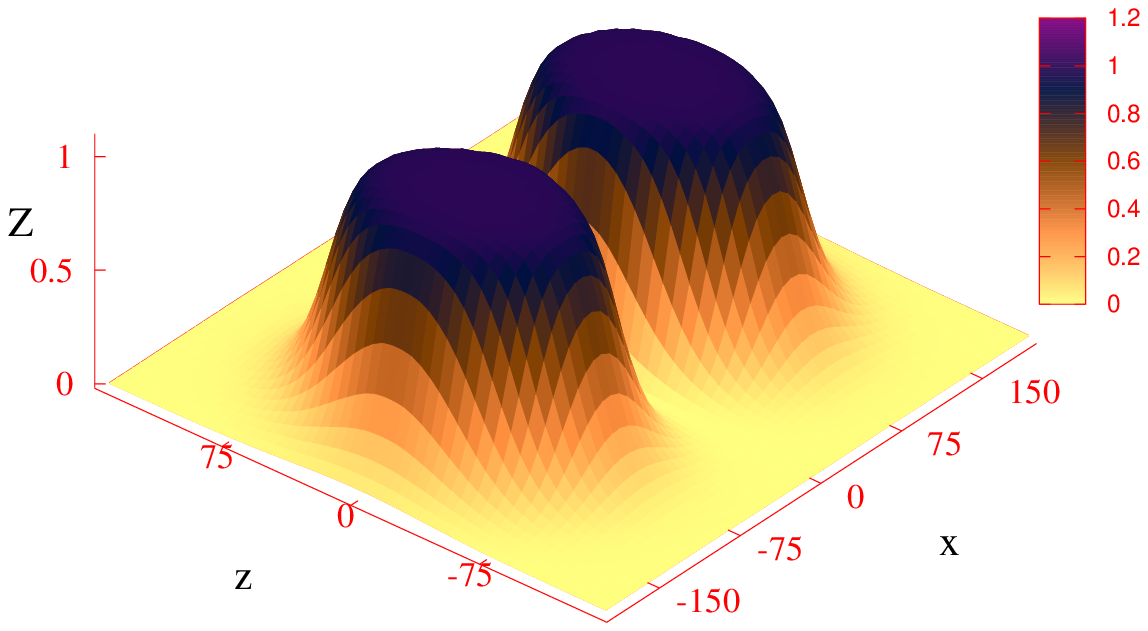,width=7.5cm}}
\put(-1,12.){\epsfig{file=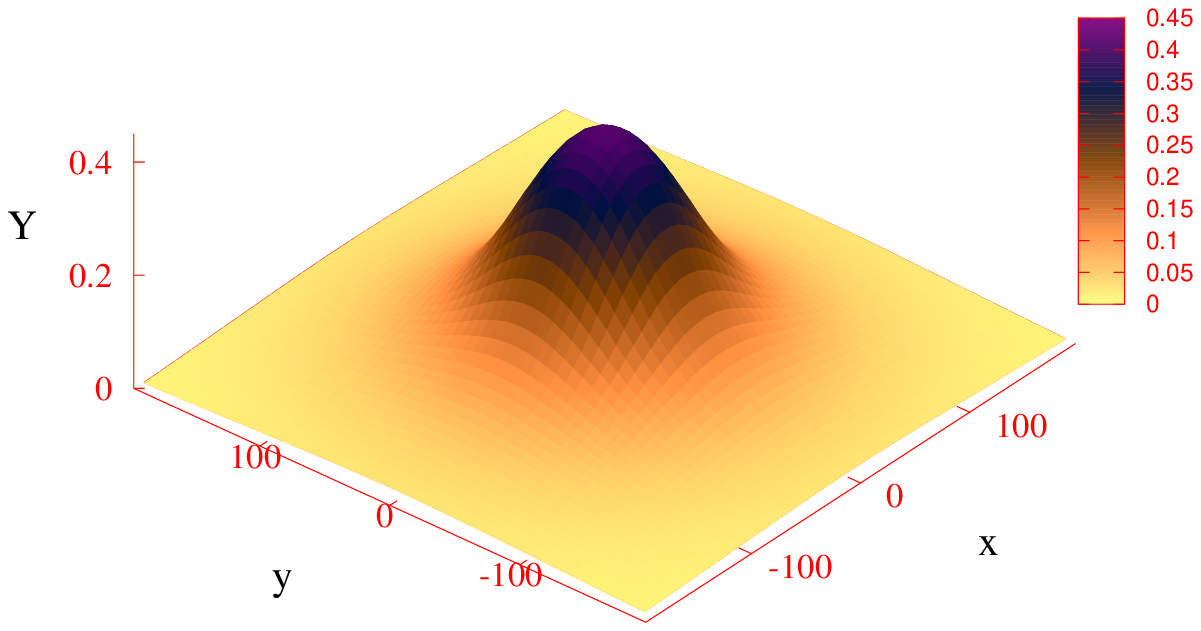,width=7.5cm}}
\put(7,12.){\epsfig{file=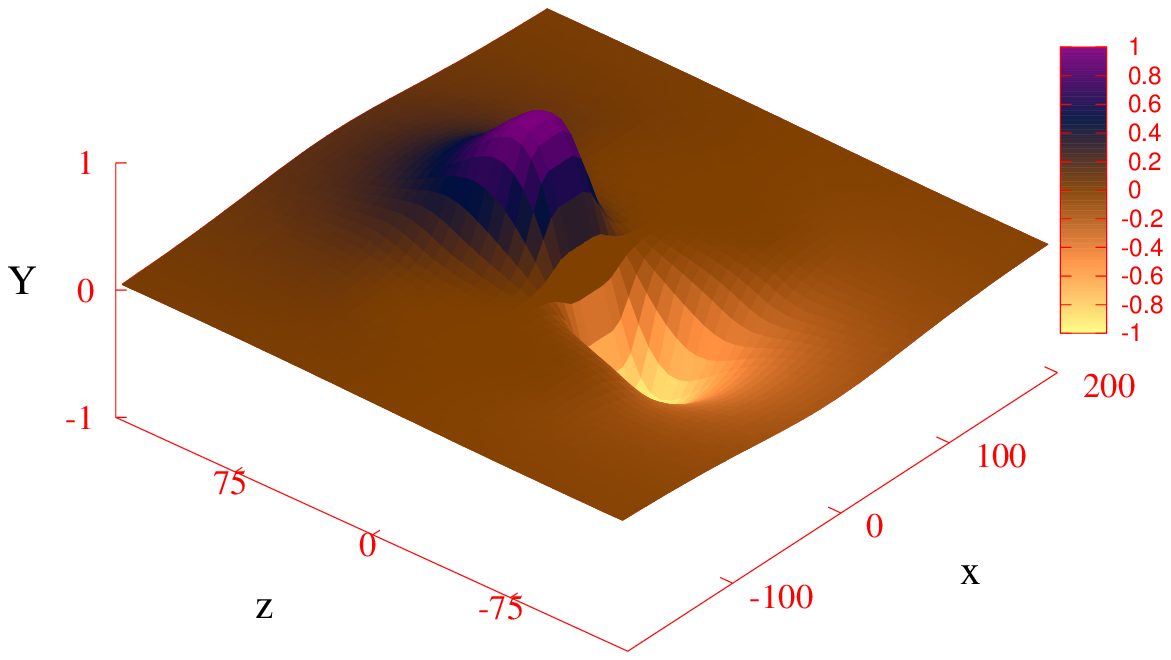,width=7.5cm}}
\put(-1,17.25){\epsfig{file=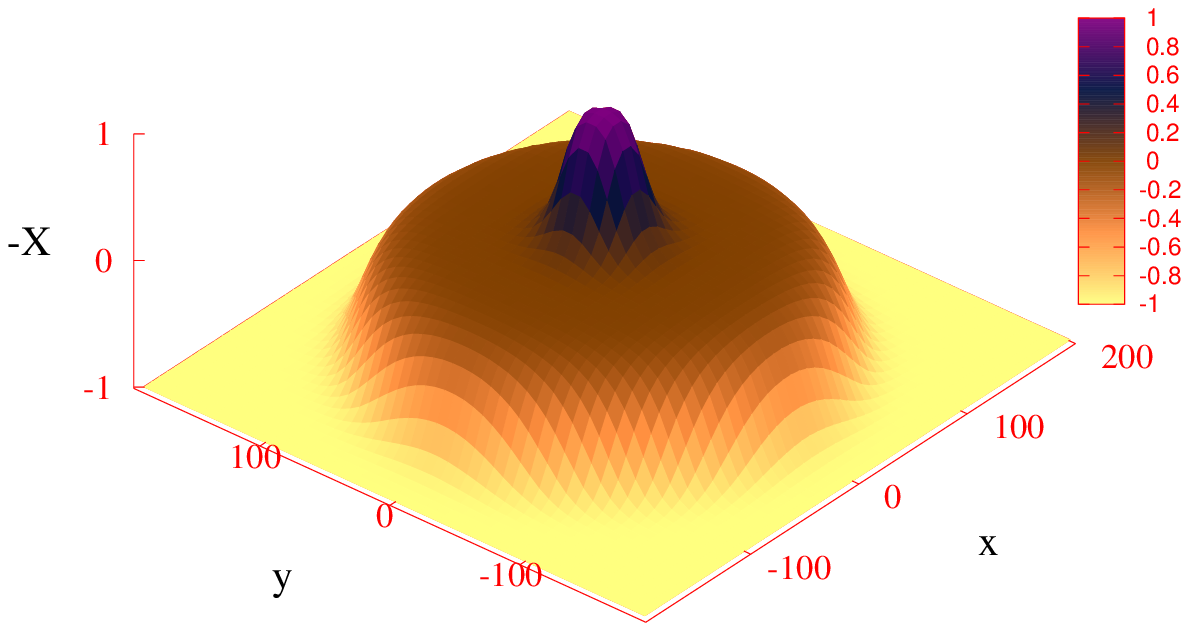,width=7.5cm}}
\put(7,17.25){\epsfig{file=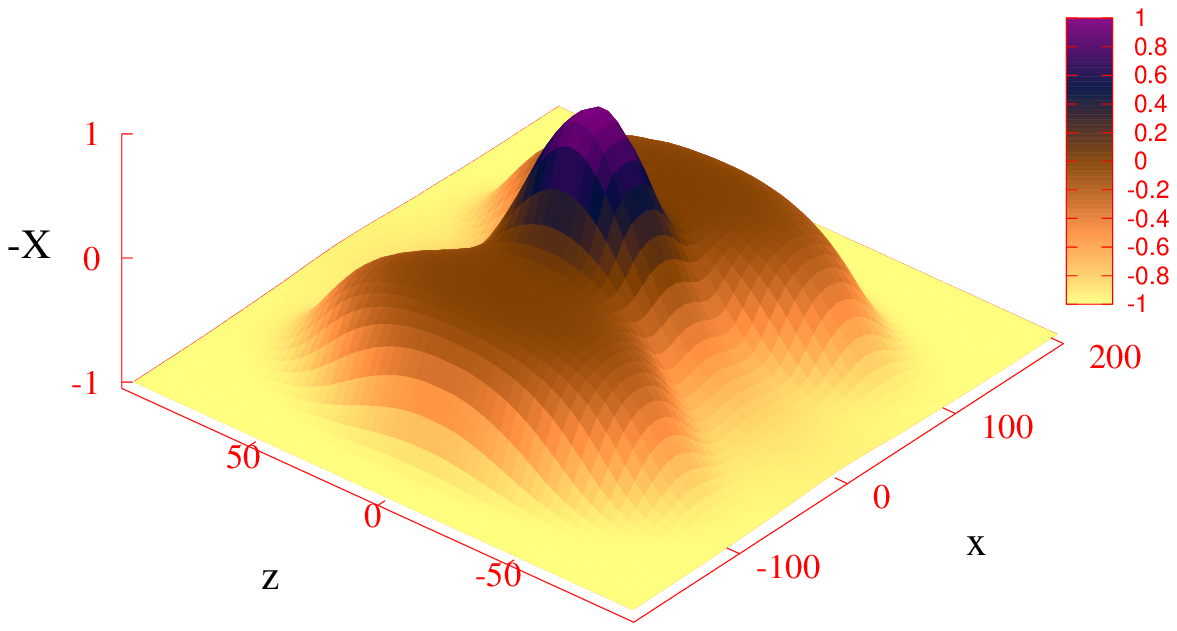,width=7.5cm}} 
 \end{picture} 
\\
{\small {\bf Figure 7.} The scalar functions $X$, $Y$, $Z$ and the amplitude $ \sqrt{X^2+Y^2}$
of the  field $\phi$ are shown for 
a typical $k=1$ vorton solution.  }  


\newpage
\setlength{\unitlength}{1cm}
\begin{picture}(15,20.85) 
\put(-1,12){\epsfig{file=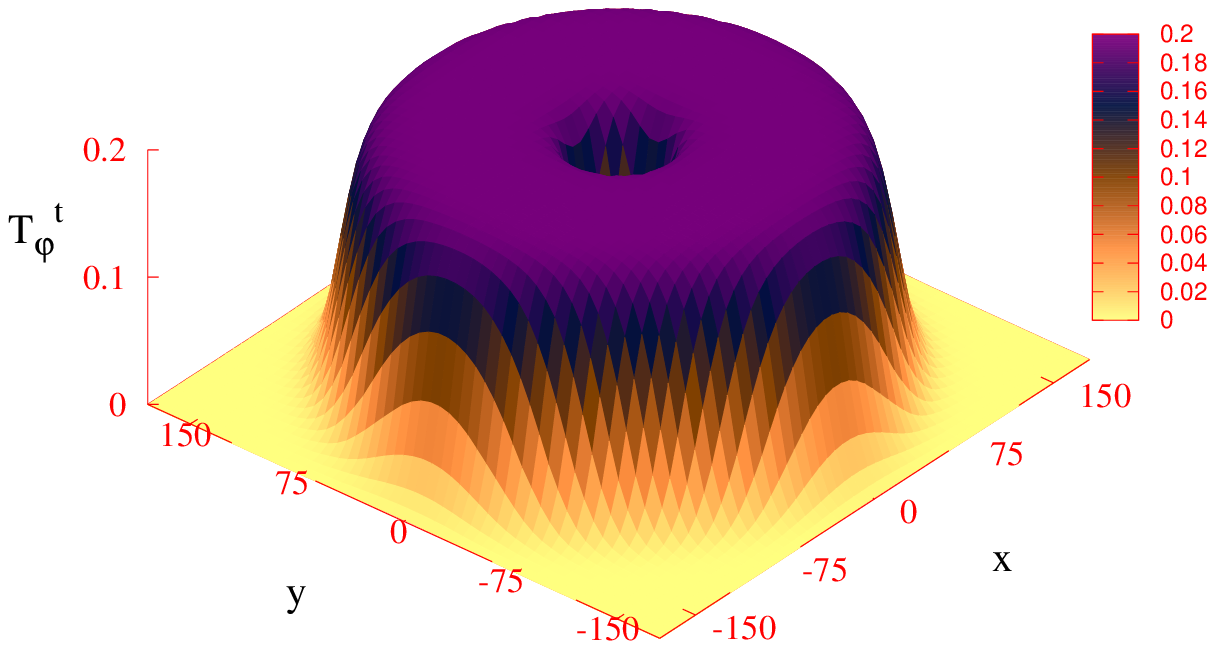,width=7.5cm}}
\put(7,12){\epsfig{file=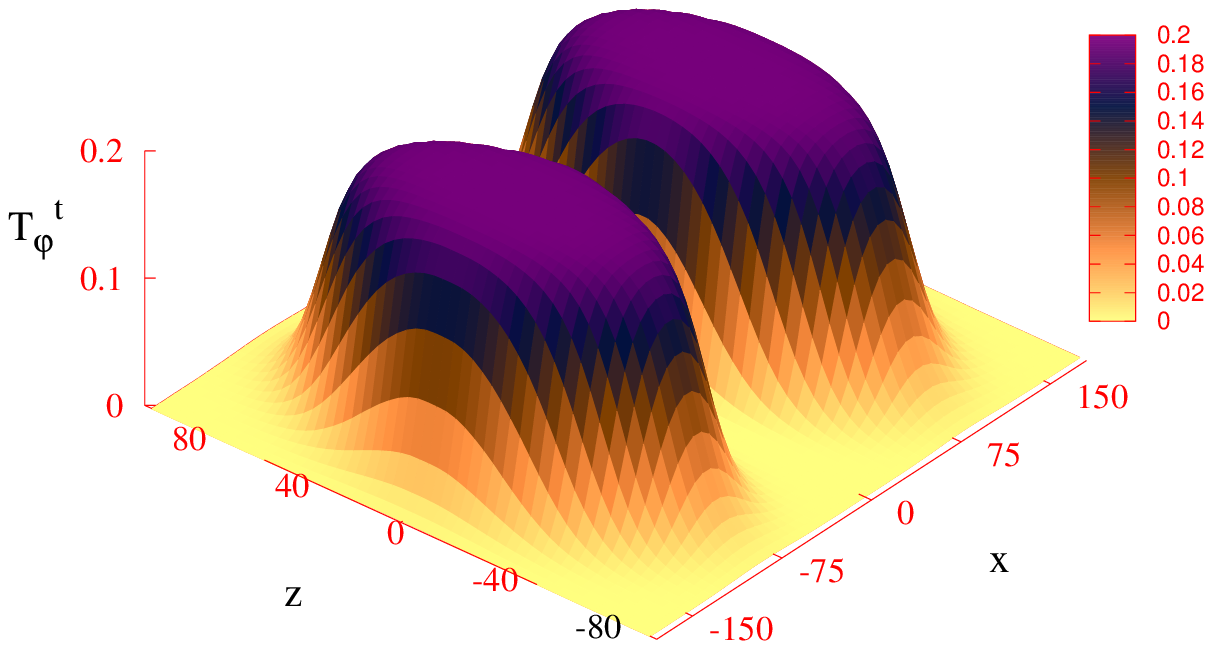,width=7.5cm}}
\put(-1,17){\epsfig{file=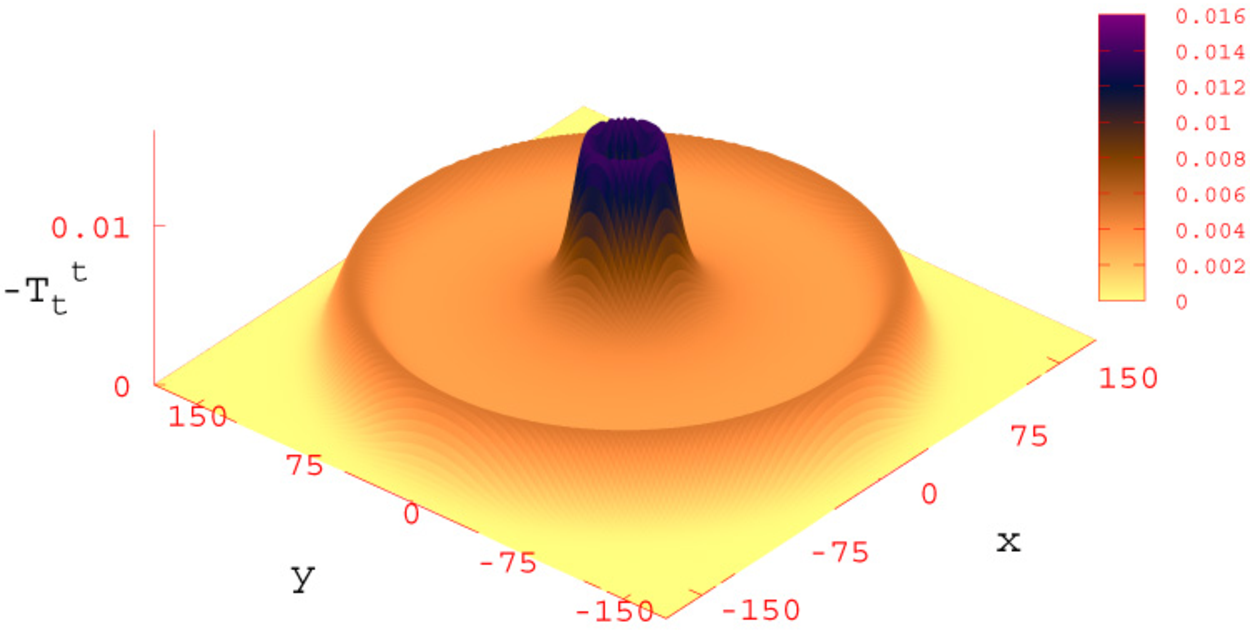,width=7.5cm}}
\put(7,17){\epsfig{file=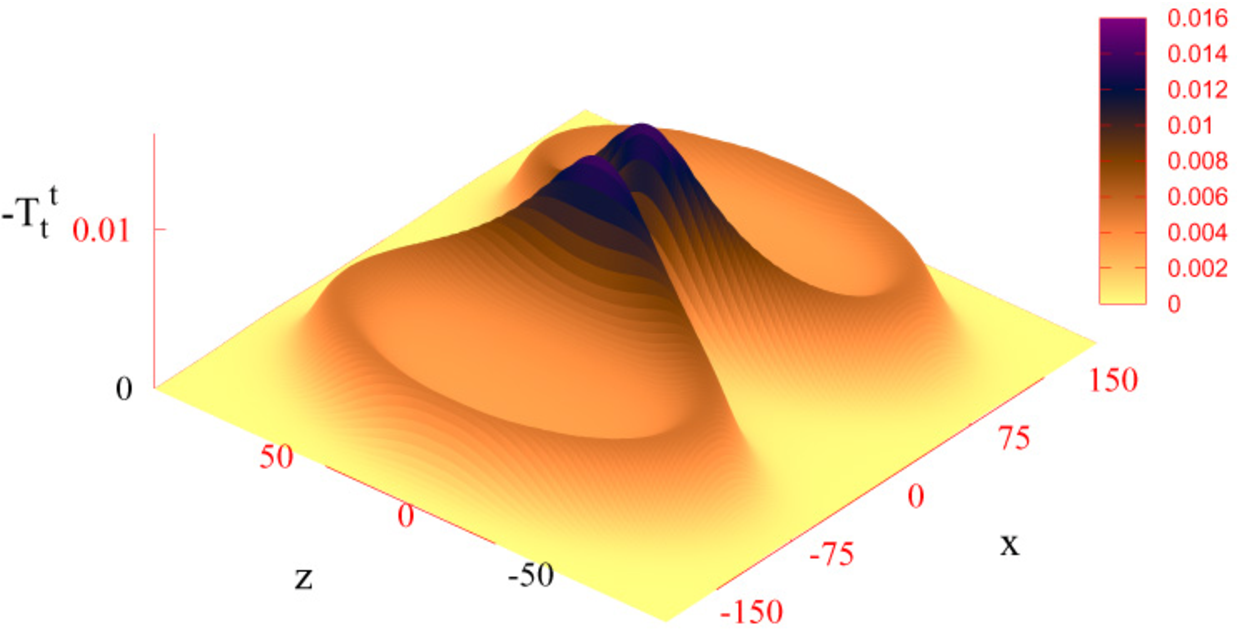,width=7.5cm}} 
\end{picture} 
 \vspace*{-11.5cm}
 \\
 {\small {\bf Figure 8.} The  energy density $-T_t^t$ and the angular momentum density $T_\varphi^t$ 
are shown  for the same solution in Figure 7.
The input parameters of the solution are
$w=0.05$,  $n=2$ and  $\beta_1=\beta_2=1,~\beta_3= 0.51$.
Here and in Figure 7, 
the left plots are for the $z=0$ plane (except for the $Y$-function, where we take $z=80$), while the right plots are for a plane containing the axis of
symmetry.
} 
\vspace{0.5cm}
\\
variation, with $Y$ approaching
its extremum. 
Then there follows an intermediate region located inside the vortex, where $X$ stays very close to
zero, while the function $Z$ is almost constant and close to its maximal 
value.
(Typically, this region is on the order of $10^2$ units of the fundamental length scale $1/M_{\phi}$.)
There, the function $X$ crosses the value zero on a circle of radius $R$ at $z=0$, with a small but non-zero first derivative,
 $\partial_r X(R,\pi/2)\neq 0$.
Finally, there is a third region outside the
vortex ring, where the fields approach their vacuum values. 
There, because of  (\ref{asympt}), the function $Y$ has a slower decay than 
the functions $X$ and $Z$.
Also, both $X$ and $Z$ show large variations
in a transition domain between the second and the third region, 
approaching afterwards very fast their 
asymptotic values.

The corresponding plots for the energy density $-T_t^t$ and  the angular momentum density $T_\varphi^t$ 
are shown in Figure 8.
One can see that, as expected, the shape of $T_\varphi^t$
copies the profile of the scalar function $Z$.
However, the energy  density distribution 
has some special features which single out vortons among the field theory solitons.
For all studied configurations, 
the energy density has a local maximum 
at the origin\footnote{Note that, 
for  spinning $Q$-balls, the energy density vanishes at $r=0$.
However, this is not the case for vortons,  due to the nonzero
contribution to $T_t^t$ of the $\phi$-field potential.}, 
whose value decreases with $n$.
Most of the vorton's energy is concentrated in a shell associated
with a circular domain wall of cross-section $r_0$. 
(The value of $r_0$ gives the transition between the 
second and third characteristic 
regions as defined above.)
The center of this shell is located at $r = R, \theta = \pi/2$.
Then, for the typical solutions in this work, 
the shape of the vorton energy density resembles a hollow tube 
(although there is always an almost constant nonzero value of $T_t^t$
inside that tube).
When the azimuthal winding number $n$ increases, the position of 
this tube
moves outwards.
More relevant plots for typical solutions are given in Ref. \cite{Radu:2008pp}
(see also the profiles in Figures 15, 16).

The vorton topology for a solution with $k=1$ is illustrated by the diagrams in Figure 9.
On the left picture we plot the lines of constant amplitude for the scalar $\phi$
(closed curves encircling the center of the vortex
where $\phi$ vanishes) and phase (radial lines emanating from the center of the vortex)
for a typical vorton
 
 \newpage
\setlength{\unitlength}{1cm}
\begin{picture}(8,6)
\put(-0.5,0){\epsfig{file=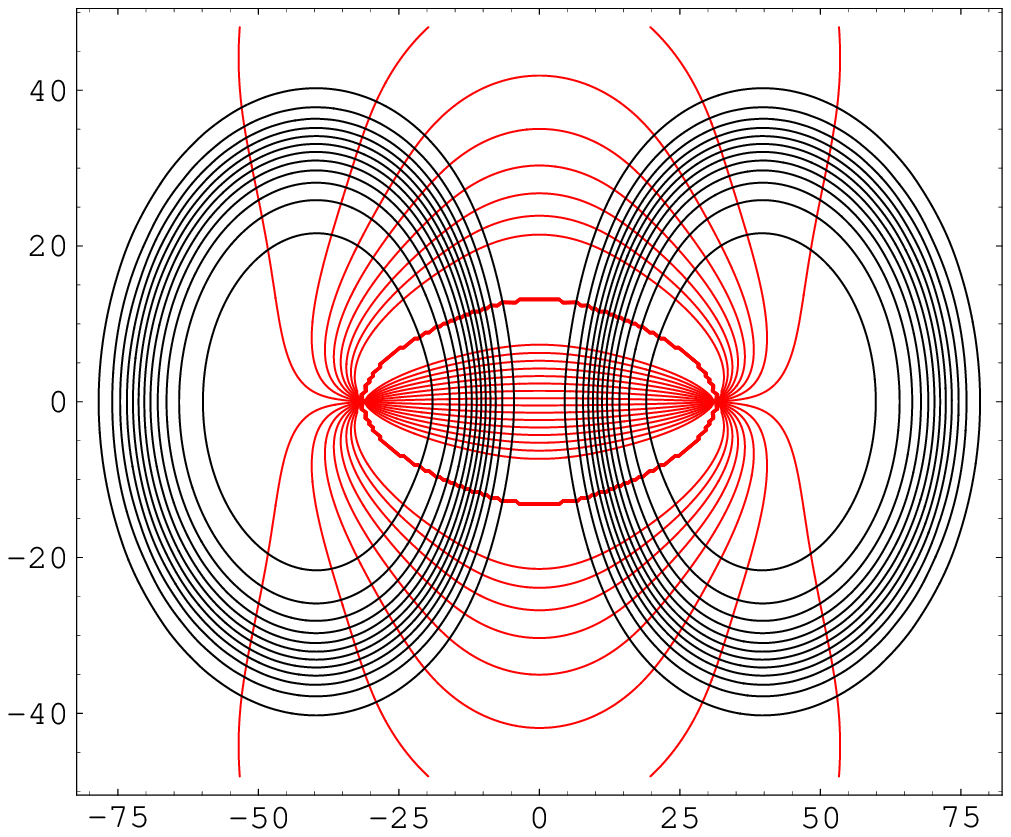,width=7cm}}
\put(8,0){\epsfig{file=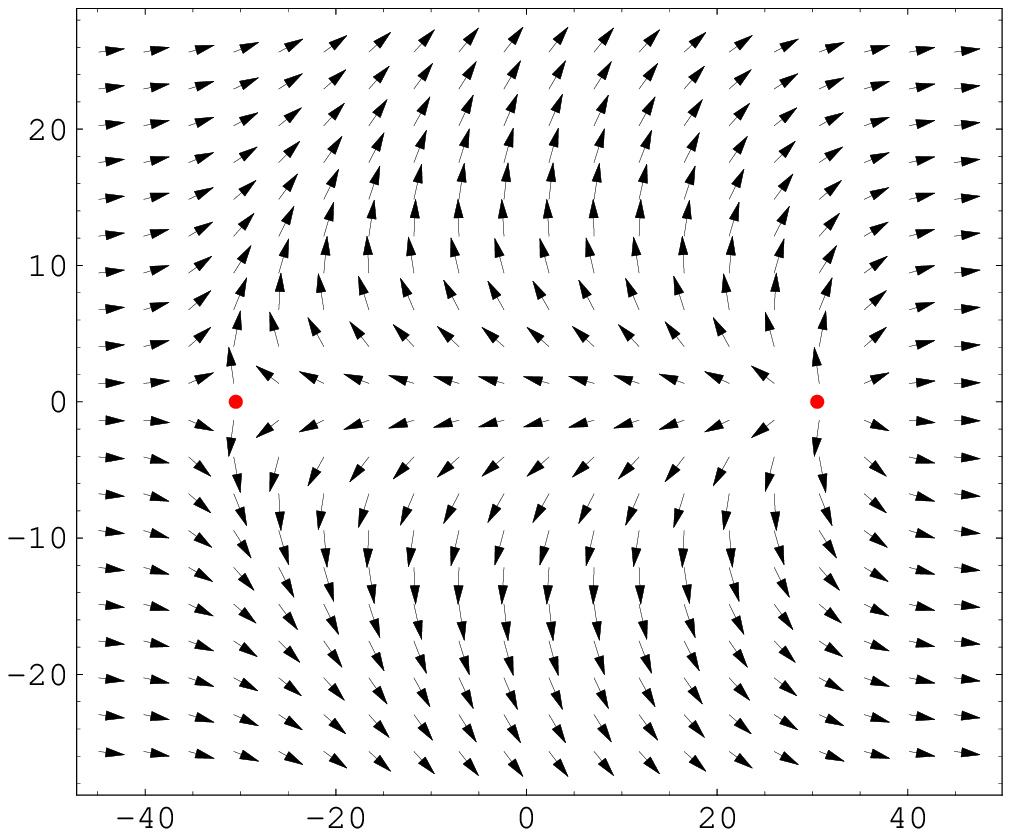,width=7cm}}
\end{picture}
\vspace*{0.3cm}
\\
\\
{\small {\bf Figure 9.} The behaviour of the field $\phi$ is displayed
for a typical $k=1$ vorton solution with $\beta_1=1$, $\beta_2=0.97$, $\beta_3=0.55$,
$w=0.13$ and $n=2$.
In the left panel we plot levels of constant amplitude $|\phi(x,0,z)|$ (closed lines) and phase $\psi(x,0,z)$ (radial lines).
In the right panel, the horizontal and vertical length  of each arrow is proportional
to the real and imaginary part of $\phi$ ($X$ and $Y$, respectively).
The red points there indicate the position of the ring.
   }
 \vspace*{0.75cm}
\\
 solution.
The same picture is shown on the right panel in a different
representation.
There, the horizontal and vertical length  of each arrow is proportional
to the real and imaginary part of $\phi$ ($X$ and $Y$, respectively).
One can see, for instance,
that the function $Y$
takes significant values only  in a region with a radius 
of about $50/M_\phi$, surrounding the vorton's core.

A systematic analysis of the parameter space of vorton solutions 
would be a very difficult task that we did not aim at here.
Also, unfortunately, some regions of the parameter space 
are not accessible within our numerical approach.
The solutions investigated so far cover a domain centred around the point\footnote{The thin vorton solutions in \cite{Battye:2008mm}
have been constructed for a choice of the parameters of 
the potential  
$\beta_1=1$, $\beta_2=\beta_3=2/3$.
} 
\{$\beta_1 = 1$,  
$\beta_2 = 0.95$,
$\beta_3 = 0.5$\},
with maximal deviations of $\beta_i$ around these values of about 7 percent.
Also, all configurations reported here (including the gravitating
solutions discussed in Section 4.2 below)
represent thick vortons. Thus 
the radius of the vortex ring does not differ significantly from the thickness
of the vortex.
The constructions of thin vortons seems to require
large values for the winding number $n$.
However, we did not yet manage to find within our numerical approach
solutions with $n>6$ with high numerical accuracy.

We also mention that these vortons  have been constructed by descending from solutions
of the sigma model limit  (\ref{Ls-sigma}),
where the constraint $X^2+Z^2+Y^2=1$ is satisfied.
However, most of the solutions in this work exhibit large deviations 
from this value, with $|\phi|^2+|\sigma|^2$
 ranging typically from about 0.8 to about 1.1. 
 
 The picture we have found when attempting to vary the input parameters
seems to be rather similar to the one found in  the $k=0$ case. 
(Although,  for true vortons, the study of the limiting solutions
is much more difficult.)  
In particular, the dependence of the solutions on the 
frequency looks qualitatively similar to the one exhibited in Figure 4.
 The $Q-$ball type behaviour is recovered.
The solutions exist for a limited range of frequencies,
and  $M(Q)$ exhibits again two branches.
(Plots illustrating this behaviour can be found in Ref. \cite{Radu:2008pp}.)

Provided both $k=0$ and $k=1$ solutions exist for a given set of parameters, 
then the
mass and angular momentum of the $k=1$ solutions 
are found to be usually larger than those of the
corresponding $k = 0$ configurations. (Note, that we could construct
 $k=0$ `semitopological' vortons for a much larger set of values 
 of the potential parameters $\beta_i$ than in the $k=1$ case.)

An interesting feature of the $k=1$
solutions which, to our knowledge, 
has not been reported so far in the literature is their nonuniqueness. 
This means that for the same input parameters $(\beta_i,n,w)$ we have found 
two vorton solutions with different global charges.
The situation is illustrated in Figure 10, where we plot a
\newpage
\setlength{\unitlength}{1cm}
\begin{picture}(15,20.85) 
\put(-0.5,10){\epsfig{file=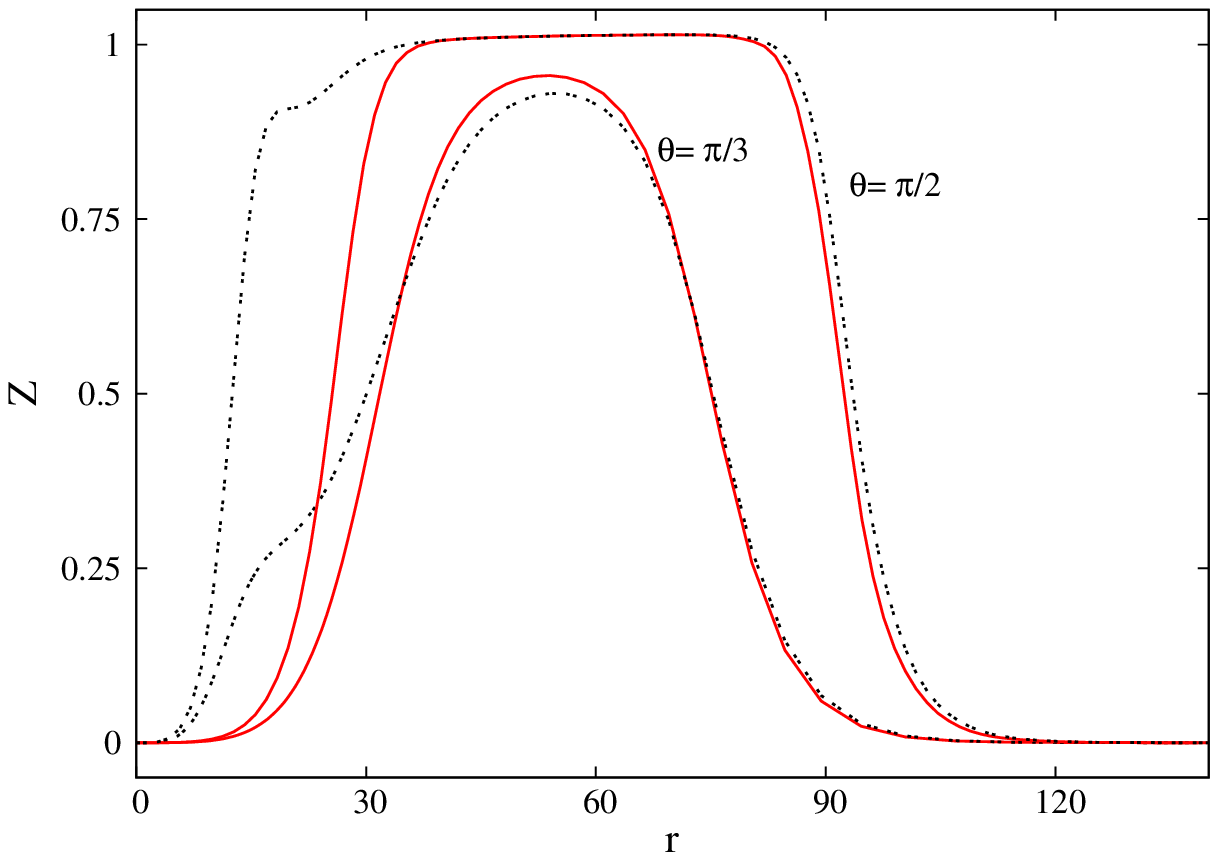,width=7.5cm}}
\put(8,10){\epsfig{file=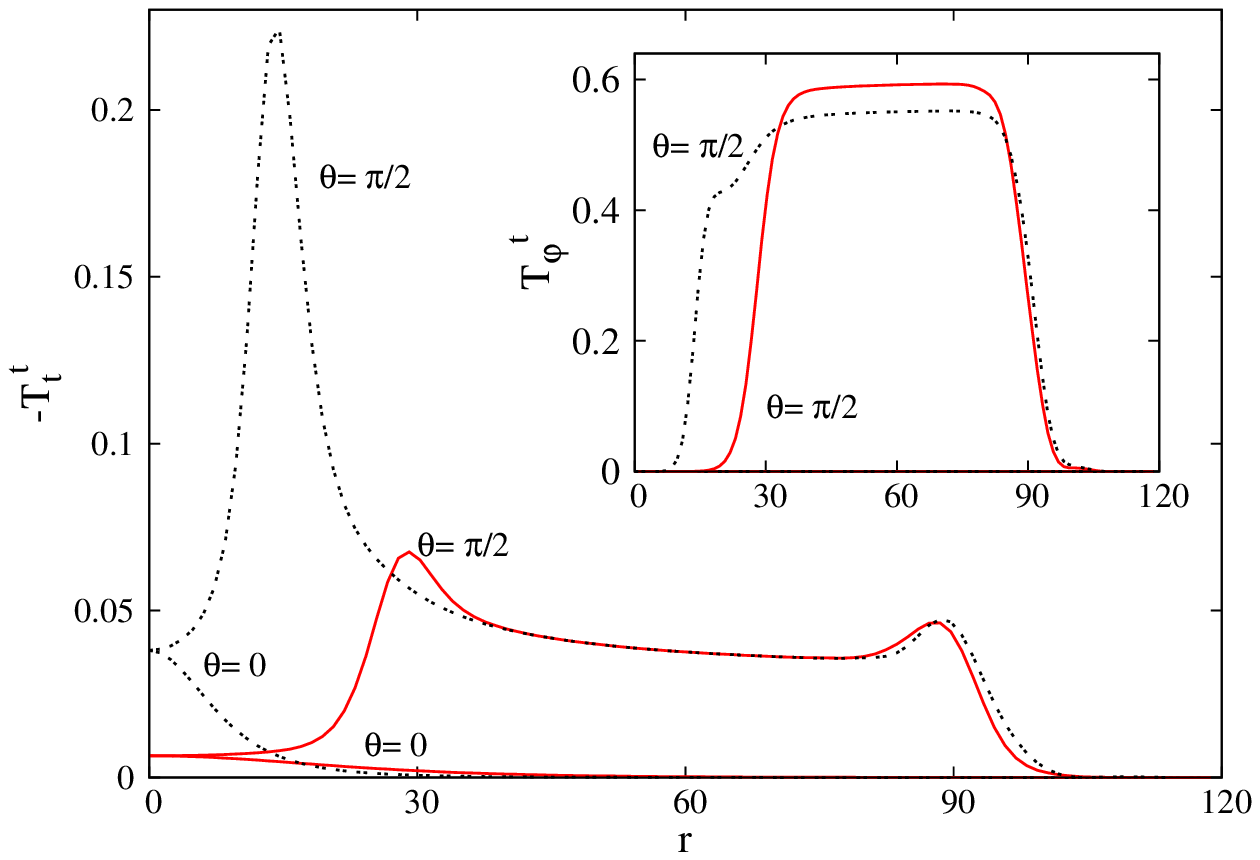,width=7.5cm}}
\put(-0.5,16){\epsfig{file=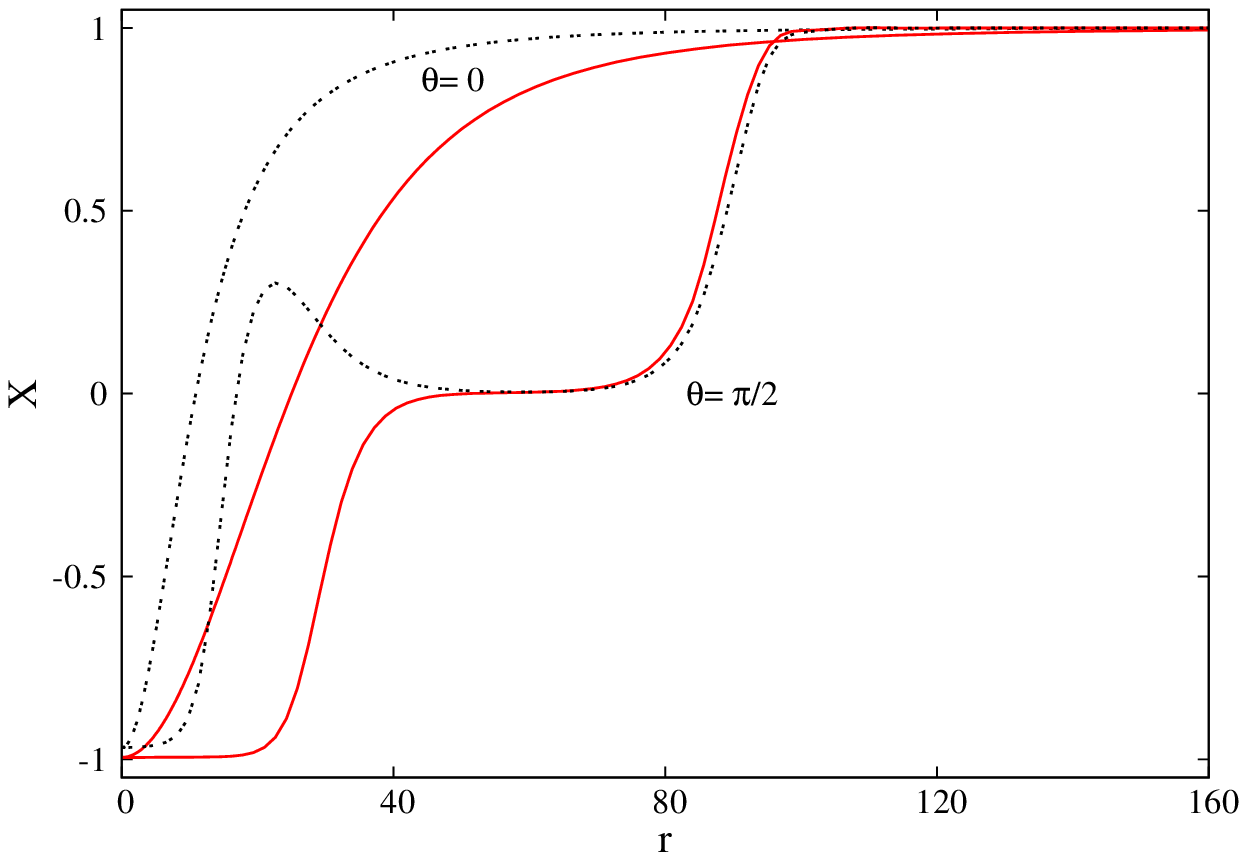,width=7.5cm}}
\put(8,16){\epsfig{file=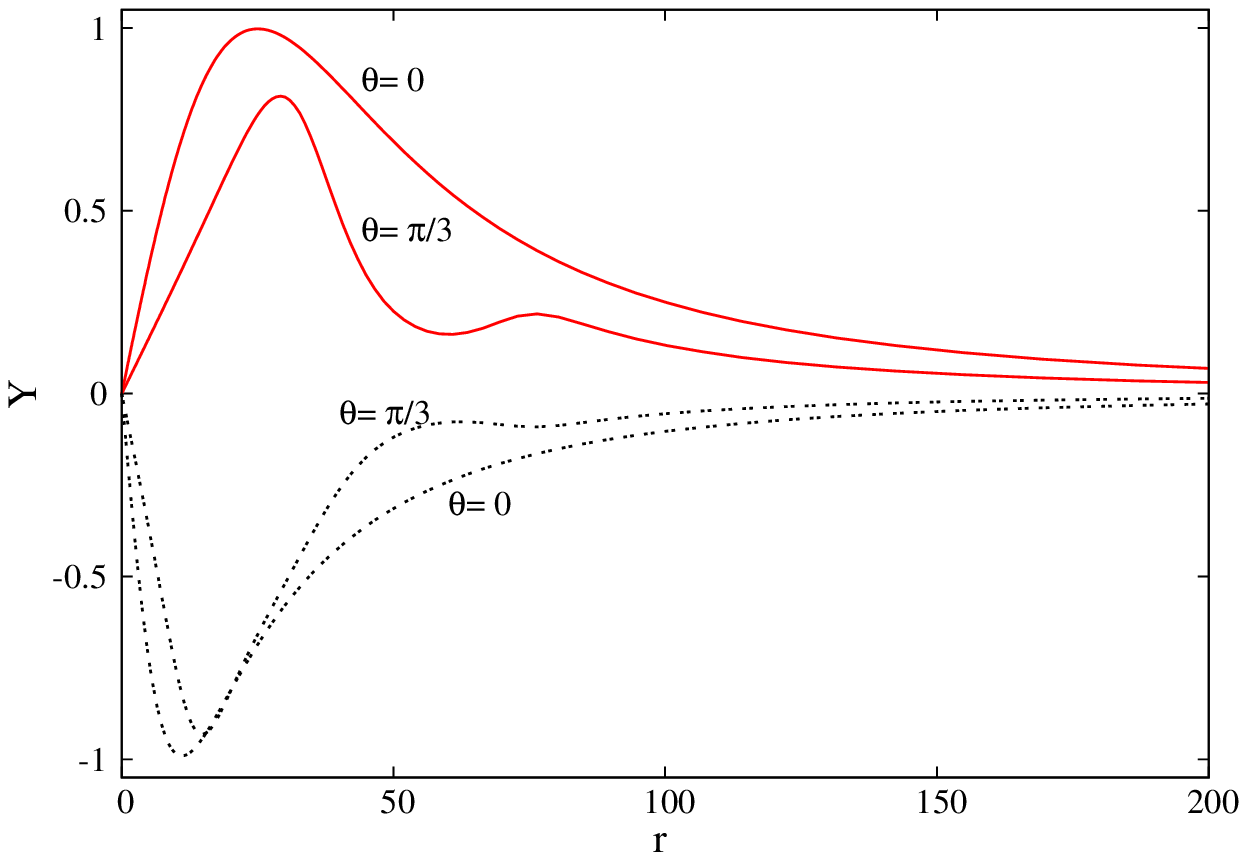,width=7.5cm}} 
\end{picture} 
\vspace*{-9.cm}
\\
{\small {\bf Figure 10.}
The profiles of the scalar functions $X(r,\theta)$, $Y(r,\theta)$, $Z(r,\theta)$
and of the energy density $-T_t^t$
are shown for two different $k=1$ vorton solutions
with the same input parameters: $\beta_1=1$, $\beta_2=0.97$, $\beta_3=0.54$, $w=0.135$ and $n=4$.  
The solution with dotted lines has higher mass and angular momentum and is an excitation 
of the fundamental solution, shown with solid lines. 
}
\vspace{0.5cm}
\\
 fundamental solution,
with lower mass (continuous line) together with the corresponding excited one 
(dotted line).
 The difference between the masses and charges of these two solutions increases with the 
winding number 
$n$.
For the parameters in Figure 10 one finds $e.g.$
$M^{({\rm fund.})}/M^{({\rm excit.})}\simeq 0.98$ for $n=2$, while 
$M^{({\rm fund.})}/M^{({\rm excit.})}\simeq 0.9$  for $n=5$.
Interestingly, the solutions with higher mass have a smaller radius that the lower mass (and thus fundamental) configurations.
Also, the excited configurations display a more complicated behaviour of the function $X$ at $\theta=\pi/2$, exhibiting always an intermediate local extremum there.
 The profiles of the energy densities for the  fundamental and excited solutions look also very different,
 see Figure 10. (Note also, that in Figures 7, 8, 9 we have displayed fundamental solutions; 
this holds also for
 all configurations reported in \cite{Radu:2008pp}). 

 These excited states are expected to be unstable.
They are much more difficult to construct
than the fundamental solutions\footnote{The excited solutions
have been found as limiting configurations of
a secondary branch of gravitating solutions,
see the discussion in Section 4.2.}.
For the input parameters investigated so far, all fundamental solutions
appear to possess excited configurations.
However, the generic picture is likely to be more complicated, 
and other excited solutions may exist as well.
For example,
the nonuniqueness of solutions for the same input parameters
and the same topological sector
 has been noticed in 
 \cite{Kunz:2006ex,Kunz:2007jw}
 for the case of Yang-Mills-Higgs theory, with a Higgs field   in
  the adjoint
 representation.
 As shown there, new branches of solutions appear at critical values of
 the strength of the Higgs self-coupling parameter $\lambda$.

\newpage
\setlength{\unitlength}{1cm}
\begin{picture}(15,20.85) 
\put(-0.5,10){\epsfig{file=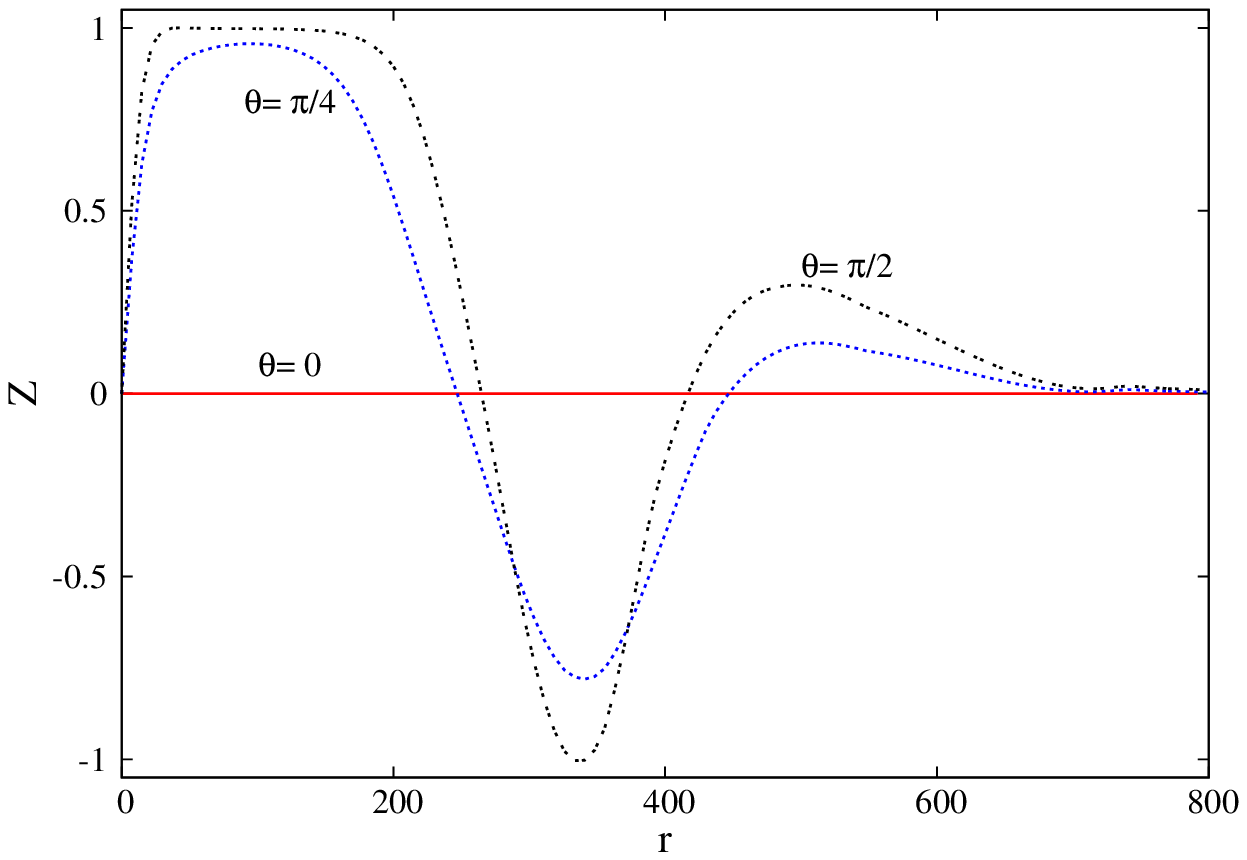,width=7.5cm}}
\put(8,10){\epsfig{file=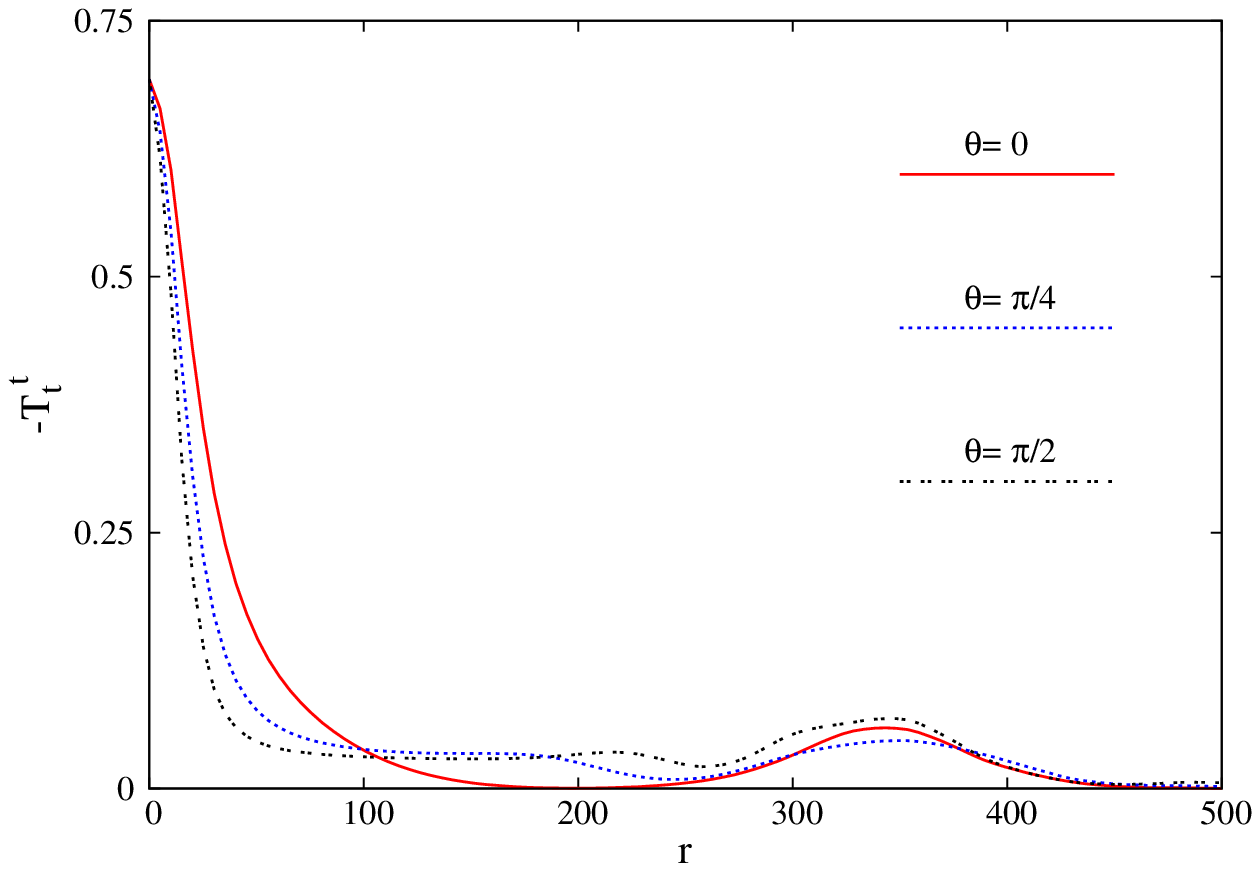,width=7.5cm}}
\put(-0.5,16){\epsfig{file=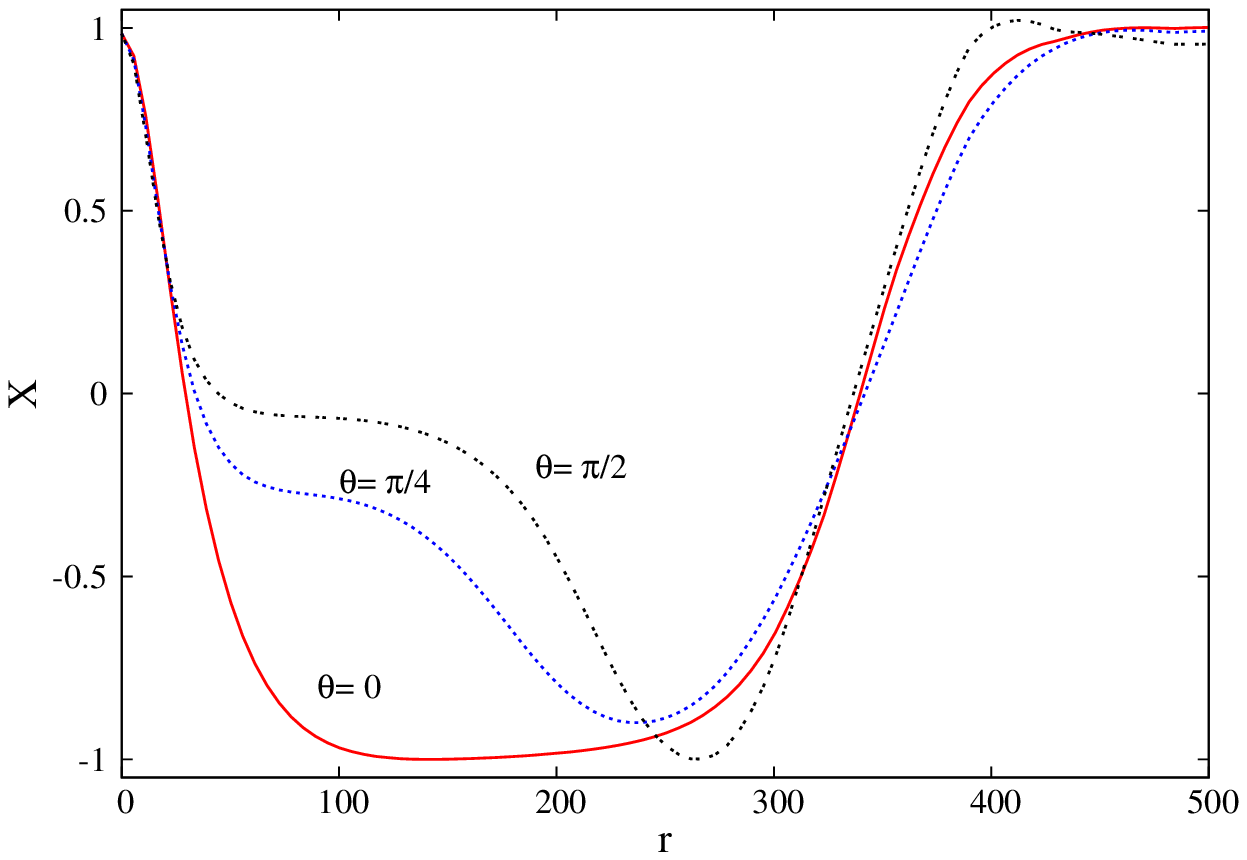,width=7.5cm}}
\put(8,16){\epsfig{file=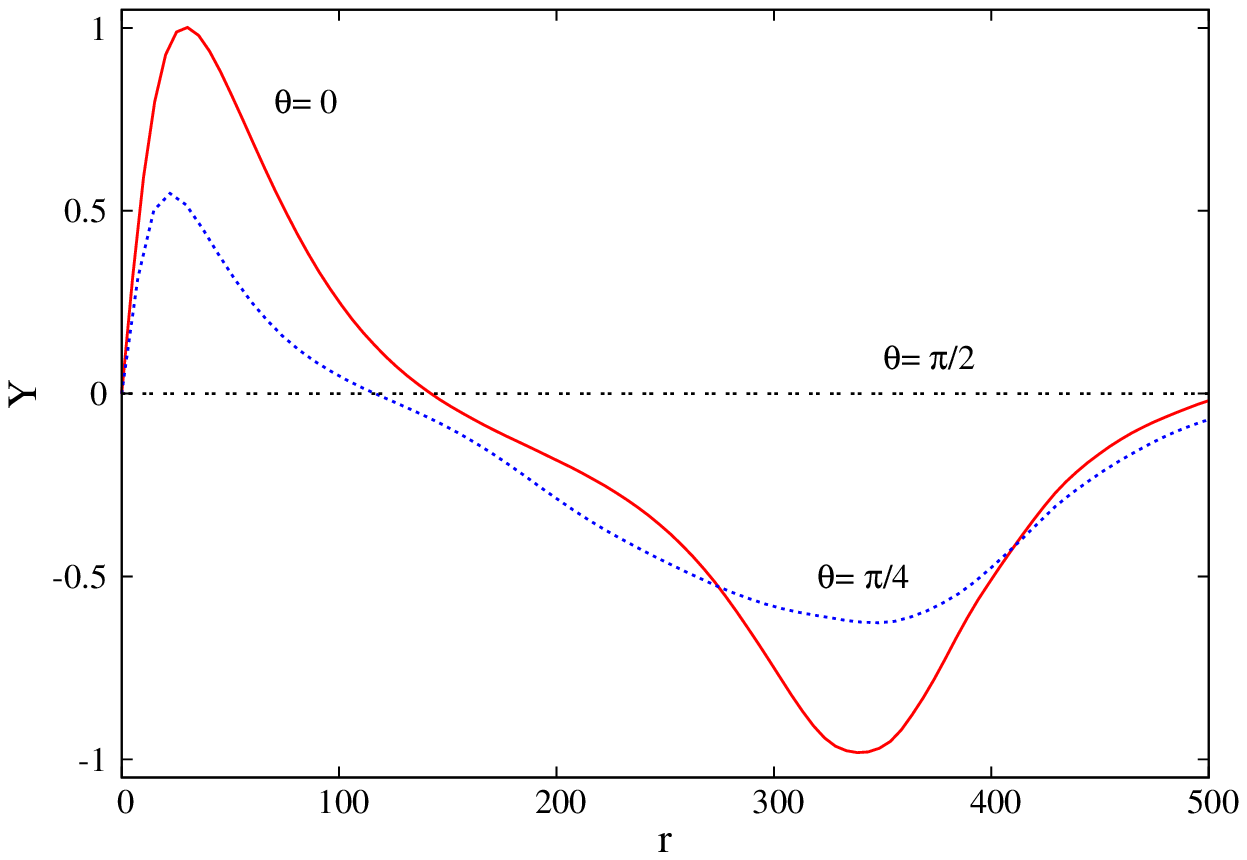,width=7.5cm}} 
\end{picture} 
 \vspace*{-9.cm}
\\
{\small {\bf Figure 11.}
The profiles of the scalar functions $X(r,\theta)$, $Y(r,\theta)$, $Z(r,\theta)$
and of the energy density $-T_t^t$
are shown for a typical di-vorton with $\beta_1=\beta_2=1$, $\beta_3=0.5$, $w=0.023$, $n=1$. 
}
\vspace{0.5cm}
\\
 In this work we have only studied the dependence on the 
frequency of the  excited solutions for fixed $\beta_i,n$.
The results show that 
 the picture  found in \cite{Radu:2008pp} for the  fundamental configurations holds also in this case.
 
Also, it is not clear to us how to interpret the  nonuniqueness
result in terms of the heuristic
construction discussed in Section 3.1.
This would imply the existence of excited vortex states
for the same input data. (The corresponding ansatz is given by (\ref{vortices}) with $k=1$.)
However, so far we could not construct such configurations.
 
\subsection{Composite configurations. The di-vortons}
As noted in the introduction, there is a striking analogy between the heuristic
construction of
a vorton (starting with a vortex) and that of a black ring (starting with
a black string).
However, in the gravity case, a variety of solutions describing composite configurations 
($e.g.$ a Saturn \cite{Elvang:2007rd} and a di-ring \cite{Iguchi:2007is})
have been constructed inspired by rather similar arguments.
Thus it is reasonable to expect 
that the global Witten model (and, in general, any field theory model admitting vorton solutions)
would also possess composite solutions--the analogs of $d=5$ gravitating Saturn, di-ring  and further multi-component objects.

In what follows we give some preliminary numerical
evidence for the existence of 
a solution describing two concentric vortons, $i.e.$
a di-vorton. 
In a heuristic construction, this configuration is found by taking finite pieces of two parallel
vortices and bending them in the same plane, around a common center.
Again, the resulting configuration would be supported against
collapse by the centrifugal force.

The profile of a typical di-vorton solution is shown in Figure 11.
There one can see that $|\phi|$ vanishes in the equatorial plane $\theta=\pi/2$,
for two different values of $r$.
These values correspond to the positions of the two concentric rings.
Also, the functions $X, Y, Z$ show a much more complicated dependence
on $r$ and $\theta$ than in the single vorton case. 
One can see $e.g.$ that,
different from the single vorton case, 
the function $X$ starts from a positive value  at $r=0$ (which is close to one, 
for all solutions we could construct), takes negative values for some range of $r$
and approaches the $v.e.v.$ $X=1$ asymptotically.
The mass and angular momentum of these solutions are much larger
than the mass of two (usual) 
vortons with the same values of the input parameters.
(Note also, that all di-vorton solutions we could construct so far  correspond to
configurations close to the sigma-model limit.)
The angular momentum of the solutions is concentrated in two tori centred around the
locations of the rings\footnote{Different
from the case of a  Saturn \cite{Elvang:2007rd} or of a di-ring \cite{Iguchi:2007is}
in $d=5$ Einstein gravity,
both components of a di-vorton
carry an angular momentum density with the same sign, see the flat spacetime limit of the relation (\ref{Qc}).}, 
while the mass density exhibits a local extremum at $r=0$. 
The two radii of the di-vorton are also much larger than in the single vorton case.

The numerical accuracy was much lower in this case,  which has prevented us from
a systematic study of di-vortons.
However, we expect them to share the basic properties
of the usual  vortons. In particular, they should exist for a limited range of frequencies.

\section{Gravitating vortons}
When $\alpha$ is increased from zero, while keeping $(\beta_i;n,w)$ fixed,
a branch of  rotating solutions emerges from the flat spacetime
 configurations.
 Since the dependence on $\alpha$ is different for $k=0,1$
 solutions, we shall discuss them separately.

\subsection{Gravitating $k=0$ `semitopological' vortons}
Not entirely surprising, some properties of the gravitating $k=0$
solutions are similar to those of the even-parity rotating boson stars in a model with a 
single complex scalar field \cite{Kleihaus:2005me}.

The behaviour of a typical $k=0$
solution is exhibited in Figure 12.
These two-dimensional plots exhibit the  small-$r$ dependence
for three angles $\theta=0$, $\pi/4$ and $\pi/2$.
 (We recall that these solutions are symmetric $w.r.t.$
an equatorial plane reflection.) 
One can notice, for example, that the functions $f,l$ and $m$
exhibit a strong angular dependence, with $l \neq m$ (except on the $z-$axis).
Also, the metric functions are completely regular 
and the line element shows no sign of a horizon.
However, the shape of the scalar functions $X,Z$
is similar to that found for  solutions in a flat spacetime background.

Considering the spatial structure of the solutions, we have found that the scalar fields $X,Z$
typically give  rise 
to a torus-like energy density, just as for the
flat spacetime solutions. 
(This torus-like shape becomes apparent by considering surfaces
of constant energy density.)
 Again, the maximum of the energy density of the solutions increases
and shifts towards larger values of the coordinate $\rho = r \sin \theta$, when $n$ (and thus the angular
momentum) increases.  

To demonstrate the effects of gravity, we exhibit in Figure  13
the mass and angular momentum of typical solutions as a function of frequency.
Three different values of $\alpha$ are considered there, for the same 
parameters $\beta_i$ of the potential and a winding number $n=1$.
Similar to  the case of a flat spacetime background, the solutions
exist for a limited range of frequencies, $w_{min}<w<w_{max}$, with $w_{max}$ given by (\ref{max-w}).
However, as expected, the divergencies of the global charges at the limits of the $w-$interval
which were found for $\alpha=0$ (see Figure 6)
are regularized by the effects of gravity.
Moreover, for $\alpha \neq 0$, the solutions exhibit a complicated pattern after approaching $w_{min}$.
There one observes 
an inspiralling behaviour of the $k=0$ solutions, 
towards a limiting
solution at the center of the spiral, for a frequency
$w_{lim}>w_{min}$. (This behaviour could be studied more precisely for spherically symmetric
solutions ($n=0$), which we do not display here.)
As a result, when the mass $M$ is considered as a function of the scalar charge $Q$,
we observe a cusp structure, as illustrated in Figure 13 (right).
This is essentially the behaviour noticed in \cite{Kleihaus:2005me}
for rotating boson stars.
The branches starting from the maximal value of the frequency
and ending at the maximal values of the mass are expected to be stable,
in analogy to boson stars.

The dependence on the parameter $\alpha$
is shown in Figure 14. (A similar behaviour has been found for spherically symmetric solutions, $n=0$.)
One can see that the solutions exist for arbitrarily large values of 

\newpage
\setlength{\unitlength}{1cm}
\begin{picture}(15,20.85) 
\put(-1,0.5){\epsfig{file=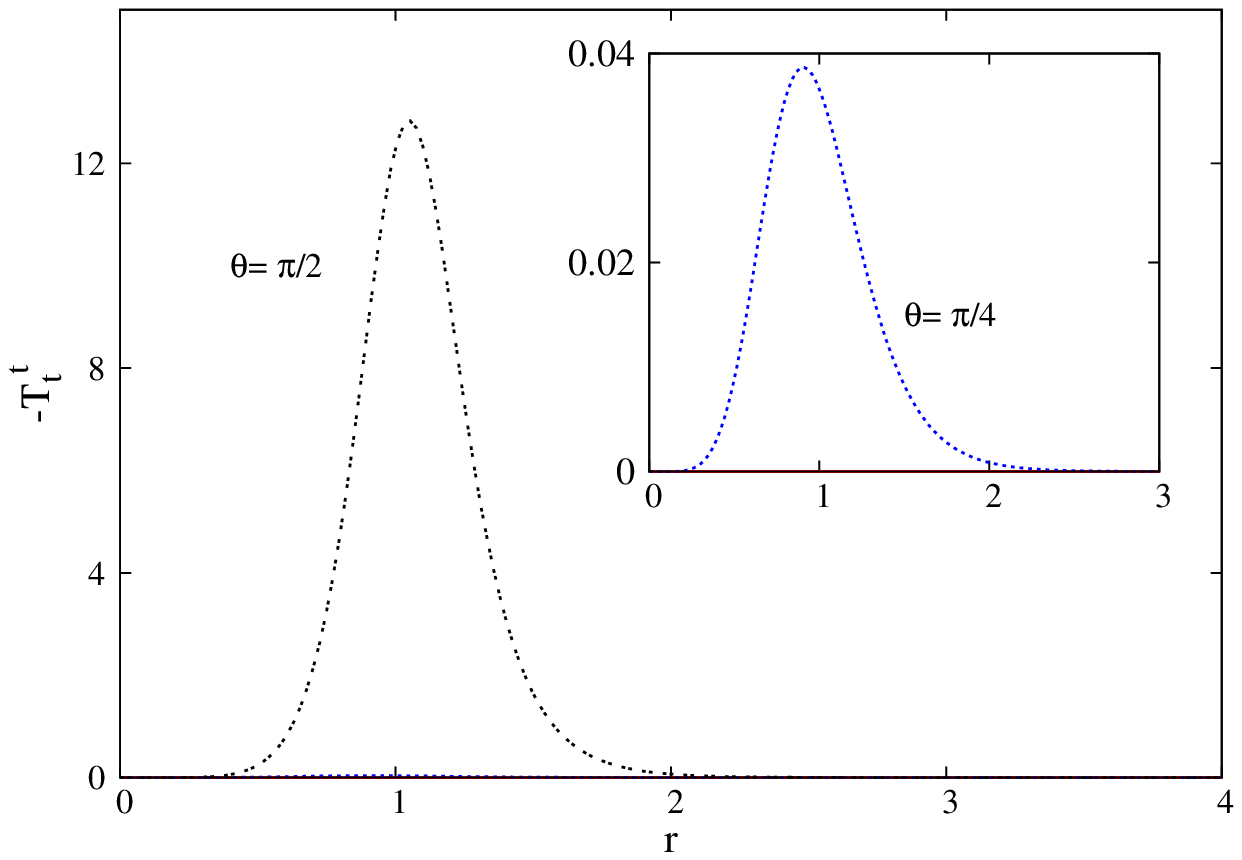,width=7.1cm}}
\put(7,0.5){\epsfig{file=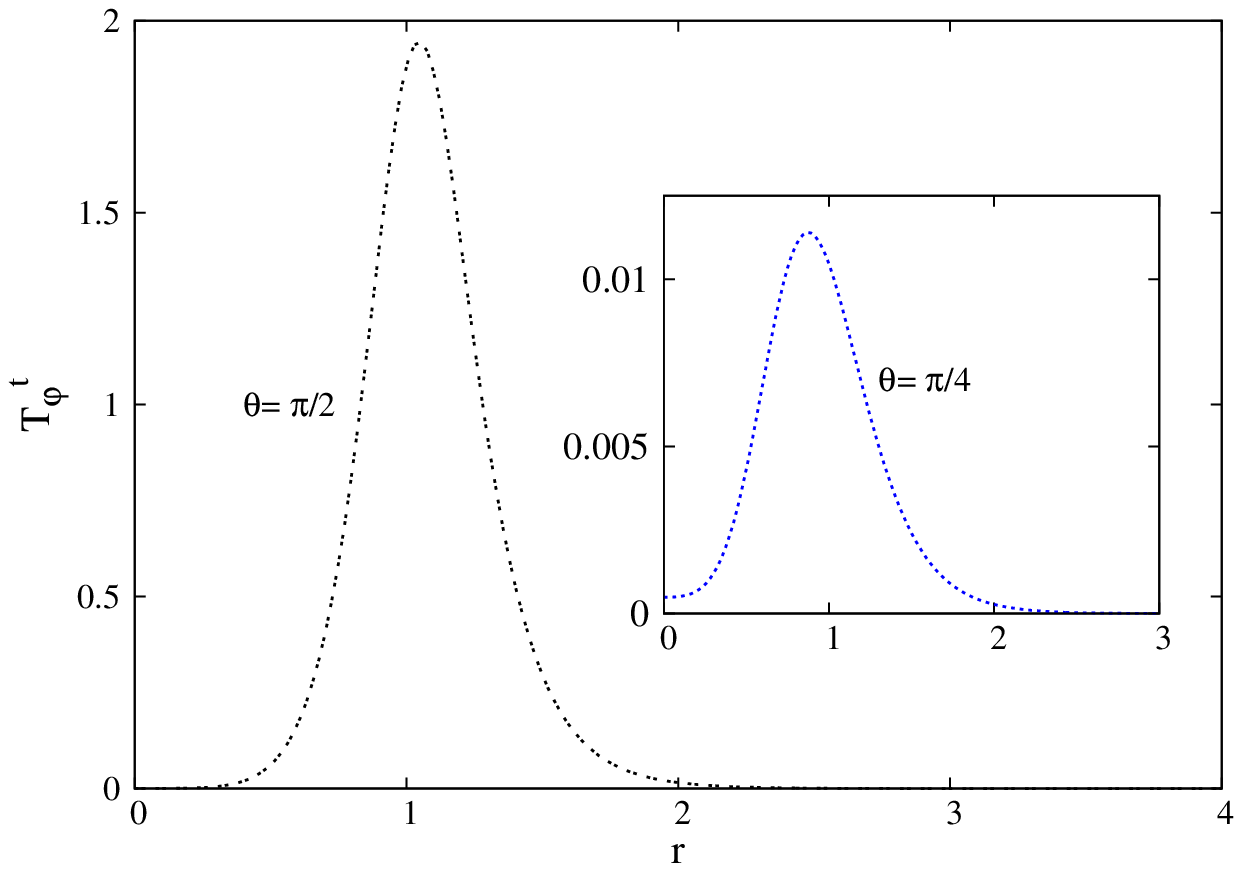,width=7.1cm}}
\put(-1,5.75){\epsfig{file=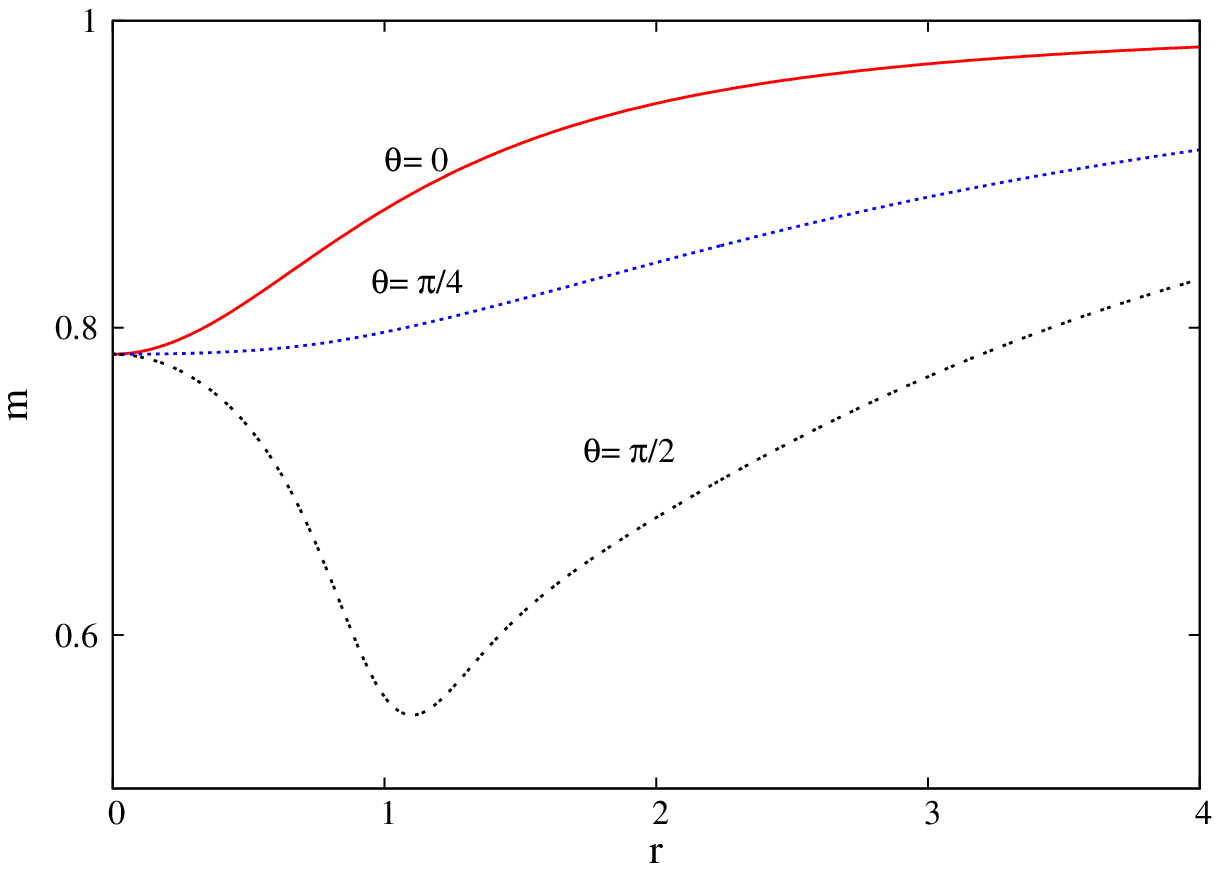,width=7.1cm}}
\put(7,5.75){\epsfig{file=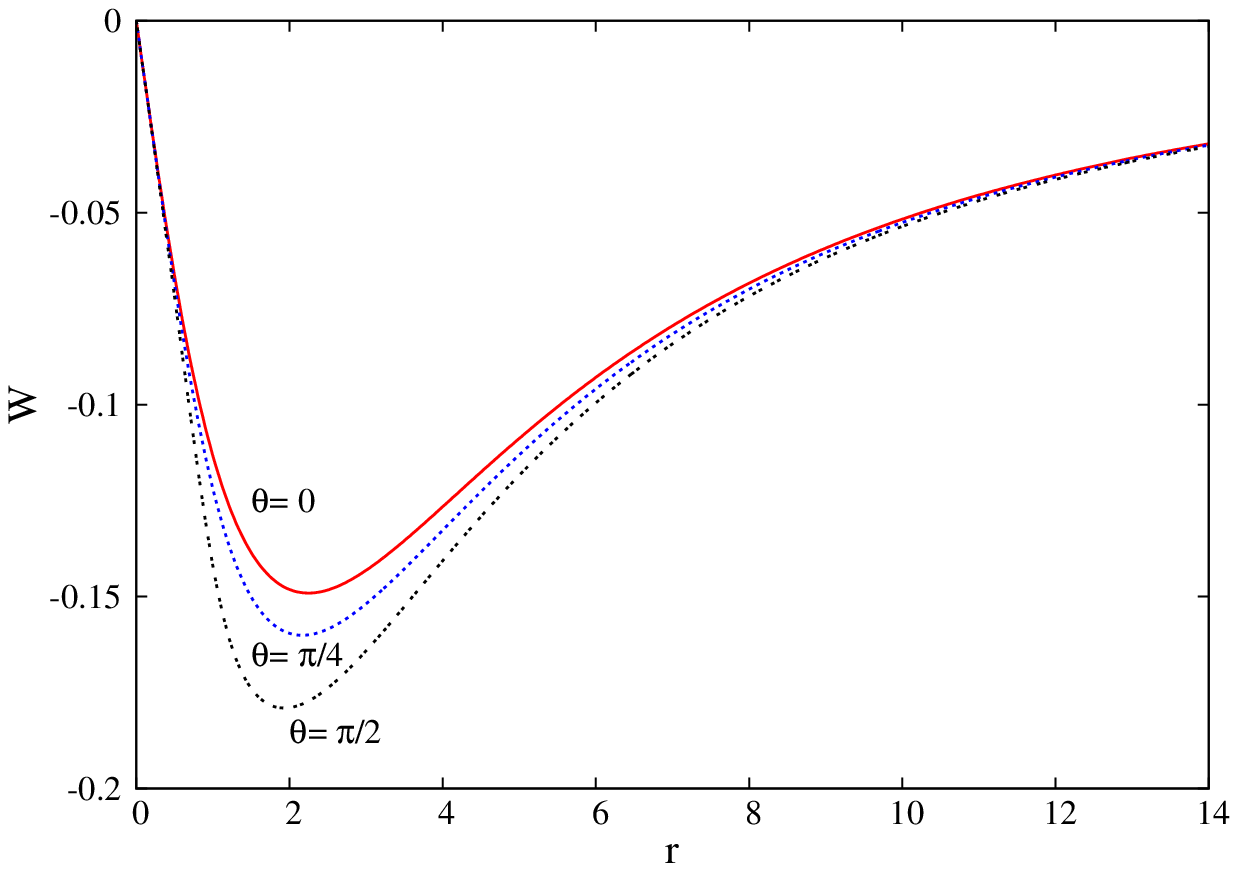,width=7.1cm}}
\put(-1,11){\epsfig{file=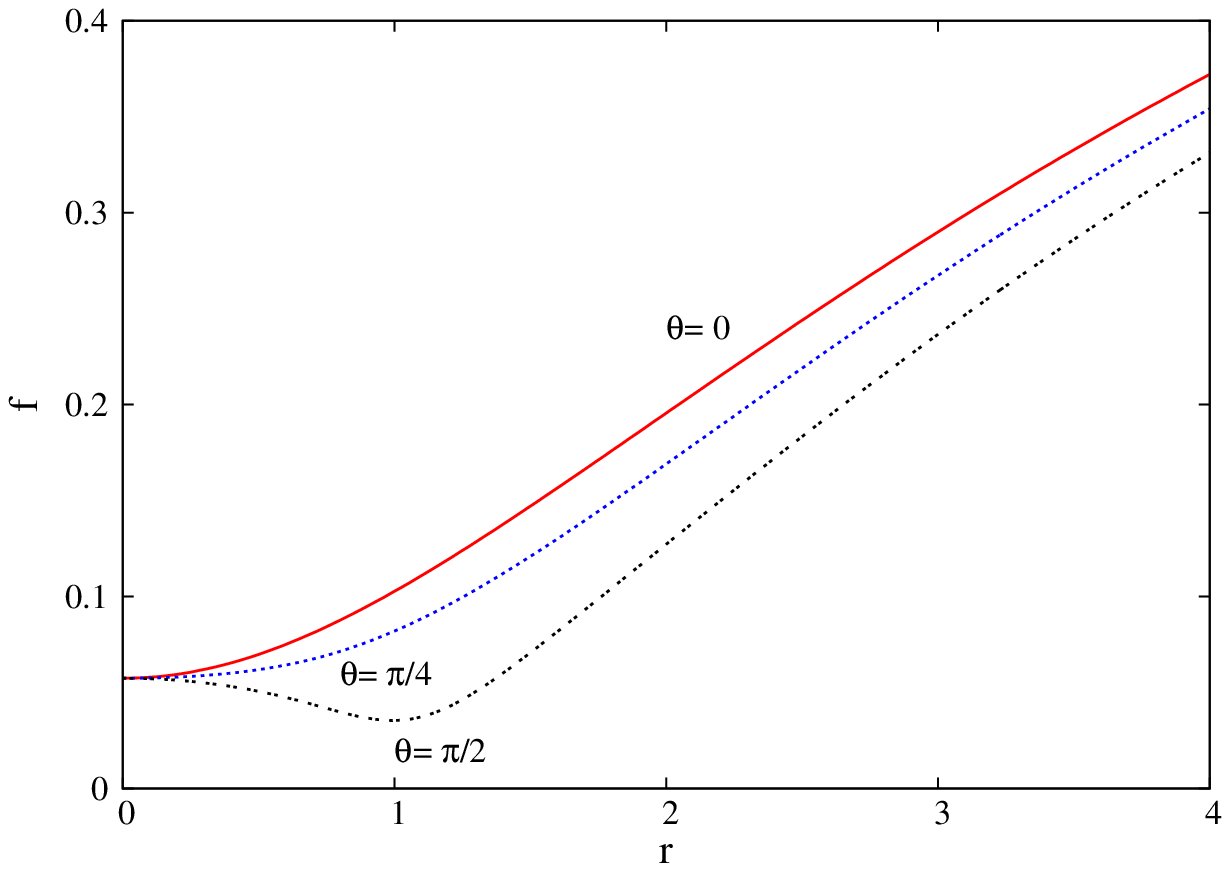,width=7.1cm}}
\put(7,11){\epsfig{file=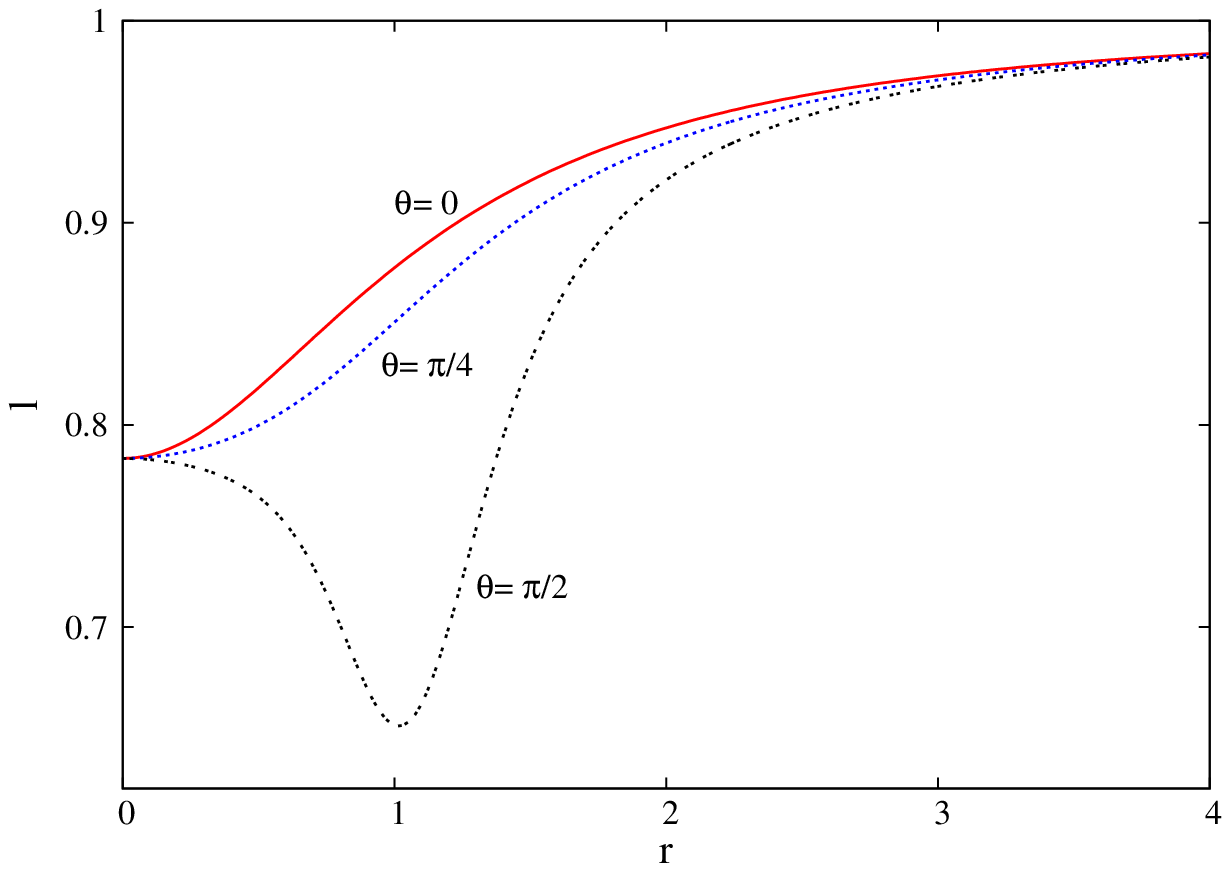,width=7.1cm}}
\put(-1,16.25){\epsfig{file=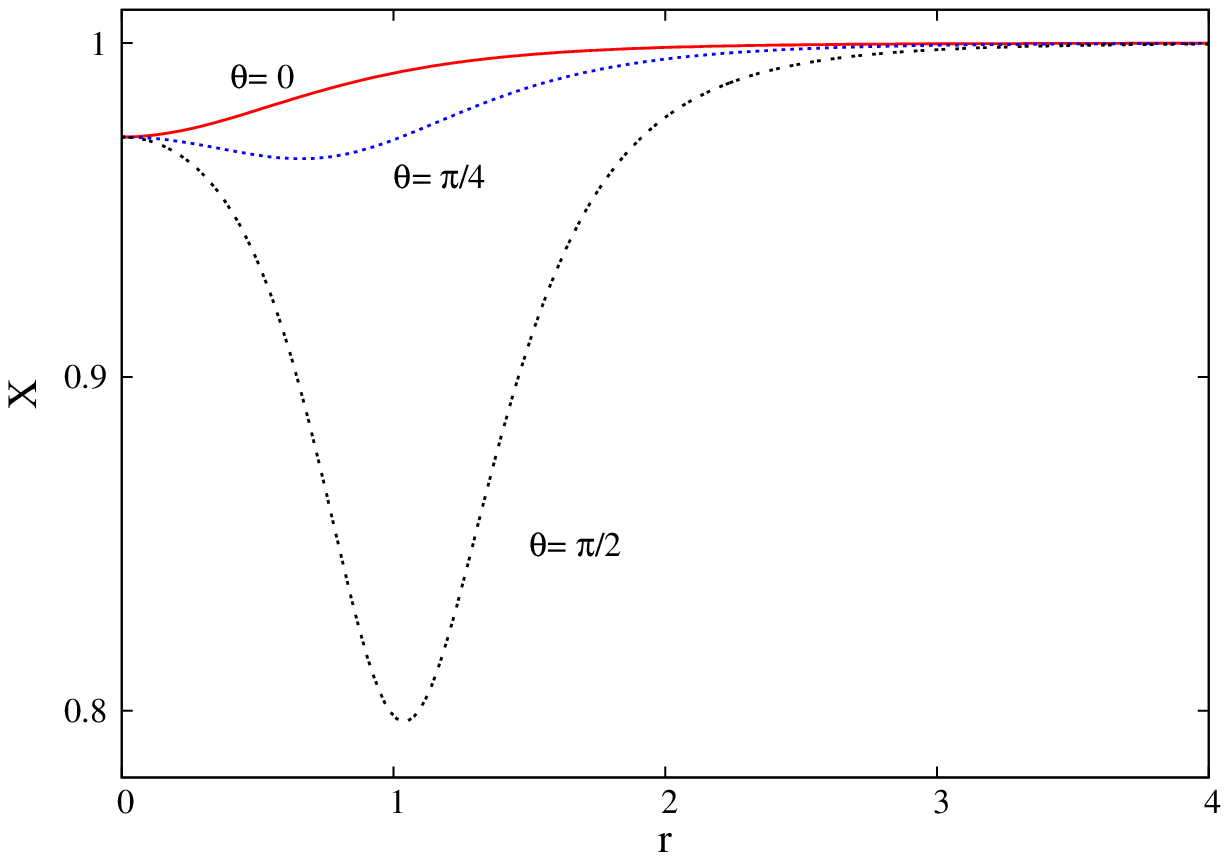,width=7.1cm}}
\put(7,16.25){\epsfig{file=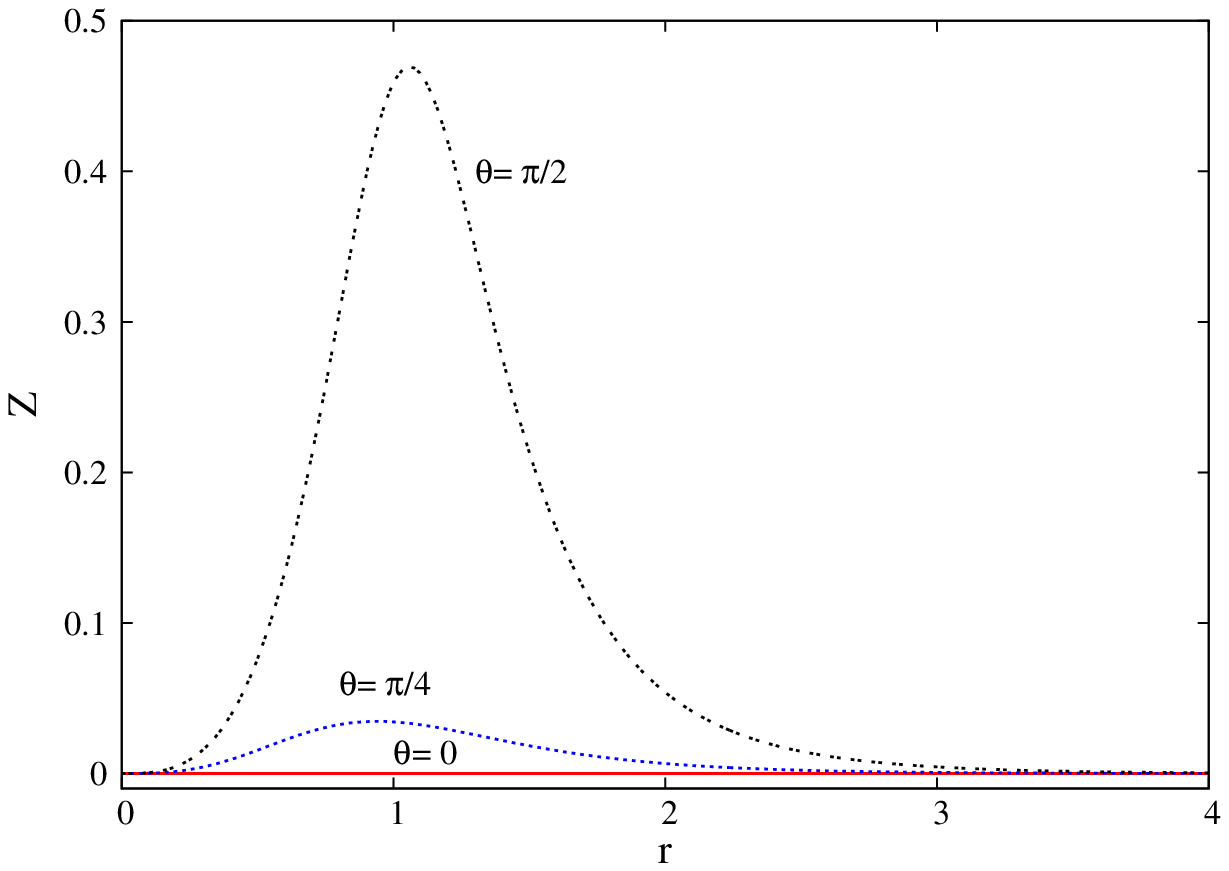,width=7.1cm}} 
 \end{picture} 
\\
{\small {\bf Figure 12.} The functions $(X,Z;f,l,m,W)$ 
and the  components
$T_t^t$ and $T_\varphi^t$ 
of the energy-momentum tensor
are shown for 
a  $k=0$ gravitating solution with 
$\beta_1=0.035$,
$\beta_2=1$,
$\beta_3=2.5$,
$w=0.79$,
$n=3$
and $\alpha=1$.  }  


\newpage
\setlength{\unitlength}{1cm}
\begin{picture}(8,6)
\label{fig-XZ-w}
\put(-0.5,0.0){\epsfig{file=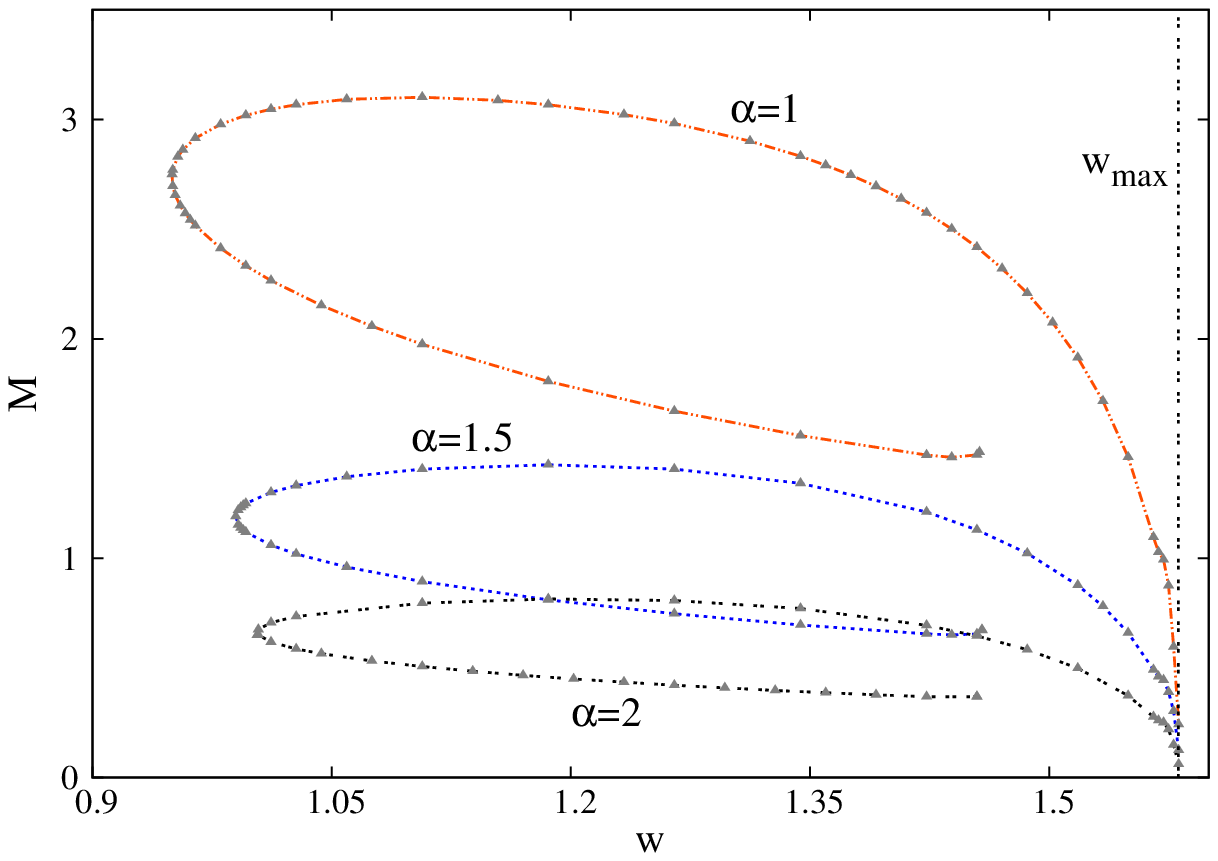,width=8cm}}
\put(8,0.0){\epsfig{file=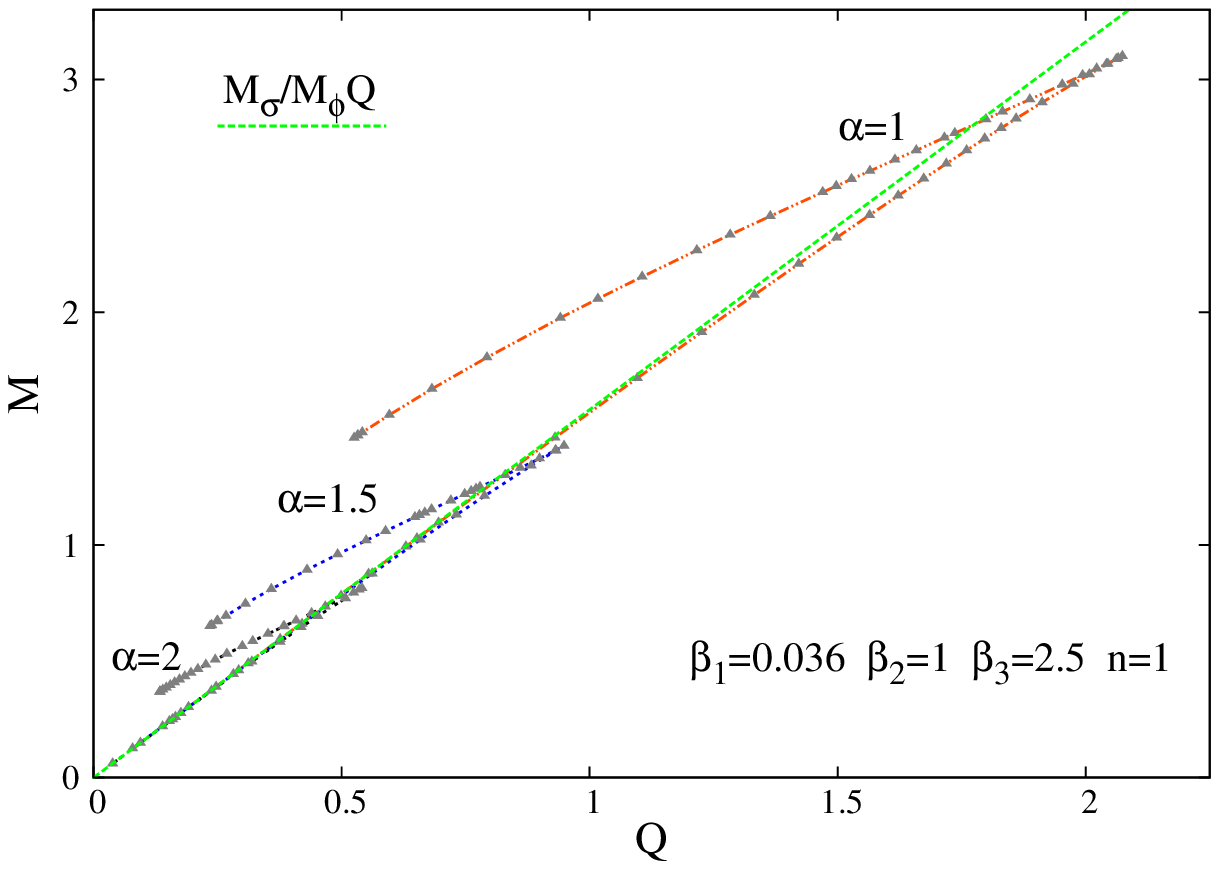,width=8cm}}
\end{picture} 
\\
\\
{\small {\bf Figure 	13.} {\it Left:} The mass  of gravitating $k=0$ `semitopological' vortons
is shown as a function of the frequency for several values of the coupling constant $\alpha$.
The vertical line corresponds to the maximal value of the frequency as given by (\ref{max-w}).
Other input parameters of these configurations are $\beta_1=0.03,~\beta_2=1,~\beta_3=2.5$, $n=1$. 
{\it Right:} The mass-charge diagram for the same solutions.
Also we show  the mass
for $Q$ free bosons, $M = M_\sigma/M_\phi Q$ in the units employed here.
} 
\vspace{0.5cm}

 \setlength{\unitlength}{1cm}
\begin{picture}(8,6)
\put(-0.5,0.0){\epsfig{file=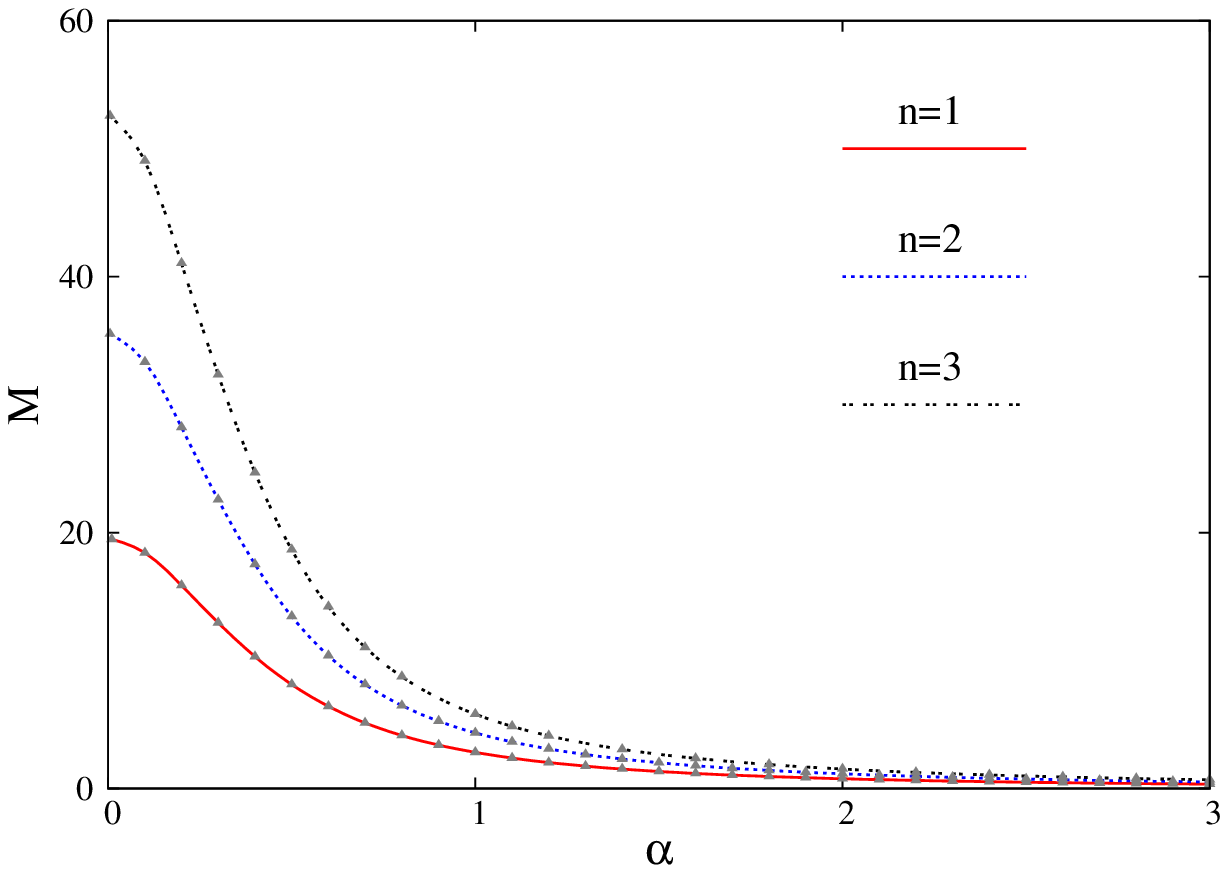,width=8cm}}
\put(8,0.0){\epsfig{file=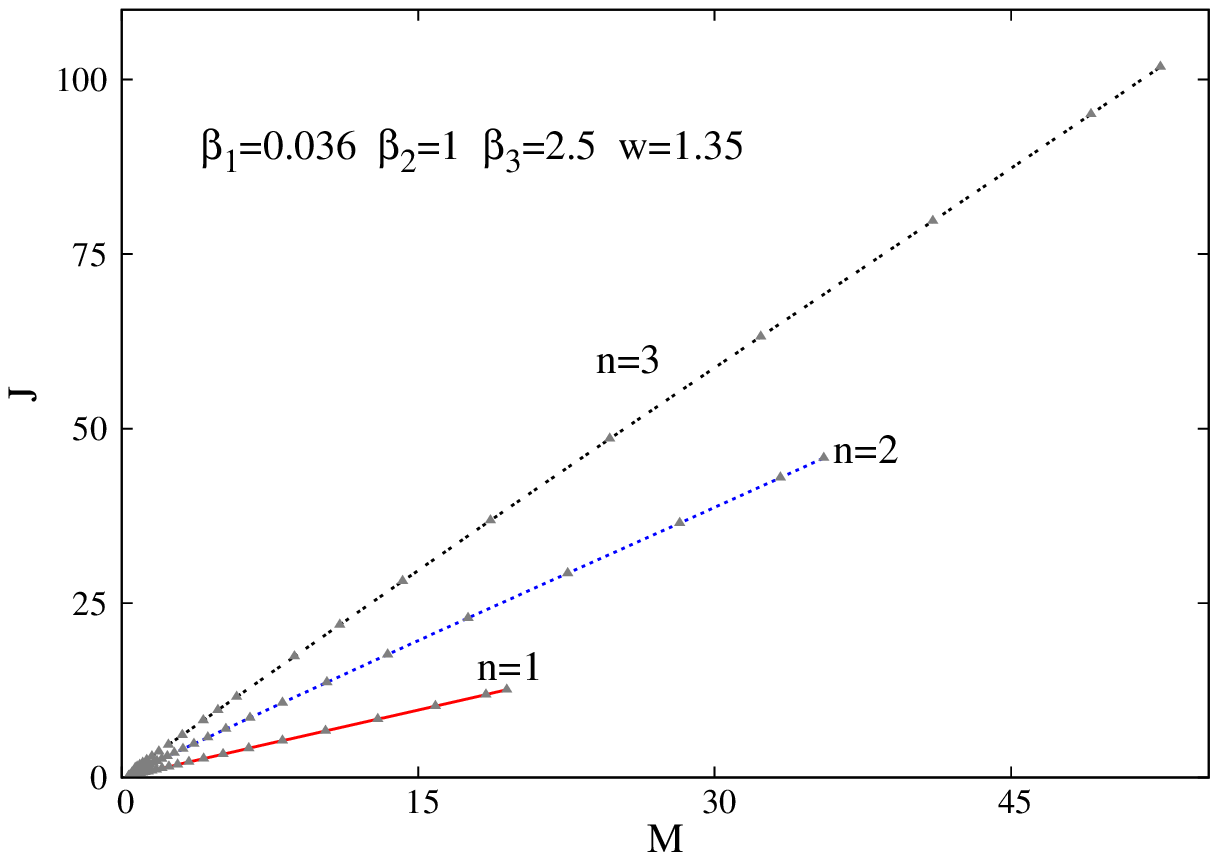,width=8cm}}
\end{picture}
\\
\\
{\small {\bf Figure 14.} The mass and angular momentum 
are shown as  functions of $\alpha$ for gravitating $k=0$ `semitopological' vortons.}
\vspace{0.5cm}
\\
$\alpha$
($i.e.$ an arbitrarily large $v.e.v.$ of the scalar field $\phi$).

This strongly contrasts with the picture found for other models 
featuring scalar fields with a symmetry breaking potential,
 $e.g.$ monopoles or sphalerons, in which 
case a maximal value of $\alpha$
 is found
  (see the review \cite{Volkov:1998cc}).  
  In the large $\alpha$ limit, the function $Z$ vanishes, while
  the function $X$ approaches a constant value everywhere, $X\to 1$, corresponding to the
 $v.e.v.$ of the scalar field.

To understand this behaviour, we consider the scaled functions
\be                                 \label{scale-k0}
\hat X=\frac{X-1}{\alpha}, ~~{\rm and}~~\hat Z=\frac{Z}{\alpha},~~
\ee
 Then, after making this replacement in the field equations and taking the limit $\alpha \to \infty$,
 the functions $\hat X,\hat Z$ 
\newpage
\setlength{\unitlength}{1cm}
\begin{picture}(15,20.85) 
\put(-1,2){\epsfig{file=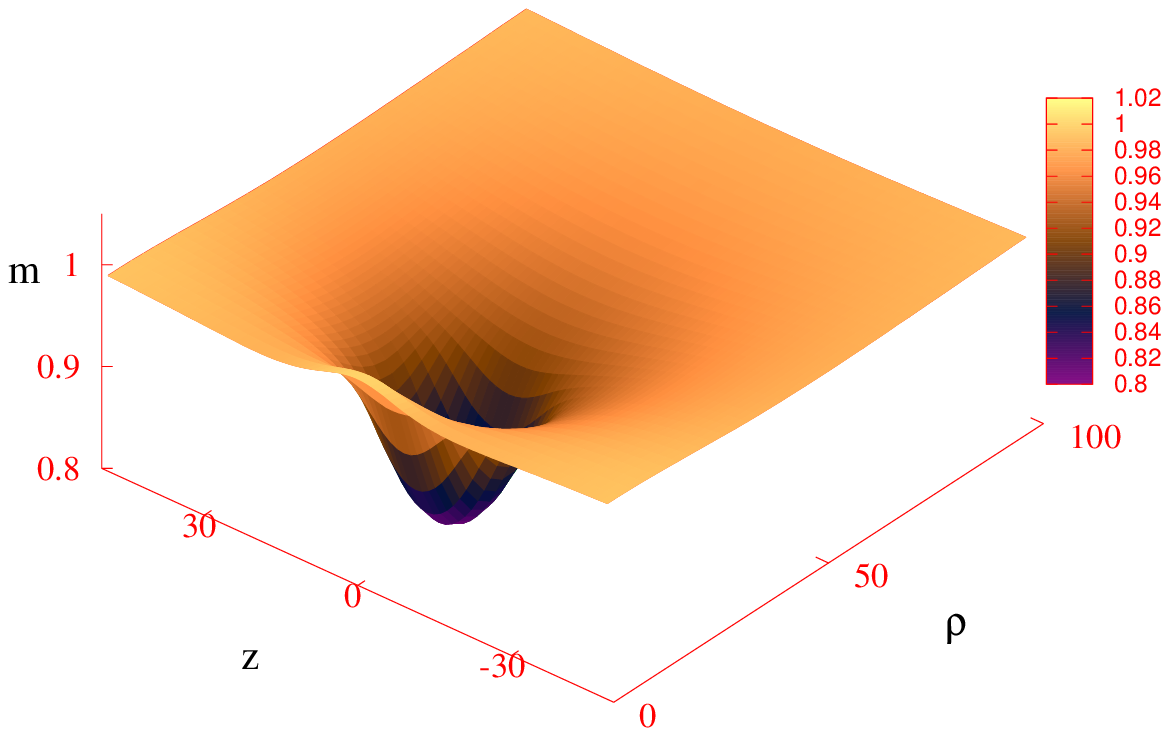,width=7.5cm}}
\put(7,2){\epsfig{file=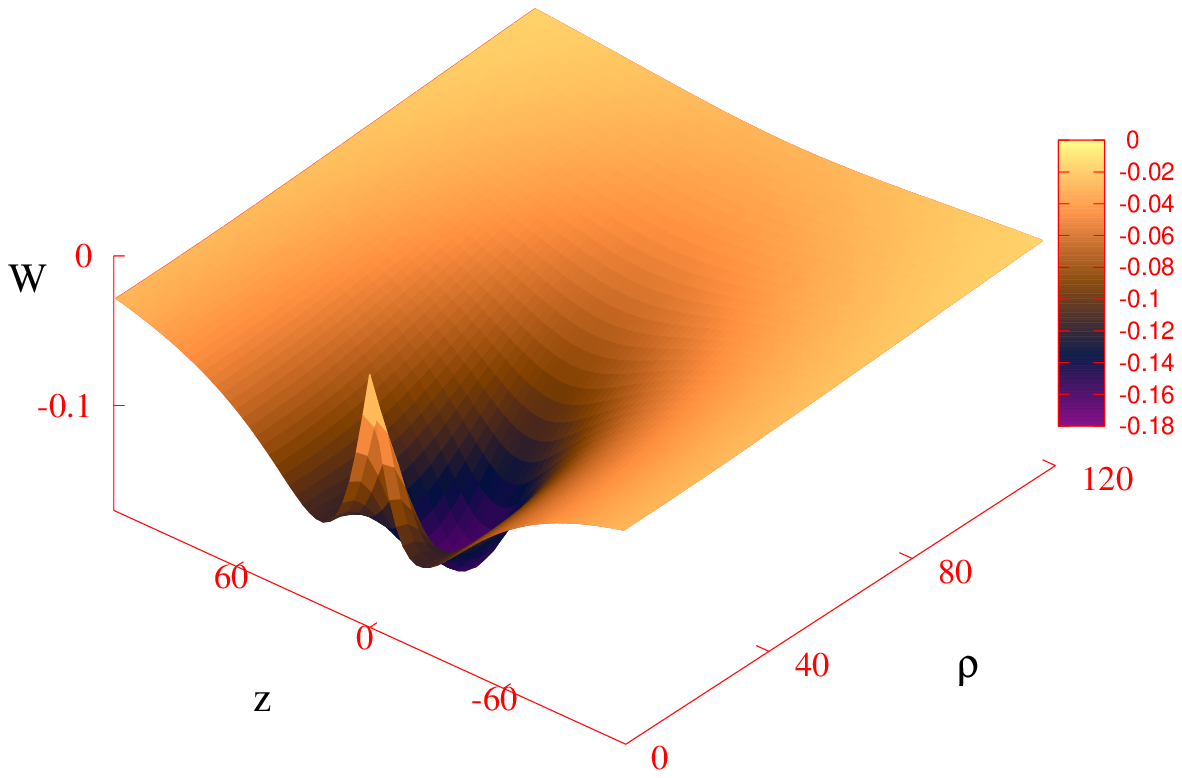,width=7.5cm}}
\put(-1,7){\epsfig{file=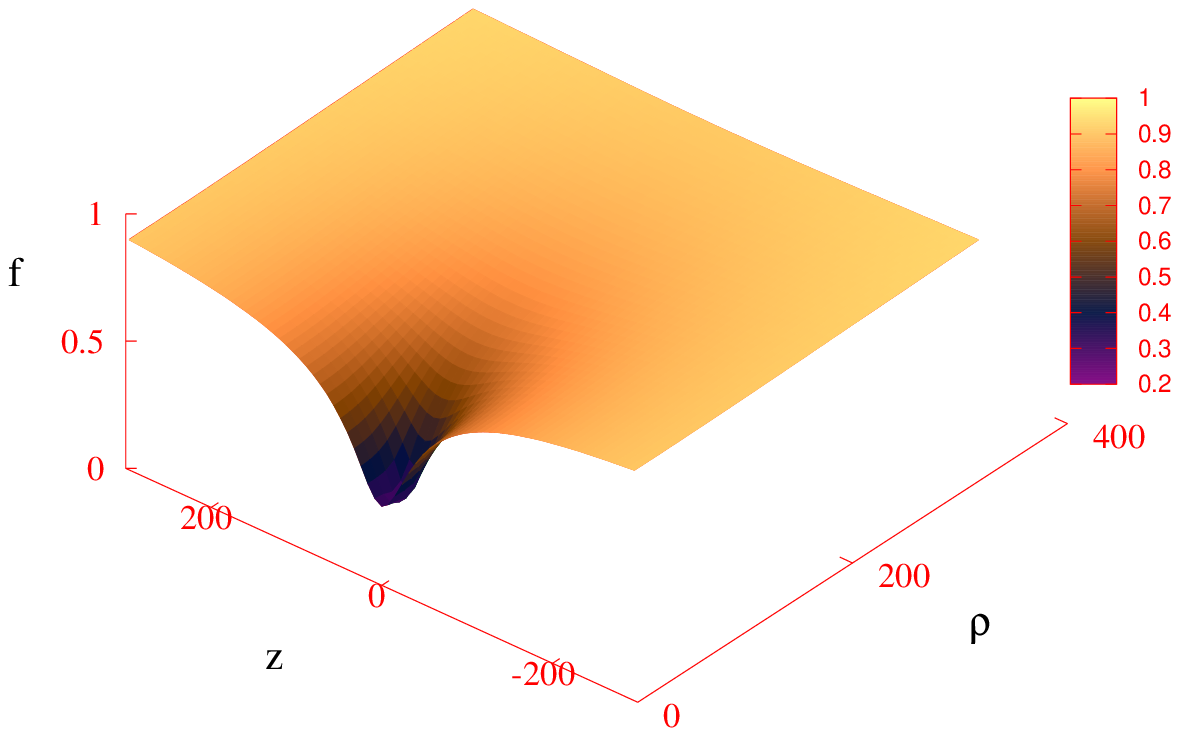,width=7.5cm}}
\put(7,7){\epsfig{file=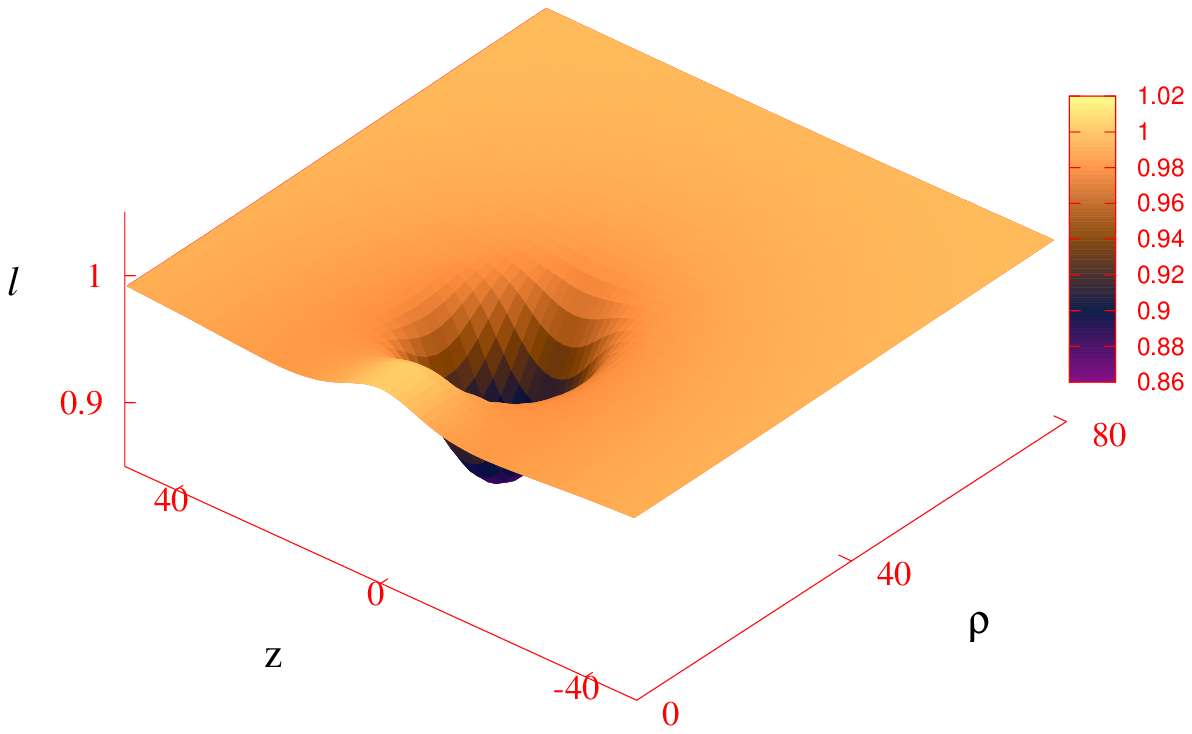,width=7.5cm}}
\put(-1,12){\epsfig{file=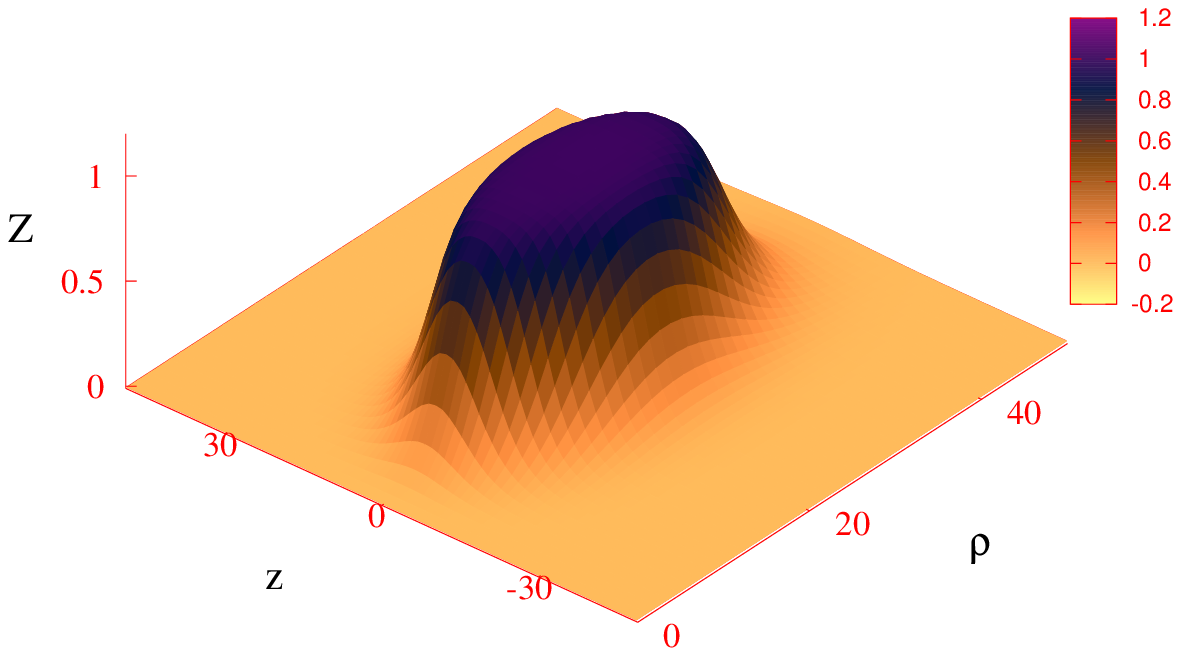,width=7.5cm}}
\put(7,12){\epsfig{file=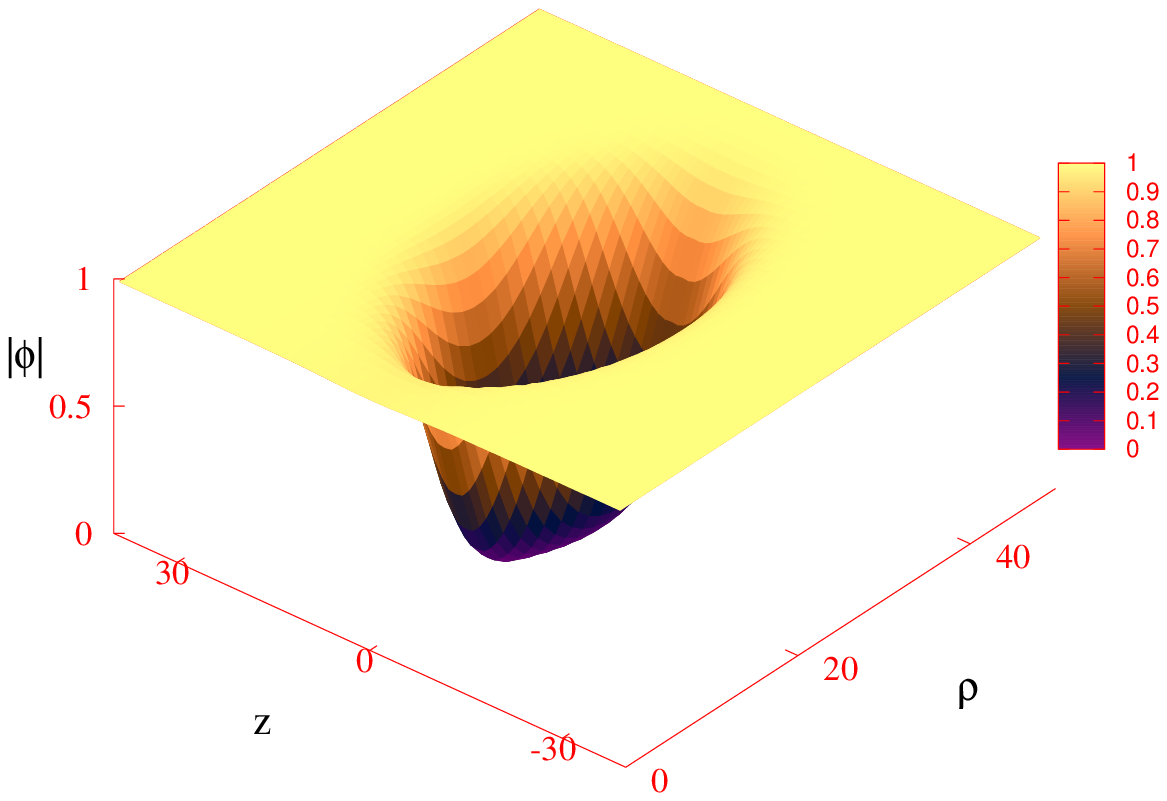,width=7.5cm}}
\put(-1,17){\epsfig{file=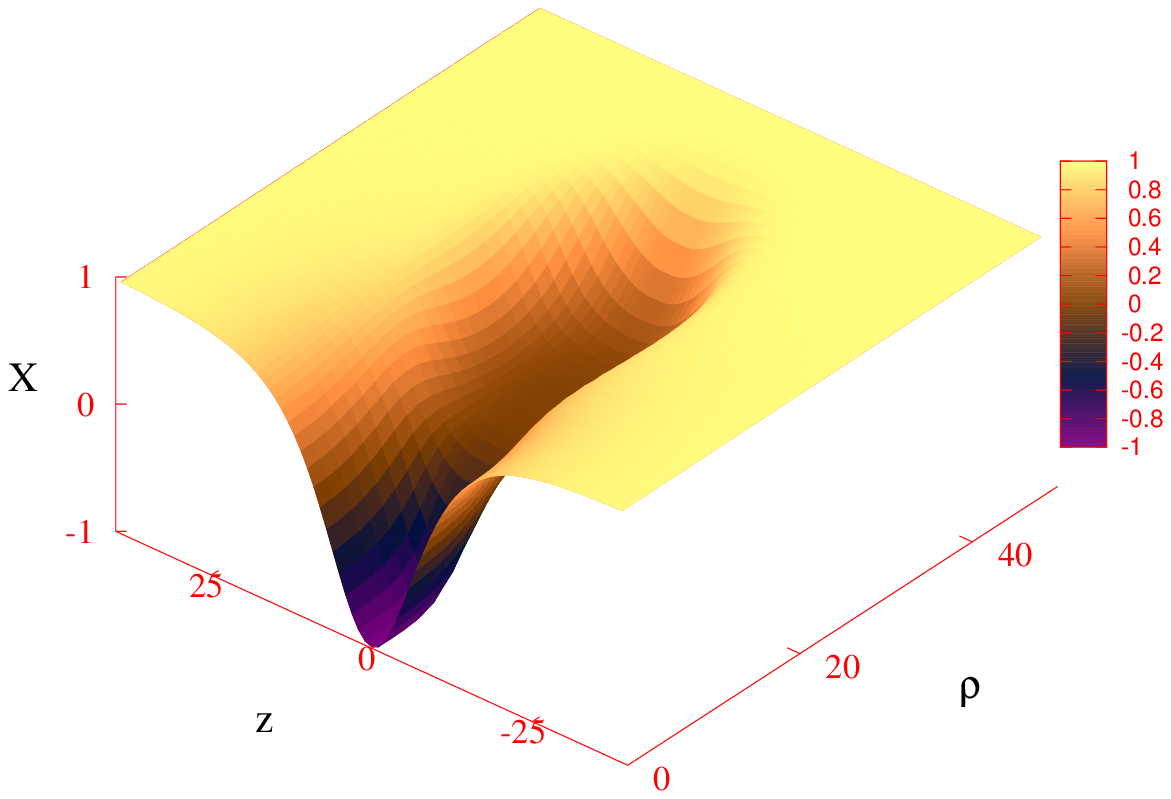,width=7.5cm}}
\put(7,17){\epsfig{file=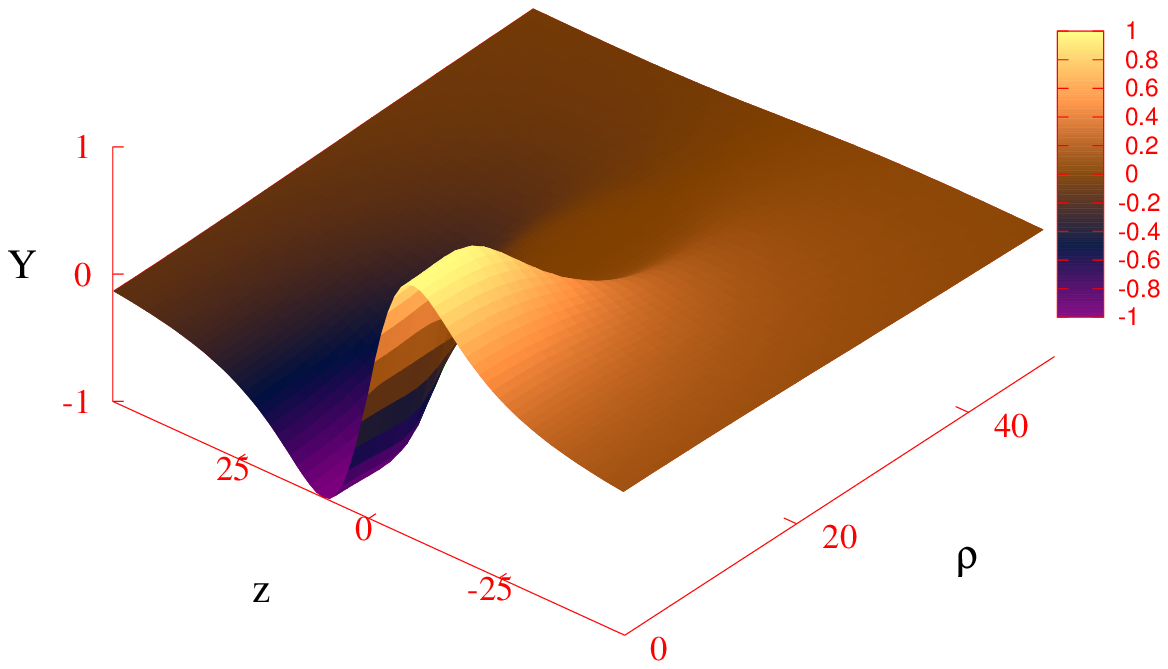,width=7.5cm}} 
\end{picture} 
\vspace{-1.cm}
\\
{\small {\bf Figure 	15.}  
The scalar functions $X$, $Y$, $Z$, the amplitude $\sqrt{X^2+Y^2}$
of the scalar field $\phi$
and the metric functions $f,l,m,W$
are shown for a $k=1$ gravitating vorton solution with $\beta_1=1$, $\beta_2=0.97$,
$\beta_3=0.54$, $w=0.11$, $n=3$ and $\alpha=0.2$.
The coordinates here are $\rho=r \sin \theta$ and $z=r \cos \theta$.
} 
\vspace{0.5cm}

\newpage
\setlength{\unitlength}{1cm}
\begin{picture}(8,6)
\put(-0.5,0.0){\epsfig{file=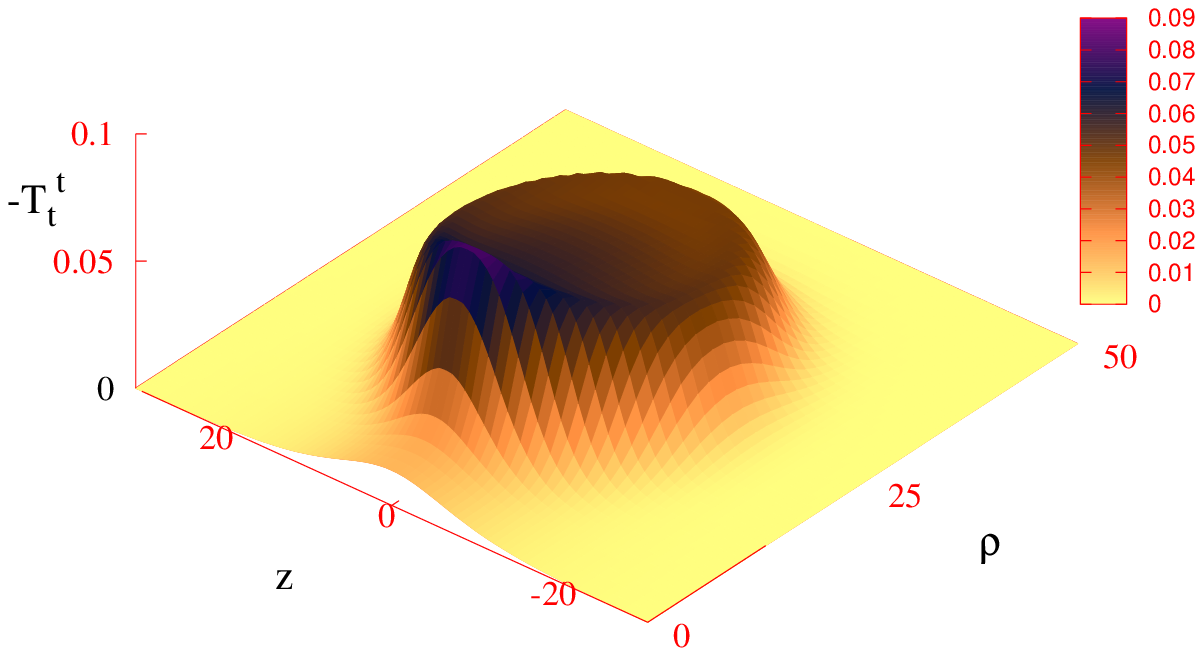,width=8cm}}
\put(8,0.0){\epsfig{file=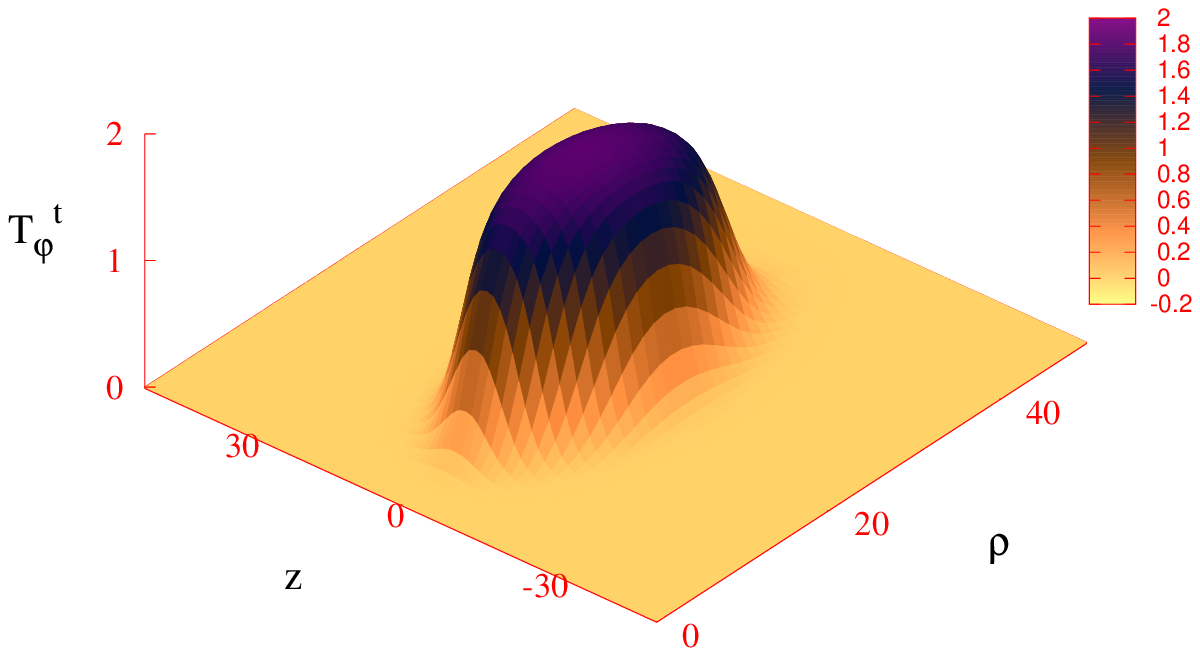,width=8cm}}
\end{picture} 
\\
\\
{\small {\bf Figure 	16.}  
The energy density $-T_t^t$ and  the angular momentum density $T_\varphi^t$
are shown for the same solution as in Figure 15.
} 
\vspace{0.5cm}

\setlength{\unitlength}{1cm}
\begin{picture}(8,6)
\put(-0.5,0.0){\epsfig{file=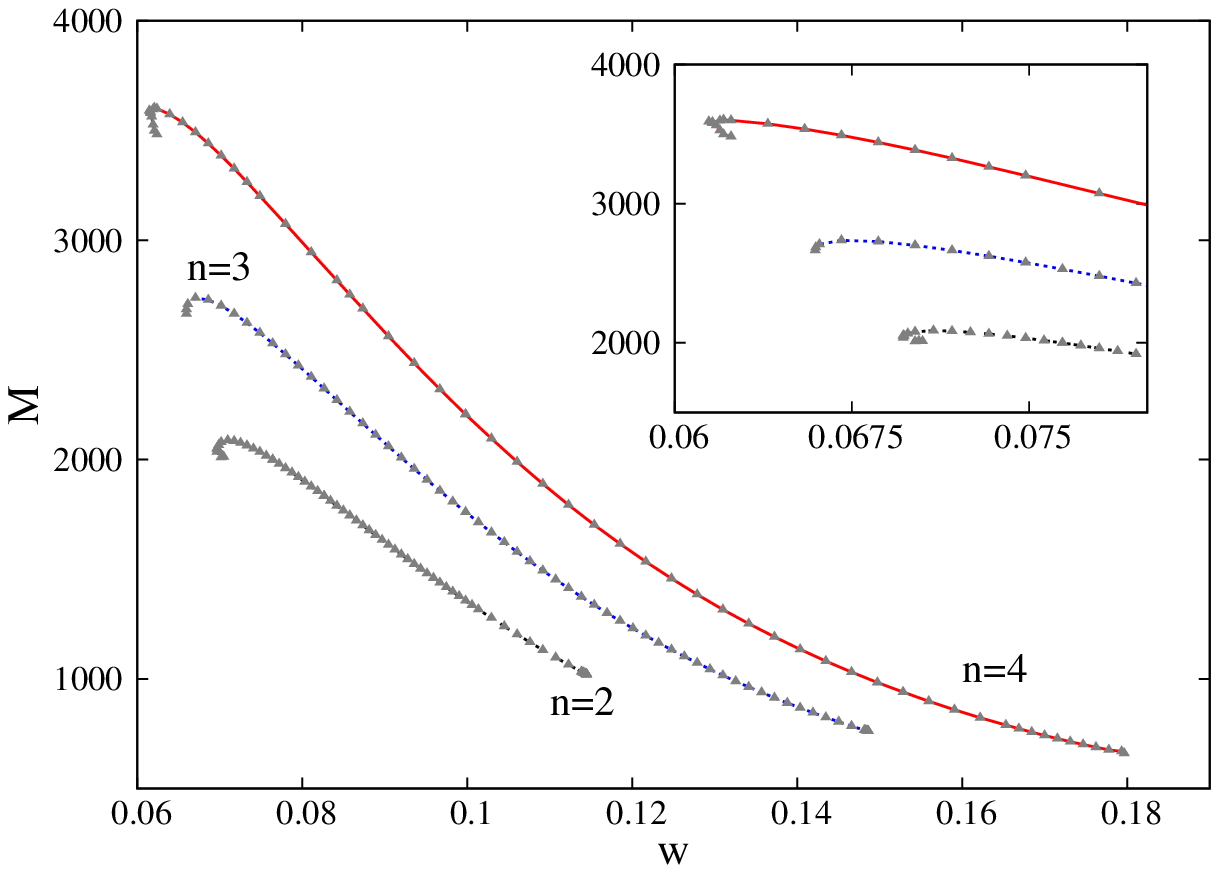,width=8cm}}
\put(8,0.0){\epsfig{file=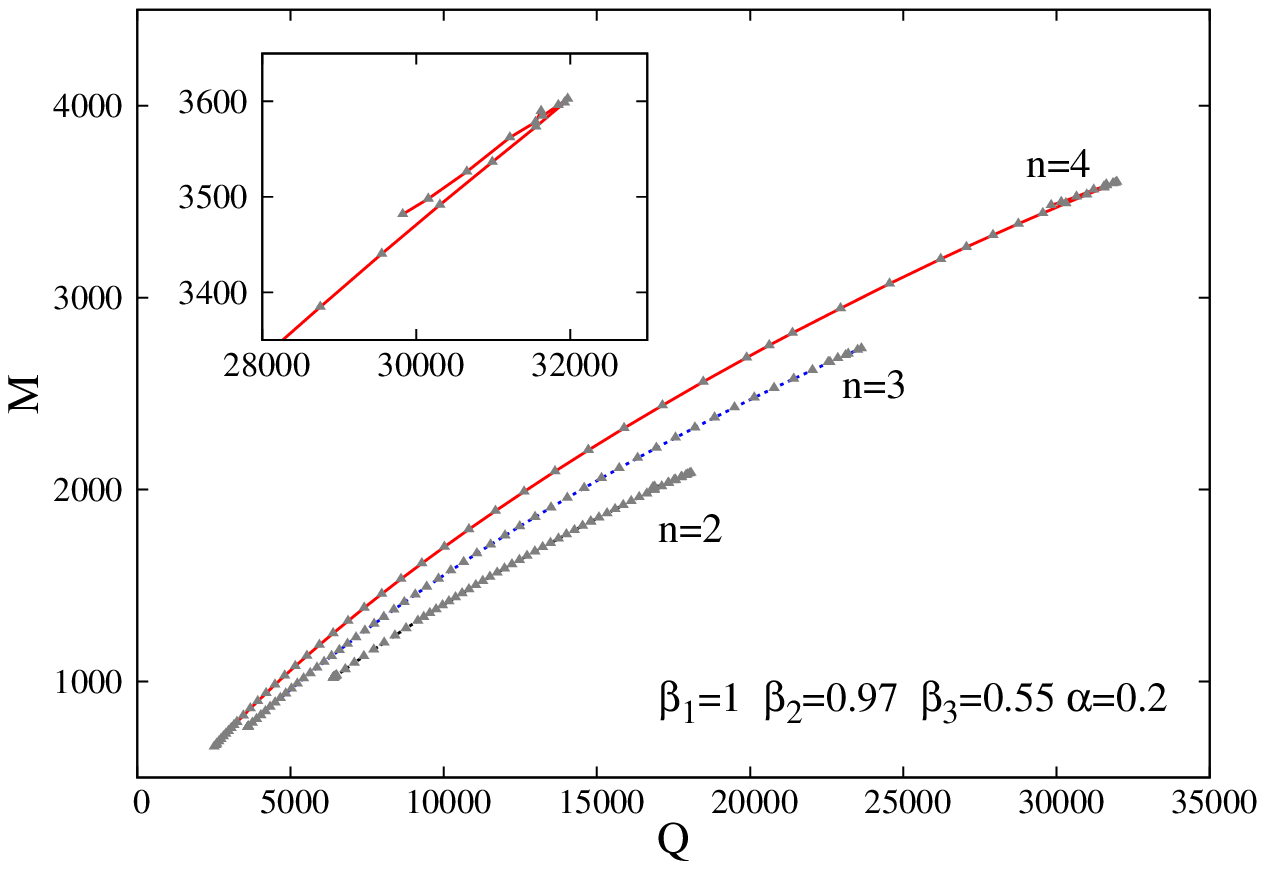,width=8cm}}
\end{picture}
\\
\\
{\small {\bf Figure 17.} {\it Left:} The mass  of gravitating $k=1$  vortons
is shown as a function of the frequency for several values of the winding number  $n$.
Other input parameters of these configurations are $\beta_1=1,~\beta_2=1,~\beta_3=0.54$
and $\alpha=0.2$. 
{\it Right:} The mass-charge diagram for the same solutions.
   }
\vspace{0.5cm}
\\ 
decouple, 
 the scaled matter fields being described by the Lagrangian                          
\begin{eqnarray} 
\label{scaled-Ls}
 L_s=\partial_\mu \hat X \partial^\mu \hat X + 
\partial_\mu \hat\sigma^\ast \partial^\mu \hat\sigma + M_\sigma^2|\hat\sigma|^2,
\end{eqnarray}
(with $\hat \sigma=\hat Z e^{i(n\varphi+w t)}$).
It is clear that $\hat X=0$ is the only physical solution, and we recover a boson star model
for the complex scalar field $\hat \sigma$ possessing a quadratic potential.
A similar behaviour has been found in 
\cite{Kleihaus:2005me} for gravitating $Q$-ball solutions
with higher order terms in the potential:
for large values of the coupling constant $\alpha$,
all higher order terms in the scalar field potential
become irrelevant.

\subsection{Gravitating $k=1$ vortons}

Again, these solutions  smoothly emerge from configurations in a fixed Minkowski background.
The backreaction is included by slowly increasing
the value of $\alpha$.

The 3D profiles of a typical solution are shown in Figure 15 as functions of the (scaled-)coordinates 
$\rho=r \sin \theta$ and $z=r \cos \theta$.
One can notice the nontrivial behaviour of the metric functions,
which strongly 
 \newpage
 \setlength{\unitlength}{1cm}
\begin{picture}(8,6)
\put(-1.,0.5){\epsfig{file=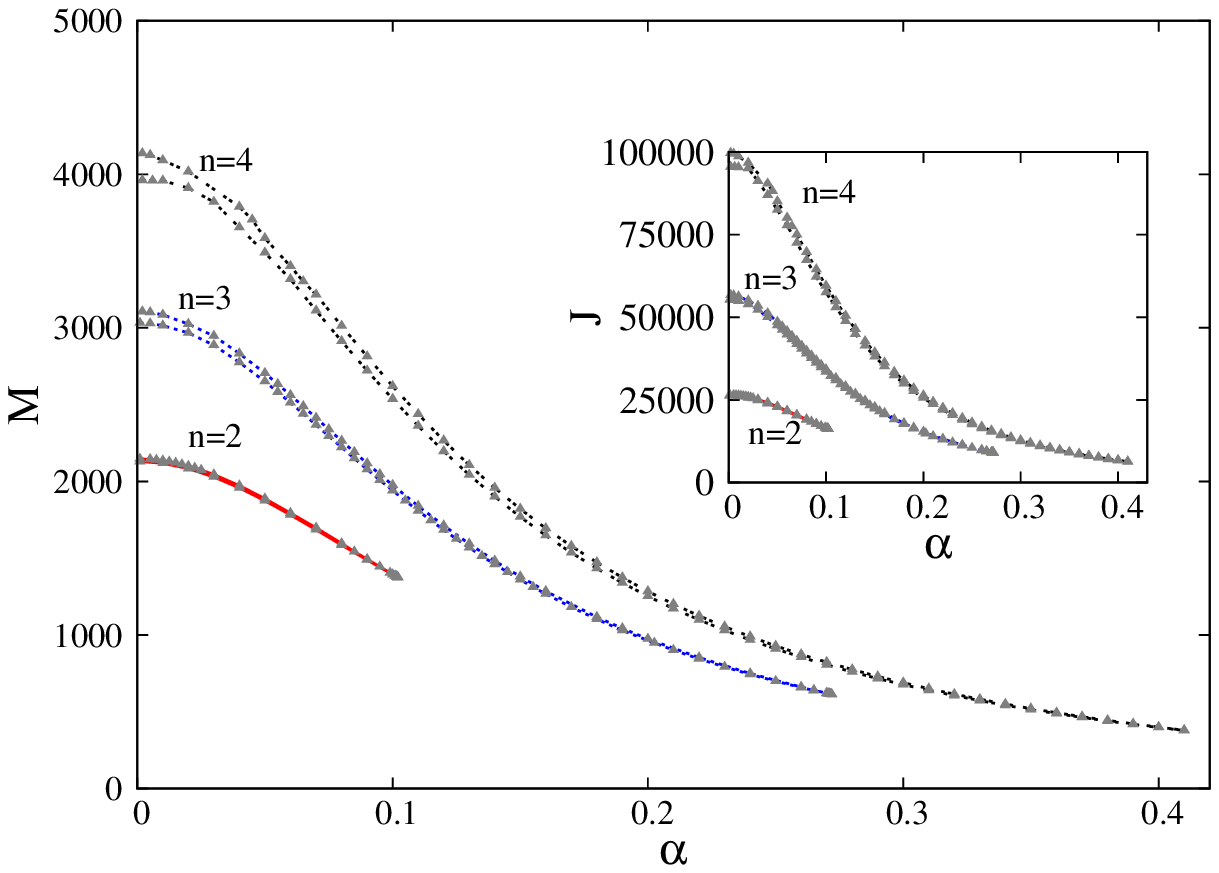,width=8cm}}
\put(7.5,0.5){\epsfig{file=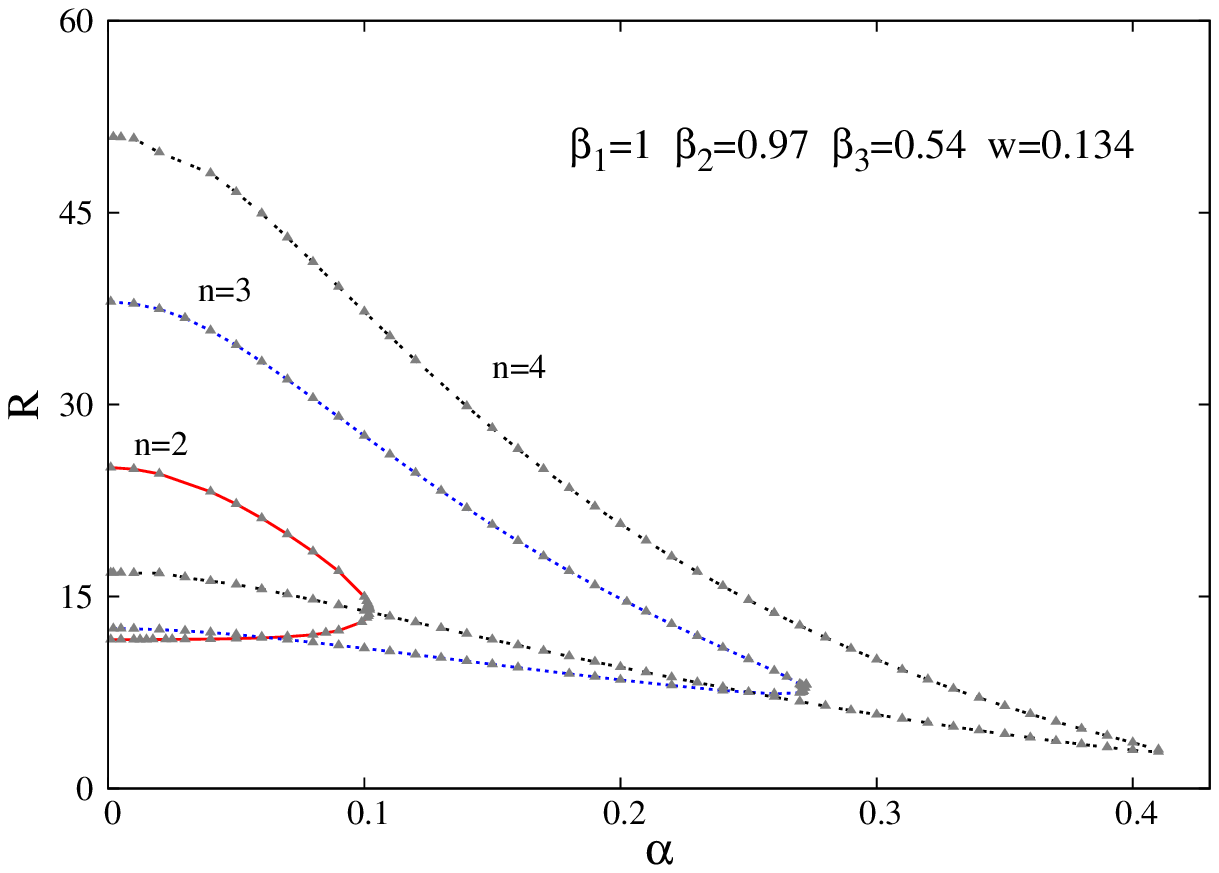,width=8cm}}
\put(-1.,-5.75){\epsfig{file=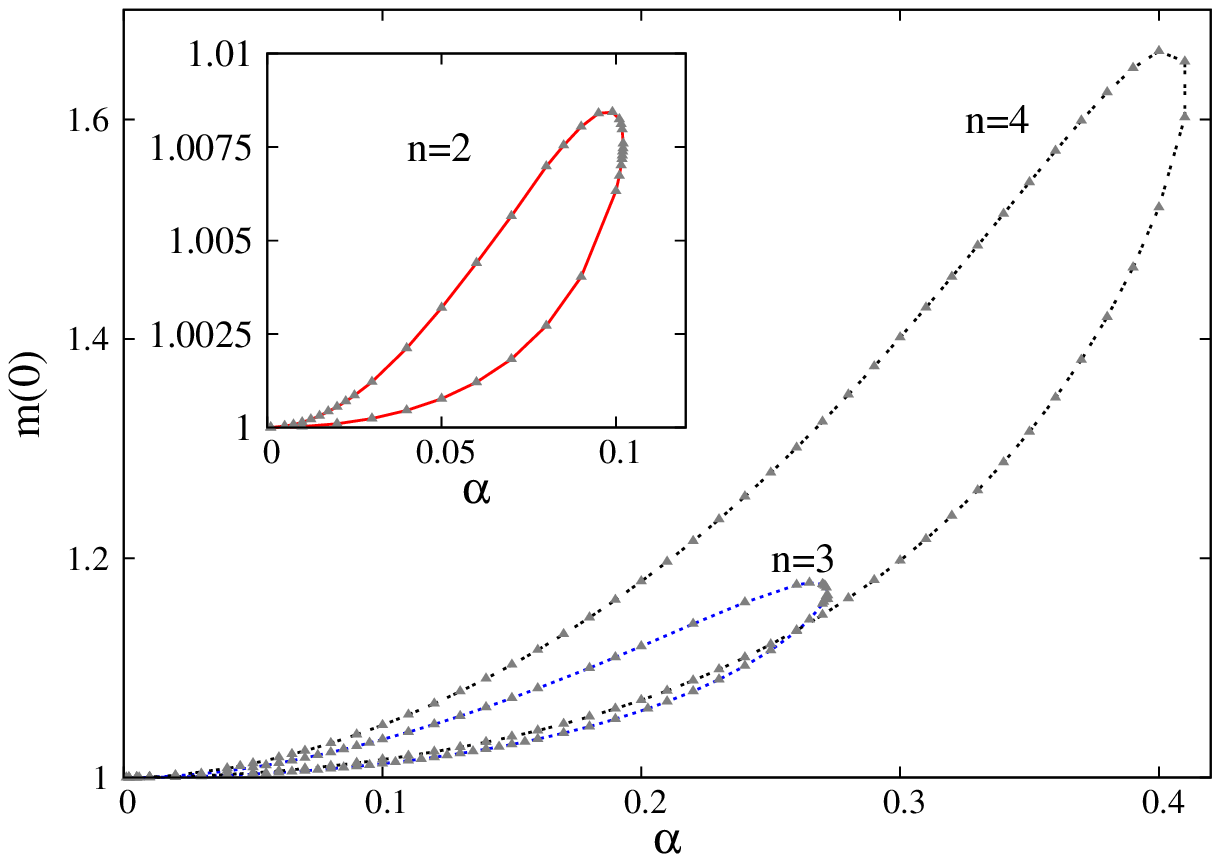,width=8cm}}
\put(7.5,-5.75){\epsfig{file=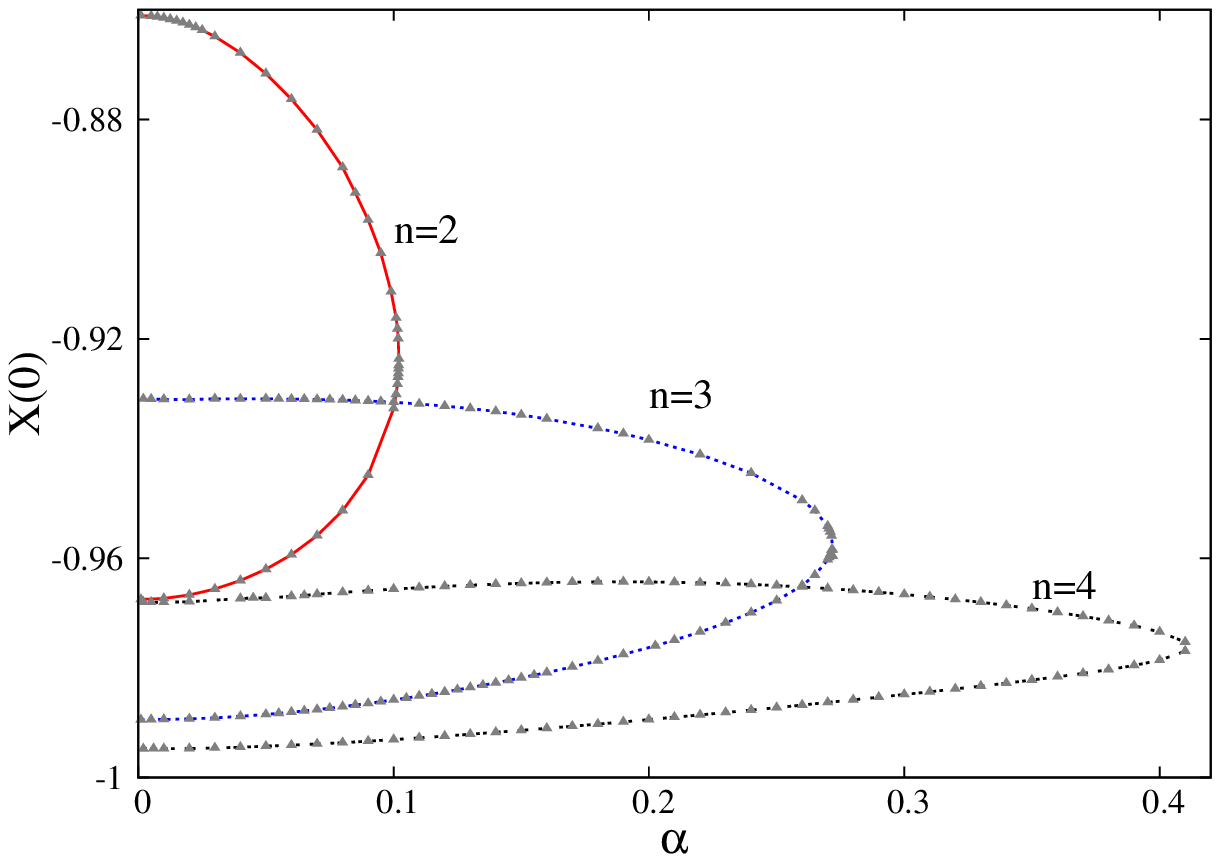,width=8cm}}
\end{picture}
\vspace*{5.75cm}
\\
\\
{\small {\bf Figure 18.}  The mass $M$, 
radius of the ring $R$ and
the values at the origin of the metric function $m$ and of the scalar function $X$  
are shown as functions of the parameter $\alpha$
 for  gravitating vorton solutions with $\beta_1=1,~\beta_2=0.97,~\beta_3=0.54$, $w=0.13$ and several values of the winding number $n$.
 One can notice the existence of two branches of solutions connecting different
 flat spacetime ($\alpha=0$) limiting configurations.   
} 
 \vspace{0.5cm} 
\\
 deviate from the flat spacetime values.
 The functions $f,l$ and $m$ approach their minimal  value in the equatorial plane at the location of
 the ring where $X=0$. 
For all solutions
studied so far, 
we did not notice a special behaviour for the functions  $X,Y,Z$
in the gravitating case as compared to the flat spacetime limit.

The same holds for the distribution of the energy
and angular momentum densities.
In particular, as seen in Figure 16,  
the vorton energy density still resembles a hollow tube. (Note, that the local peak 
at $r=0$ of the energy density
is much smaller for this $n=3$
solution than for the configuration with $n=2$
exhibited in Figure 8.) 

As expected, new effects occur when considering the vortons'
dependence on the frequency.
This is shown in Figure 17 for a  typical family of gravitating
solutions emerging from a flat spacetime fundamental configuration.
  One can see that some features found for $k=0$ `semitopological' solutions
are likely to hold also in this case.
For example, 
the range of $w$ is bounded,  
both $M$ and $J$ approaching a maximum close to a minimal value of the frequency $w_{min}$.
Furthermore, as for the $k=0$ solutions,
we notice the existence  at $w_{min}$ 
of a backbending
towards larger values of $w$,
leading apparently also to an inspiralling
of the solutions towards a limiting 
configuration.
Unfortunately, despite our efforts, numerical accuracy currently does not allow 
us to continue the construction of the (expected) spiral in a reliable way. 

This holds also for the issue of a maximal frequency,
 in which case we could not 
 construct the counterparts of the curves in Figure 13,
 with a fast decreasing mass and charge.
 The situation here is puzzling, since,
 for all considered sets of input parameters $(\beta_i;n;\alpha)$, 
 we could not construct $k=1$ gravitating vortons 
with frequencies below a critical value $w_c$.
 As $w\to w_c$, the solver diverges, or, more often,
provides a solution of the problem
 with $Y=0$. 
 At the same time, we do not notice any special features of the $k=1$ configurations close to $w_c$;
 in particular the metric functions do not exhibit any pathological behaviour.
 Interestingly, the $k=0$ configuration which is found in this way 
 possesses almost the same 
 mass and angular momentum as the $k=1$  vorton solution with $w\to w_c$
 (although other relevant parameters ($e.g.$
 the value of the metric function $f$ at the origin) are different).
 Then one may speculate that the $k=1$
 branch 
 would continue for $ w> w_c$  with a branch of  
 $k=0$ configurations towards $w_{max}$ given by (\ref{max-w}), 
 where the vacuum state is approached\footnote{Note also that, since $|\phi|$
 vanishes on a circle in the equatorial plane, 
the vortons cannot approach  in a continuous way the vacuum state with $X\equiv 1, Y\equiv 0$.}.
 However, due to limitations, resulting from our numerical treatment of the problem,
 we could not clarify this point.


A distinct feature of the $k=1$ gravitating vortons is their dependence on $\alpha$.
The emerging picture which 
results from the study of several different
sets of solutions
can be summarized as follows\footnote{However, we do
not exclude that a different picture may exist for
other sets of parameters, not considered in this work.}: 
given a fundamental
flat spacetime configuration with a set of input parameters $(\beta_i;n,w)$,
the curved spacetime configurations are found by increasing the parameter $\alpha$.
The emerging branch of solutions
exists up to a maximal value of $\alpha$ (which depends on the input parameters).

As $\alpha \to \alpha_{max}$ all relevant quantities stay finite.
(In particular, we do not see any sign for the emergence of an event horizon.)
At $\alpha_{max}$  a second branch of solutions is found, 
extending backwards in $\alpha$.
As $\alpha\to 0$, this branch approaches the excited flat spacetime vorton discussed in Section 3.3.
These features are illustrated in Figure 18.
(We mention that the ring radius $R$, the absolute value of the scalar function $X$
and the value of the metric function $m$
at the origin
are smaller for solutions on the excited branch.)

The fact that there is an upper bound on $\alpha$ for $k=1$
solutions can be understood as follows. A similar scaling to (\ref{scale-k0})
(this time also with $\hat Y=Y/\alpha$) shows that the limit   $\alpha \to \infty$ would imply necessarily  
$\hat  X \equiv 0$ (thus  $X \equiv 1$), $\hat  Y\equiv 0$.
However, these values cannot be continuously approached, since $|\phi|$
should always vanish on a circle in the equatorial plane.
In other words, since $k=0$ and $k=1$
correspond to different `topological' sectors of the theory, no continuous transition 
between them is possible\footnote{It may be interesting to note, that we could not construct solutions 
in which the function $Y$ takes very small values and can be considered as a perturbation around a configuration
featuring only the $X,Z$ functions.
Related to that, for all solutions with $Y\neq 0$, the function $X$ possesses a node in the equatorial plane.}.

As a result, the solutions exhibit instead a maximal value of the parameter $\alpha$, with a backbending
towards a second $\alpha=0$ solution in a flat spacetime background\footnote{We would like to mention
that the physical meaning of the second limiting solution is not obvious {\it apriori}.
Since
 $\alpha \to 0$
 can be reached also when taking $\eta_\phi \to 0$ (see (\ref{alpha})),
 the limiting solution might represent a scaled boson star in a model featuring only the scalar field
 $\sigma$ coupled to gravity.
 However, as seen in Figure 18, this is not the case,
 at least for the solutions in this work.}.

Thus we have reached the important conclusion\footnote{One should also mention that,
different from most of the other gravitating solutions
with a scalar field featuring a $v.e.v.$,
we did not find any indication for the existence of black hole solutions.
(Typically, they emerge from the globally regular configurations for a critical value of $\alpha$ \cite{Volkov:1998cc}.)
This property should be attributed to the Q-ball features of the vortons induced by the field $\sigma$.
In fact, for the simplest case $k=0,~n=0$,
 one can prove the absence of black hole solutions following the arguments in \cite{Pena:1997cy}.
A similar result may hold also in the spinning case.
Note, that this is not obvious, since the approach used in \cite{Pena:1997cy} 
cannot be easily generalized to the rotating case.
In fact, spinning black holes with complex scalar fields possesing a harmonic time dependence
 have been constructed recently in \cite{Dias:2011at} for a five dimensional anti-de Sitter 
 spacetime background.
 (They require a fine tuning between the frequency of the field and the event horizon velocity.)
} that the effects of gravity 
impose an upper bound on the $v.e.v.$ $\eta_\phi$ of the scalar $\phi$ 
for  solutions to exist.
This upper bound depends on the parameters of the model, with a general expression 
\be                                 \label{bound}
\alpha^2=\frac{16\pi}{\lambda_\phi}(\frac{M_\phi}{M_{Pl}})^2<c,
\ee
with $c$ a parameter of order one, whose precise value depends on the other input data.

Rotating objects in general relativity may possess ergoregions.
However, unlike the case of black holes, the
ergoregions of regular objects like those in this work would signal the presence of an instability
\cite{cardoso}.

Within the metric ansatz (\ref{metric}),
the ergosurface is defined by the condition
\begin{eqnarray}
\label{ergo-cond}
g_{tt}=-f+\frac{l}{f}W^2\sin^2 \theta=0.
\end{eqnarray}

\newpage
 \setlength{\unitlength}{1cm}
\begin{picture}(8,6)
\put(-1.5,0.0){\epsfig{file=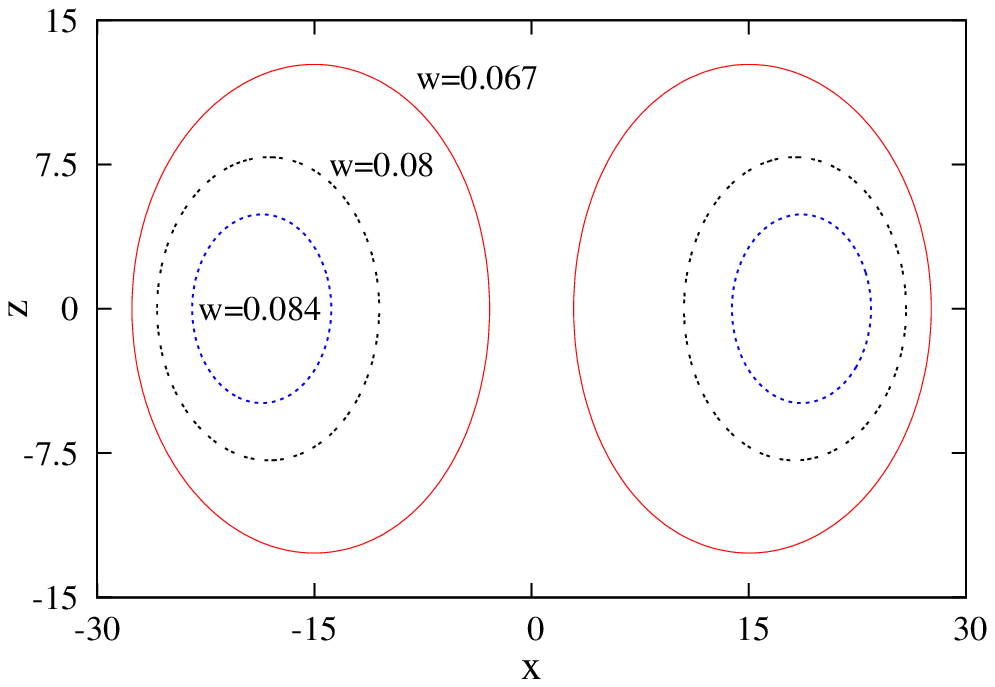,width=10cm}}
\put(7,0.0){\epsfig{file=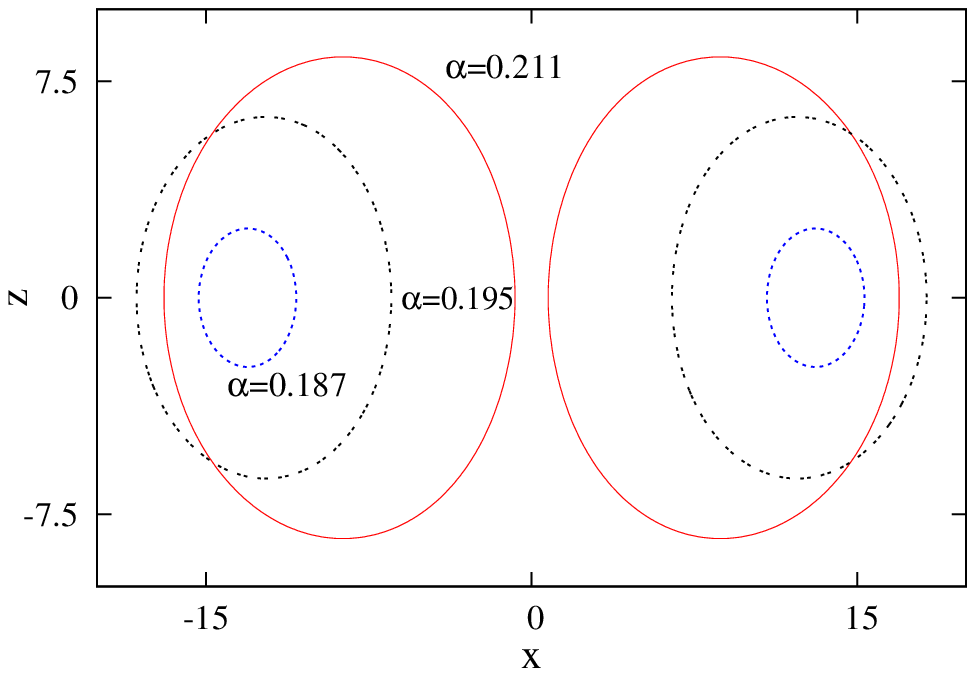,width=10cm}}
\end{picture}
\\
{\small {\bf Figure 19.} The location and shape of the vortons' ergoregion 
is exhibited for  solutions with three different frequencies (left) and three different values
of $\alpha$ (right).
The parameters of the potential are 
$\beta_1=0.14,~\beta_2=0.974,~\beta_3=0.54$ while  $\alpha=0.2$, $n=3$ (left)
and $w=0.07$, $n=2$ (right). } 
\vspace{0.5cm}
 
Then the ergoregion resides inside this ergosurface.
Examining the condition (\ref{ergo-cond}),
 we find that this is not satisfied  for  most of the gravitating vorton solutions.
For example, when varying the frequency, 
the ergoregion typically occurs for a set of solutions close to the backbending point at $w_{min}$,
in particular, for secondary branch configurations.
The solutions with large enough frequencies (in particular those close enough to $w_{max}$) 
have $g_{tt}<0$.
(This is again similar to the situation encountered for boson stars \cite{Kleihaus:2005me}.)

When studying instead the influence of the parameter $\alpha$,
we find that an ergoregion may occur for a set of solutions close to $\alpha_{max}$ (and with values of $w$
not too far from $w_{min}$).
That is, for the parameter choice in this work, the solutions
with small values of $\alpha$ (which are expected to be relevant in a realistic context)
do not exhibit an ergoregion.
 Some results displaying these features
are shown in Figure 19.
One can see that the vortons' ergoregions also possess a toroidal shape, whose size and location
depend on the input
 parameters.
 Without entering into details, we mention the existence of similar results for  $k=0$ 
`semitopological' vortons,  where some of these solutions  also possess  an ergoregion. 


We close this part by remarking that the di-vortons should also possess
gravitating generalizations.
However, due to the severe numerical
difficulties, we could construct such solutions for very small values of $\alpha$ only.
Their systematic study might possibly require a different numerical approach.

\section{Discussion and conclusions}

More than 20 years after vortons have been first proposed,
they continue to provide an interesting field of research.
Posing a severe numerical challenge, 
the systematic quantitative study of such configurations 
has started only recently 
\cite{Radu:2008pp,Battye:2008mm},
within the global version of Witten's 
$U(1)\times U(1)$ theory.

The first purpose of this work was to give an overview of the  properties of vortons
in a flat spacetime background.
The heuristic construction of these objects as reviewed in Section 3.1
suggests to view the usual vortons as
the second member ($k=1$) of 
a family of solutions indexed by the number of nodes $k$
of the vortex field $\phi$.
Then the case $k=0$
would correspond to `semitopological' vortons.
These solutions could be studied to a greater extent and 
seem to exhibit most of the basic properties of the usual vortons ($k=1$). 
A new result reported in Section 3.3 is the  nonuniqueness of the vortons  
 for the same input parameters.
 There we have given numerical evidence for the existence of 
 excited solutions, which play in important role when including
 the gravity effects.
Another new type of configuration  describing two concentric rings, the di-vorton, has been
reported in Section 3.4.
Here we remark that one expects that Witten's model 
 possesses a variety of other composite configurations.
 For example, it would be interesting to find a solution describing a `Saturn'
 $i.e.$
 a configuration with a central
 concentration of the energy density  surrounded
 by a ring (this would correspond to different zeros of the scalar field $\phi$).

Another aim of this work was to present a first discussion
of how gravity affects the properties of 
$k=0$ `semitopological' vortons and $k=1$ `true' vortons. 
Our results can be summarized as follows.
First,  we have shown that, when varying the frequency, 
gravity imposes an upper bound on the mass
and angular momentum of these objects.
The coupling to gravity gives rise to a spiral-like frequency 
dependence of the mass and charge.
Second, we have noticed a different behaviour of the $k=0$
and $k=1$ solutions as  functions of the parameter $\alpha$,
describing the coupling to gravity.
The $k=0$ solutions are essentially `dressed' Q-balls in a
more complicated version of the FLS
model \cite{Friedberg:1976me}. (This holds also in the flat spacetime limit.)
Different from true vortons, the limit $\alpha \to \infty$ is allowed in this case.
For $k=1$, one finds instead two branches of solutions joining at a maximal value of $\alpha$.
Finally, we have noticed that, similar to the boson star case,
vortons may possess an ergoregion, with the associated instability.
Here one should say that
a systematic study of the dependence of the properties of the vortons
on the various input parameters is still missing
and continues to present a numerical challenge.

As avenues for future research, perhaps the most
compelling task would be the study of
vortons in  models including gauge fields as well.
The simplest case here would be again the local version of  
 Witten's 
$U(1)\times U(1)$ theory.
The corresponding gauged vortons could be constructed by using a similar
approach to that described in this work and will 
be reported elsewhere.

However, vortons should also exist in other more complex models,
$e.g.$ possessing non-Abelian fields apart from scalars.
An interesting question to ask in this context is 
the possible existence of `springs' \cite{Peter:1993pz},
$i.e.$ vortex loops stabilized by  magnetic fields
instead of rotation\footnote{Note, that
the static dipole black rings can be balanced by immersing them in a background magnetic field
\cite{Ortaggio:2004kr}.}.
While `spring' solutions are unlikely to exist in  Witten's 
$U(1)\times U(1)$ theory,
such configurations have been found in other models featuring non-Abelian fields, see $e.g.$
\cite{Kleihaus:2003xz,Kleihaus:2008gn}.

Then
it would be interesting to reconsider some of the known toroidal solitons
based on the heuristic construction of a vorton/`spring' starting with pieces of strings.
For example, vortex ring solutions are known to exist in 
the Yang-Mills-Higgs model with the Higgs field in the adjoint representation \cite{Kleihaus:2003xz}.
However, these configurations 
seem to correspond\footnote{We thank F.~Navarro-L\'erida
and T.~Tchrakian for discussions on that and for 
sharing with us their unpublished results.} to loops made of Yang-Mills-Higgs
vortices discussed in \cite{NavarroLerida:2009dm}.
Finally, the possible existence of vortons and `springs' in the electroweak sector of the Standard Model
is perhaps the most exciting open problem (see 
\cite{Kleihaus:2008gn,Volkov:2006ug,Radu:2008ta,Kleihaus:2008cv}
for some work in this direction).

\vspace*{0.7cm}
\noindent{\textbf{~~~Acknowledgements.--~}} 
E. R. would like to thank M. S. Volkov for introducing him to the
subject of vortons and for fruitful collaboration on this topic.
We also thank  J. Garaud and M. S. Volkov for sending us their
unpublished results on global vortons.
We would like to thank B. Kleihaus and  T. Tchrakian
for discussions and remarks during various stages of this work.
 We gratefully acknowledge support by the DFG,
in particular, also within the DFG Research
Training Group 1620 ``Models of Gravity''.

\begin{appendix}
\setcounter{equation}{0}
\section{System of differential equations}

The metric functions $f,l,m$ and $W$
which enter the Ansatz (\ref{metric})
satisfy the following set of coupled non-linear 
partial differential equations:
\begin{eqnarray} 
\label{eqf}
&&
\nabla^2 f
-\frac{1}{f}(\nabla f)^2
+\frac{1}{2l}(\nabla f)\cdot (\nabla l)
-\frac{l}{f}r^2 \sin^2\theta (\nabla (\frac{W}{r}))^2
\\
\nonumber
&&{~~~~~~~~~~~~~~~~~~~~~~~~~~~~~~~~~~~~~~~~~~~~~~~~~~~~~~~~~}
+16 \pi G m
\bigg(
-\frac{2(w+n \frac{W}{r})^2Z^2}{f}+U(X,Y,Z)
\bigg)
=0,
\end{eqnarray} 
\begin{eqnarray} 
\\
\label{eql}
&&
\nabla^2 l
-\frac{1}{2l}(\nabla l)^2
+\frac{\partial_r l}{r}
+\frac{\cot \theta}{r^2}\partial_{\theta} l
+32 \pi G m
\bigg(
\frac{n^2 Z^2}{r^2 \sin^2 \theta}
-(w+\frac{nW}{r})^2\frac{lZ^2}{f^2}
 +\frac{l}{f}U(X,Y,Z)
  \bigg)=0,~~{~~}
\\
\label{eqm}
&&
\nabla^2 m+\frac{m}{2f^2}(\nabla f)^2
-\frac{1}{m}(\nabla m)^2
-\frac{3lm}{2f^2}r^2\sin^2 \theta (\nabla (\frac{W}{r}))^2
-\frac{\partial_r m}{r}
-\frac{\cot \theta}{r^2}\partial_\theta m
\\
\nonumber
&&
{~~~~~~~~~~~}+16 \pi G m
\bigg(
(\nabla X)^2+(\nabla Y)^2+(\nabla Z)^2
-(w+\frac{n W}{r})^2\frac{m}{f^2}Z^2
-\frac{ m}{l}\frac{n^2 Z^2}{r^2\sin^2 \theta}
+\frac{m}{f} U(X,Y,Z)
\bigg)
=0,
\\
\label{eqW}
 &&
 \nabla^2 W
 -\frac{2}{f}(\nabla f)\cdot (\nabla W)
 +\frac{3}{2l}(\nabla l)\cdot (\nabla W)
 +\frac{2}{r^2}
               \left(
               \partial_r W
               +\frac{\cot \theta}{r}\partial_\theta W
               -\frac{W}{r}
               \right)
\\
\nonumber
&&
{~~~~~~~~~~~~~~~~~~~~~~~~~~~~~~~~~~~~~~~~~~~~~~~~~~~~~~~~~~~~~~~~~~~~~~~~~~~~~}               
 -32 \pi G \frac{m}{l}\frac{nZ^2}{r^2 \sin^2 \theta}(w+\frac{nW}{r})=0.
\end{eqnarray}
They are equivalent to the following combinations of the Einstein equations:
$E_t^t-E_r^r-E_\theta^\theta-E_\varphi^\varphi+\frac{2W}{r}E_\varphi^t=0$ 
(for (\ref{eqf})), 
$E_r^r+E_\theta^\theta=0$ 
(for (\ref{eql})),  
$ E_\varphi^\varphi-\frac{2W}{r}E_\varphi^t=0$ 
(for (\ref{eqm})),
and
$E_\varphi^t=0$ 
(for (\ref{eqW})), 
multiplied with suitable factors ($e.g.$ $-m$ for  (\ref{eqf})).

The equations for the scalar functions result directly from (\ref{scalar-eqs}) and read: 
\begin{eqnarray}
\nonumber
&&
\nabla^2 X
+\frac{1}{2l}(\nabla l)\cdot (\nabla X)
+\frac{m}{f}X 
\left (
\frac{1}{2}\lambda_\phi (\eta_\phi^2-X^2-Y^2)-\gamma Z^2
\right)
=0,
\\
\label{eq2}
&&
\nabla^2 Y
+\frac{1}{2l}(\nabla l)\cdot (\nabla Y)
+\frac{m}{f}Y 
\left (
\frac{1}{2}\lambda_\phi (\eta_\phi^2-X^2-Y^2)-\gamma Z^2
\right)
=0,
\\
\nonumber
&&
\nabla^2 Z
+\frac{1}{2l}(\nabla l)\cdot (\nabla Z)
+\frac{m}{f}Z 
\left (
\frac{1}{2}\lambda_\sigma (\eta_\sigma^2-Z^2 )
-\gamma (X^2+Y^2)
-\frac{n^2}{r^2\sin^2 \theta}\frac{f}{l}
+\frac{(w+\frac{nW}{r})^2}{f}
\right)
=0.
\end{eqnarray}
In the above relations
we have defined
\begin{eqnarray}
&&
\nabla^2 A=\frac{1}{r^2}\frac{\partial }{\partial r}(r^2 \frac{\partial A}{\partial r})+
\frac{1}{r^2 \sin \theta}\frac{\partial }{\partial \theta}(\sin \theta \frac{\partial A}{\partial \theta} ),
\\
\nonumber
&&
(\nabla A)\cdot (\nabla B)= \frac{\partial A}{\partial r} \frac{\partial B}{\partial r}+ \frac{1}{r^2}\frac{\partial A}{\partial \theta} \frac{\partial A}{\partial \theta}.
\end{eqnarray}
Also, $U(X,Y,Z)$ is the potential of the scalar fields,
which results directly from (\ref{potential}) after substituting the scalar field ansatz (\ref{scalar-ansatz})
\begin{eqnarray} 
\label{U-XYZ}
U(X,Y,Z)=\frac{1}{4}\lambda_\phi(X^2+Y^2-\eta_\phi^2)^2
             +\frac{1}{4}\lambda_\sigma Z^2(Z^2-2\eta_\sigma^2)
             +\gamma(X^2+Y^2)Z^2.
\end{eqnarray}

These equations are supplemented with the following constraints
\begin{eqnarray}
\label{constr1}
&&
\frac{1}{l}\left(\partial_{r}^2l-\frac{1}{r^2}\partial_{\theta}^2l \right)
-\frac{1}{lm}\left(\partial_{r} l \partial_{r} m-\frac{1}{r^2} \partial_{\theta} l \partial_{\theta} m \right)
-\frac{1}{2l^2}\left( (\partial_{r} l)^2 -\frac{1}{r^2} (\partial_{\theta} l)^2 \right)
+\frac{1}{f^2}\left( (\partial_{r} f)^2 -\frac{1}{r^2} (\partial_{\theta} f)^2 \right)
\\
\nonumber
&&
-\frac{l\sin^2\theta}{f^2}\left( (\partial_{r} W)^2 -\frac{1}{r^2} (\partial_{\theta} W)^2 \right)
+\frac{1}{rl}\left (\partial_{r} l-\frac{2\cot \theta }{r}\partial_{\theta} l \right)
-\frac{2}{rm}\left (\partial_{r} m-\frac{\cot \theta }{r}\partial_{\theta} m \right)
+\frac{W l\sin^2 \theta }{f^2 r^2}(-W+2 r \partial_{r} W)
\\
\nonumber
&&
{~~~~~~~~~}
+32 \pi G 
     \left(
(\partial_{r} X)^2+(\partial_{r} Y)^2+(\partial_{r} Z)^2
-\frac{1}{r^2}((\partial_{\theta} X)^2+(\partial_{\theta} Y)^2+(\partial_{\theta} Z)^2) 
   \right )
=0,
\\
\label{constr2}
&&
\frac{1}{l}\partial_{r \theta }l
+\frac{\partial_{r} f\partial_{\theta}f}{f^2}
-\frac{\partial_{r} l\partial_{\theta}l}{2l^2}
-\frac{l\sin^2 \theta}{f^2}\partial_{r} W\partial_{\theta}W
-\frac{1}{2lm}
         \left( \partial_{r}m \partial_{\theta} l+\partial_{r}l \partial_{\theta} m
          \right)
 -\cot \theta \frac{\partial_{r}m }{m}
  - \frac{\partial_{\theta}m }{ r m}     
  \\
\nonumber
&&
+\cot \theta \frac{\partial_{r}l }{l}  
+\frac{Wl\sin^2 \theta }{r f^2}\partial_{\theta}W
+32 \pi G 
\left(
 \partial_{r}X \partial_{\theta} X+\partial_{r}Y \partial_{\theta} Y+\partial_{r}Z \partial_{\theta}Z
      \right)
      =0,
\end{eqnarray}
which result from the Einstein equations $E_r^r-E_\theta^\theta=0$ for (\ref{constr1}),  
and $E_r^\theta=0$ for (\ref{constr2})).

\section{The nonvanishing components of the energy momentum tensor}
\setcounter{equation}{0}

The nonvanishing components of the  energy-momentum tensor of  gravitating vortons
are found by substituting the ansatz (\ref{scalar-ansatz}) in the general expression (\ref{Tik}) 
and read
\begin{eqnarray} 
\label{Tik1}
\nonumber
&&
 T_r^r=
 (w+\frac{nW}{r})^2\frac{Z^2}{f}
 -U(X,Y,Z)
 -\frac{1}{r^2\sin^2 \theta}\frac{n^2f }{l}Z^2
 \\
 \nonumber
 &&
 {~~~~~~~~~~}+\frac{f}{m}
 \left [
    (\partial_{r} X)^2+  (\partial_{r} Y)^2+  (\partial_{r} Z)^2
      -\frac{1}{r^2}\left( (\partial_{\theta}X)^2 +(\partial_{\theta}Y)^2 +(\partial_{\theta}Z)^2 \right)
 \right],
 \\
 &&
 T_r^\theta=\frac{2f}{r^2m}
 \left (
 \partial_{r}X \partial_{\theta}X+ \partial_{r}Y \partial_{\theta}Y +\partial_{r}Z \partial_{\theta}Z
 \right),
 \\
 \nonumber
 &&
  T_\theta^\theta=
 (w+\frac{nW}{r})^2\frac{Z^2}{f}
 -U(X,Y,Z)
 -\frac{1}{r^2\sin^2 \theta}\frac{n^2f }{l}Z^2
 \\
 \nonumber
 &&
 {~~~~~~~~~~}-\frac{f}{m}
 \left [
    (\partial_{r} X)^2+  (\partial_{r} Y)^2+  (\partial_{r} Z)^2
      -\frac{1}{r^2}\left( (\partial_{\theta}X)^2 +(\partial_{\theta}Y)^2 +(\partial_{\theta}Z)^2 \right)
 \right],
  \\
 \nonumber
 &&
 T_\varphi^\varphi=
 (w^2-\frac{n^2 W^2}{r^2})\frac{Z^2}{f}
 -U(X,Y,Z)
 + \frac{1}{r^2\sin^2 \theta}\frac{n^2f }{l}Z^2
 -\frac{f}{m}\left((\nabla X)^2+(\nabla Y)^2+(\nabla Z)^2 \right),
\end{eqnarray}
and finally, the most important components are
\begin{eqnarray} 
\nonumber
&&
-T^t_t=
\frac{f}{m}
\left(
(\nabla X)^2+(\nabla Y)^2+(\nabla Z)^2 
\right)
+U(X,Y,Z)+\frac{1}{r^2\sin^2 \theta}\frac{n^2f }{l}Z^2
+(w^2-\frac{n^2}{r^2}W^2)\frac{Z^2}{f},
\\
\label{Tik-2}
&&
T^t_\varphi=2n(w+\frac{nW}{r})\frac{Z^2}{f}~,
\end{eqnarray}
corresponding to the  energy density and the angular momentum density, respectively.

\section{Spherically symmetric solutions}
\setcounter{equation}{0}

These solutions occur as a limiting case of the general 
configurations.
Their line element is found from (\ref{metric})
for $l=m$ and $W=0$,  and reads
\begin{eqnarray} 
\label{metric-sph}
ds^2=\frac{m}{f}\left(dr^2+r^2(d\theta^2+\sin^2\theta d\varphi^2) \right)-f dt^2,
\end{eqnarray}
the metric functions $f$ and $m$ depending only on the radial coordinate $r$.
The scalar field ansatz results from  (\ref{scalar-ansatz})
for $n=0$ and reads
\begin{eqnarray} 
\label{s-sph}
\phi=X(r), ~~\sigma=Z(r)e^{i w t}.
\end{eqnarray}
The equations greatly simplify in this case.
From (\ref{eq2}), 
(\ref{eqf}), 
(\ref{eqm}) one finds
\begin{eqnarray} 
\label{XZ-eq1}
X''+(\frac{m'}{2m}+\frac{2}{r})X'-\gamma\frac{m}{f}XZ^2+\frac{\lambda_\phi}{2}\frac{m}{f}X(\eta_\phi^2-X^2)=0,
\\
\nonumber
Z''+(\frac{m'}{2m}+\frac{2}{r})Z'+\frac{w^2m}{f^2}Z
-\gamma\frac{m}{f}X^2 Z  +\frac{\lambda_\sigma}{2} \frac{m}{f}Z(\eta_\sigma^2-Z^2)=0,
\end{eqnarray}
for the scalar fields, and
\begin{eqnarray}
f''+(\frac{m'}{2m}+\frac{2}{r}-\frac{f'}{f})f'
+16\pi G m
\left(
-\frac{2w^2Z^2}{f}+U(X,Z)
\right)
=0,
\\
\nonumber
m''
+\frac{m'}{r}
-\frac{m'^2}{m}
+\frac{mf'^2}{2f^2}
+16\pi G m
\left (
 X'^2+Z'^2 
-\frac{w^2m Z^2}{f^2}+\frac{m}{f} U(X,Z)
\right )=0,
\end{eqnarray}
for the metric functions
(with $U(X,Z)$ given by (\ref{U-XYZ}) with $Y=0$),
together with the constraint (which results from (\ref{constr1}))
\begin{eqnarray}
\label{constr-XZ}
\frac{m''}{m}
+\frac{f'^2}{f^2}
-\frac{m'}{rm}
-\frac{3m'^2}{2m^2}
+32\pi G (X'^2+Z'^2)=0.
\end{eqnarray}
The approximate form of the solutions as $r\to 0$
reads (note that $Z$ does not vanish there)
\begin{eqnarray}
\label{XZ-r0}
&&
f(r)=f_0+f_2  r^2+O(r^4),~~m(r)=m_0+m_2  r^2+O(r^4)~,
\\
&&
X(r)=x_0+x_2  r^2+O(r^4),~~Z(r)=z_0+z_2  r^2+O(r^4)~.
\end{eqnarray}
All coefficients in this expansion are fixed by the 
parameters $f_0$, $m_0$, $x_0$ and $z_0$ (one finds $e.g.$
$
x_2=
\frac{m_0 x_0}{6 f_0}
\left (
\frac{\lambda_\phi}{2}(x_0^2-\eta_\phi^2)+\gamma z_0^2
\right )
$
and
$
z_2=
\frac{m_0 z_0}{6 f_0^2}
\left (
-w^2
+f_0(\gamma x_0^2- (\frac{\lambda_\sigma}{2}(z_0^2-\eta_\sigma^2)
\right )
$).
The large-$r$ expression of the scalar functions $X,Z$
results from (\ref{asympt}) by taking $A=0$.
The leading order terms in the corresponding expansion of the metric function are
\begin{eqnarray}
\label{XZ-inf}
&&
f(r)=1+\frac{f_1}{r}+\frac{f_1^2}{2r^2}+\dots,
~~
m(r)=1-\frac{f_1^2}{8r^2}+\dots, 
\end{eqnarray}
with $f_1$ a parameter fixing the ADM mass of the solutions.

Some basic properties of these configurations are rather similar 
to those found for spinning $k=0$ solutions.
In particular, we have found the same qualitative dependence on the parameters $w$, $\alpha$
as in that case.

\end{appendix}

\begin{small}

 \end{small}
 
 \end{document}